\documentclass[english,12pt]{article}
\usepackage{lmodern}

\usepackage[T1]{fontenc}
\usepackage[latin9]{inputenc}
\usepackage{amsmath}
\usepackage{amssymb}
\usepackage{graphicx}

\makeatletter

\newcommand{\lyxaddress}[1]{
	\par {\raggedright #1
	\vspace{1.4em}
	\noindent\par}
}

\usepackage{imakeidx}
\makeindex[program=makeindex,columns=2,intoc=true,options={-s index_style.ist}]

\usepackage{cancel}
\usepackage{mathrsfs}  
\usepackage{slashed}   
\usepackage{bbold}
\usepackage{url}
\usepackage{graphicx}
\usepackage[colorlinks=true,linkcolor=redLinks,citecolor=greenLinks,urlcolor=redLinks, pdfborder={0 0 1}]{hyperref}
\usepackage{xcolor}
\usepackage[numbers,sort&compress]{natbib}
\usepackage{ytableau}
\usepackage{microtype}
\usepackage[htt]{hyphenat}
\usepackage{titlesec}

\allowdisplaybreaks

\colorlet{shadecolor}{gray!15}

\definecolor{greenLinks}{rgb}{0, 0.6, 0} 
\definecolor{blueLinks}{rgb}{0, 0, 0.6}
\definecolor{redLinks}{rgb}{0.6, 0, 0}
\definecolor{tempText}{rgb}{0.55, 0.10,0.67}
\definecolor{eprintLinks}{rgb}{0.4, 0.4, 0.4}
\definecolor{journalLinks}{rgb}{0.6, 0, 0}

\titleformat{\section}[block]{\color{black}\Large\bfseries\filcenter}{\thetitle.\;}{0em}{}
\titleformat{\subsection}[block]{\color{black}\large\bfseries\filcenter}{}{0em}{}

\newcommand{\MYhref}[3][redLinks]{\href{#2}{\color{#1}{#3}}}%

\usepackage{multirow}
\let\orig@Hy@EveryPageAnchor\Hy@EveryPageAnchor
\def\Hy@EveryPageAnchor{%
    \begingroup
    \hypersetup{pdfview=Fit}%
    \orig@Hy@EveryPageAnchor
    \endgroup
}

\let\oldFootnote\footnote
\newcommand\nextToken\relax

\renewcommand\footnote[1]{%
    \oldFootnote{#1}\futurelet\nextToken\isFootnote}

\newcommand\isFootnote{%
    \ifx\footnote\nextToken\textsuperscript{,}\fi}

\usepackage[most]{tcolorbox} 

\definecolor{block-gray}{gray}{0.95}
\definecolor{darkRed}{rgb}{0.67, 0, 0}

\newtcolorbox{codeExample}{
    enhanced,
    frame hidden,
    colback=block-gray,
    boxrule=0pt,
    borderline west={2pt}{0pt}{gray!80!black}
}

\newtcolorbox{codeSyntax}{
	collower=black,
	bicolor,
	colback=gray!80!black,
	colupper=white,
	colbacklower=block-gray,
	colframe=gray!80!black,
	boxrule=2pt,
	fontupper=\ttfamily\bfseries,
	sharp corners=all
    }

\usepackage[scale=1.0]{inconsolata}

\makeatother

\usepackage{babel}
\begin{document}
\title{GroupMath: A Mathematica package for group theory calculations}
\author{Renato M. Fonseca\date{}}
\maketitle

\lyxaddress{\begin{center}
{\Large{}\vspace{-0.5cm}}Institute of Particle and Nuclear Physics\\
Faculty of Mathematics and Physics, Charles University,\\
V Holešovi\v{c}kách 2, 18000 Prague 8, Czech Republic\\
~\\
Email: fonseca@ipnp.mff.cuni.cz
\par\end{center}}
\begin{abstract}
\texttt{GroupMath} is a Mathematica package which performs several
calculations related to semi-simple Lie algebras and the permutation
groups, both of which are important in particle physics as well as
in other areas of research.

\tableofcontents{}
\end{abstract}

\section{Introduction}

Calculations involving Lie algebras and their representations are
ubiquitous in several fields of science, including high energy physics.
It is therefore worth having tools that perform this kind of computations
automatically, with fast and reliable results. That is the goal of
\texttt{GroupMath}, the Mathematica package presented in this document.

This software has its origins in another piece of code --- \texttt{Susyno}
\cite{Fonseca:2011sy} --- whose main purpose has no obvious connection
with group theory. Indeed the aim of this latter program is to calculate
the renormalization group equations for supersymmetric models. Doing
so requires generating the superpotential and the soft-supersymmetric
breaking potential in a fully expanded form, and for the user's convenience
the program calculates these quantities automatically from the model's
defining properties: its gauge symmetry group and the superfields
transforming as representations of that group. Several functions associated
to Lie algebras were created for this purpose, constituting a significant
part of the whole program.

Given that it can be useful by itself, from the very beginning the
group theory code of \texttt{Susyno} was made directly accessible
to the user, and eventually it was used in other programs (\texttt{SARAH}
\cite{Staub:2013tta}, \texttt{PyR@TE} \cite{Lyonnet:2013dna,Sartore:2020gou}
and \texttt{Sym2Int} \cite{Fonseca:2017lem,Fonseca:2019yya}). However,
over time and with successive increments to this part of the program,
it became clear that it would be best to have a separate program for
this Lie algebra code. Hence \texttt{GroupMath} was created.

The finite permutation groups $S_{n}$ are also useful to our understanding
of the representations of Lie algebras (see for example \cite{Fonseca:2019yya}),
hence \texttt{GroupMath} contains several function related to these
groups and their representations. It is also conceivable that future
versions of the program might contain code related to other finite
groups.

This is by no means the only available tool for performing group
theory calculations. A likely incomplete list of other programs includes
\texttt{LiE} \cite{LieProgram}, \texttt{GAP} \cite{GAP4}, \texttt{Affine.m}
\cite{Nazarov:2011mv} and \texttt{LieART} \cite{Feger:2012bs,Feger:2019tvk},
the latter two being Mathematica packages just like \texttt{GroupMath}.
Furthermore, the reader also has at its disposal several references
containing a large amount of pre-computed Lie algebra data \cite{Slansky:1981yr,Yamatsu:2015npn}.

The rest of this work describes the program and how to use it. The
available functions are also documented in the program's built-in
documentation, which can be accessed directly from within Mathematica.
Additionally, usage of the code requires some knowledge of Lie algebras
(but not much), therefore a brief review of this topic is included
in appendix. It contains a minimalist explanation of what are simple
algebras, Cartan matrices, Dynkin coefficients, weights and others
important concepts (further clarifications are provided directly in
the main text when each function is presented). More details can be
found in Lie algebra textbooks \cite{Slansky:1981yr,Cahn:1985wk,Hamermesh:1999,Georgi:1999,Fuchs:2003}.
It is perhaps appropriate to warn the reader that, for convenience,
Lie algebras are also referred to as Lie groups in the remainder of
the text, even though strictly speaking they are not the same thing.
Furthermore, a representation is a map from a group's elements to
some vector space, but it is also very convenient to call the vector
space itself a representation (as physicists usually do).

\section{Installation}

An up-to-date version of the program is available at
\begin{center}
\texttt{\href{https://renatofonseca.net/groupmath}{renatofonseca.net/groupmath}}
\par\end{center}

\noindent The downloaded folder \textit{GroupMath}, once unpacked,
must be placed in the \textit{Applications} subfolder of one of the two special directories defined by the Mathematica
variables \texttt{\$BaseDirectory} and \texttt{\$UserBaseDirectory}.
Normally these correspond to the paths

~

\noindent \textit{/usr/share/Mathematica/Applications}

\noindent \textit{/home/(computer name)/.Mathematica/Applications}

~

\noindent on Linux systems, and

~

\noindent \textit{C:\textbackslash ProgramData\textbackslash Mathematica\textbackslash Applications}

\noindent \textit{C:\textbackslash Users\textbackslash (computer
name)\textbackslash AppData\textbackslash Roaming\textbackslash Mathematica\textbackslash Applications}

~

\noindent on Windows machines. After placing the folder \textit{GroupMath} in one of
these locations, the program is loaded by typing
\noindent \begin{center}
\texttt{}<\hspace{0mm}<\texttt{GroupMath$\grave{\,\textrm{\,}}$}
\par\end{center}

\noindent in Mathematica's front end.

\section{Lie groups}

The format used by the program for Lie groups and its representations
will now be explained. The various computations which can be performed\texttt{}
with them are described afterwards.

\subsection{\label{subsec:LieGroups}Working with Lie groups}

In order to specify a simple Lie group, its Cartan matrix much be
provided. The user can write these matrices, but this is neither recommended
nor necessary, since they have already been associated to the variables
\texttt{SU2}, \texttt{SU3}, ..., \texttt{SU32}, \texttt{SO3}, \texttt{SO5},
\texttt{SO6}, ..., \texttt{SO32}, \texttt{SP4}, \texttt{SP6}, ...,
\texttt{SP32}, \texttt{E6}, \texttt{E7}, \texttt{E8}, \texttt{F4}
and \texttt{G2}. For example:

\begin{codeExample}

\hspace{-10mm}\includegraphics[scale=0.62]{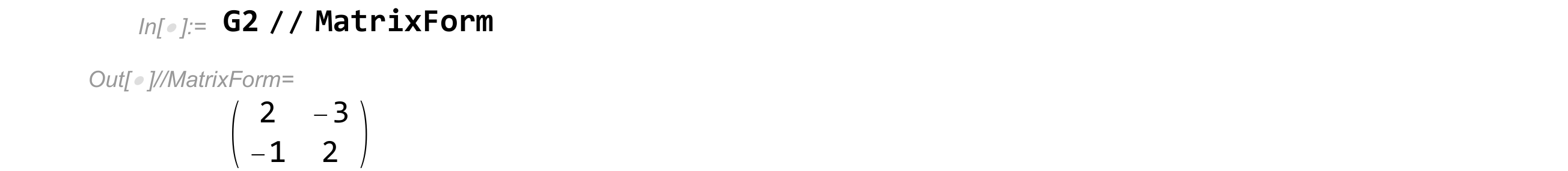}

\end{codeExample}

For even larger groups of the $SU(n)$, $SO(n)$ and $Sp(2n)$ families,
there is a function which will return the corresponding Cartan Matrix
for any $n$:\index{CartanMatrix}

\begin{codeSyntax}
CartanMatrix[<"group family name">,<family index>]
\tcblower
Returns the Cartan matrix of a simple group, given its name
\end{codeSyntax}

The $U(1)$ group is associated with an empty list \texttt{\{\}},
and for convenience this value has already been assigned to the variable
\texttt{U1}.

\begin{codeExample}

\includegraphics[scale=0.62]{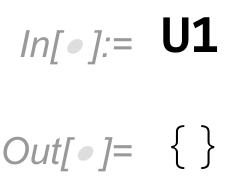}

\end{codeExample}

Several of the program's function also work with semi-simple and reductive
Lie groups, which are products of simple Lie groups and, perhaps,
$U(1)$'s. In this case, each of the factor groups should be indicated
in a list: for example \texttt{\{SU3,SU2,U1\}} represents the group
$SU(3)\times SU(2)\times U(1)$.

The format just mentioned applies not only to input but also to the
output of some functions (for instance when calculating maximal subgroups).
This makes it easy to chain several functions together. However, the
Cartan matrix notation is not ideal for human reading: it is not immediately
obvious that \texttt{\{\{\{2,-1\},\{-1,2\}\},\{\{2\}\},\{\}\}} stands
for the group $SU(3)\times SU(2)\times U(1)$. Therefore, in some
circumstances it is useful to have a function which does the reverse
of \texttt{CartanMatrix}:\index{CMtoName}\begin{codeSyntax}
CMtoName[<group>]
\tcblower
Converts a group, given in the Cartan Matrix notation, to a name
\end{codeSyntax}

\begin{codeExample}

\includegraphics[scale=0.62]{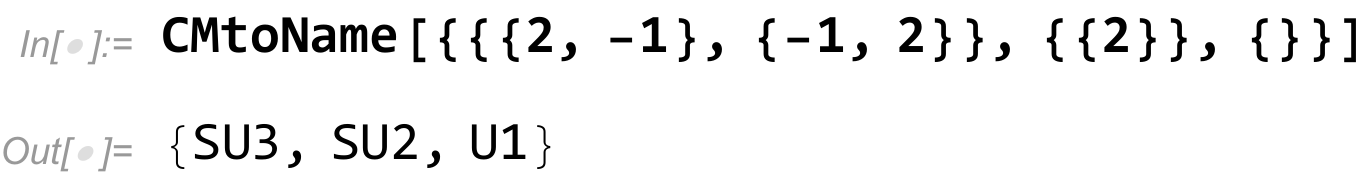}

\end{codeExample}The output is a string (or list of strings in the
case of non-simple groups).

\subsection{\label{subsec:Representations}Working with representations of Lie
groups}

An irreducible representation of a simple Lie group is specified by
its Dynkin coefficients, which are lists of $m$ non-negative integers,
where $m$ equals the group's rank. Therefore, \texttt{\{1,2,0,0\}}
is an example of a valid $SU(5)$ representation, since this group
has rank 4. The representation of a $U(1)$ is specified by a charge,
which is just a real number.

In the case of reductive Lie groups, a list of the representations
of each of the factor groups is required, so \texttt{\{\{1,2,0,0\},3,\{1\}\}}
is a valid $SU(5)\times U(1)\times SU(2)$ representation.

Just like \texttt{CMtoName} for groups, it is possible to convert
a representation to a name:\index{RepName}\begin{codeSyntax}
RepName[<group>,<representation>]
\tcblower
Computes a string-like name of a Lie group representation.
\end{codeSyntax}

\begin{codeExample}

\includegraphics[scale=0.62]{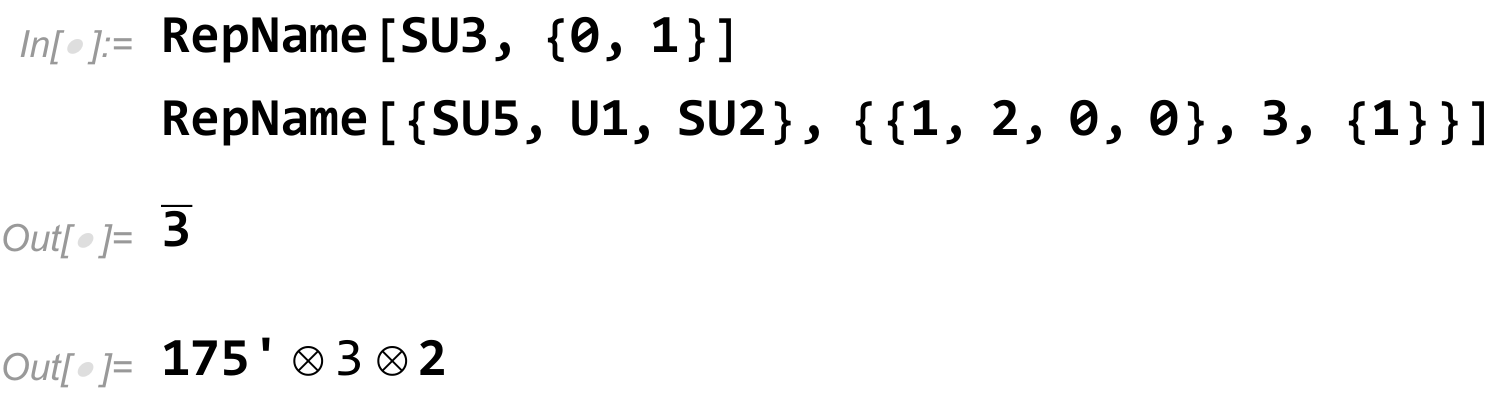}

\end{codeExample}

This is such a useful conversion that several \texttt{GroupMath} functions
containing Lie group representations in the output have an option
\texttt{UseName} which by default is \texttt{False} but, when set
to \texttt{True}, it automatically converts Dynkin coefficients into
names (several examples will be provided later). The convention used
to name representations is the one in \cite{Slansky:1981yr,Feger:2012bs},
with two caveats: \texttt{GroupMath} associates the Dynkin indices
\texttt{\{0,2\}} to the $SU(3)$ sextet $\boldsymbol{6}$ (not \texttt{\{2,0\}}),
and \texttt{\{0,0,0,2,0\}}=$\boldsymbol{126}$ for the $SO(10)$ group
(not \texttt{\{0,0,0,0,2\}}).

Having said this, in many cases there is an easier way to specify
a representation $R$ of dimension $d$. If the name of $R$ does
not contain primes nor bars, then the user may provide only its dimension
$d$; if the name of $R$ does contain a bar (but no primes), then
it is associated to the number $-d$, where $d$ is again the dimension
of the representation:

\begin{codeExample}

\includegraphics[scale=0.62]{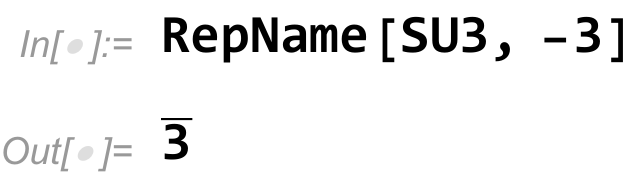}

\end{codeExample}

The convertion from this simplified input notation to Dynkin coefficients
is done internally by the following function, which the user can also
call directly:\index{SimpleRepInputConversion}

\begin{codeSyntax}
SimpleRepInputConversion[<group>,<rep in the simplified notation>]
\tcblower
Returns the Dynkin coefficients of the representation.
\end{codeSyntax}

\begin{codeExample}

\includegraphics[scale=0.62]{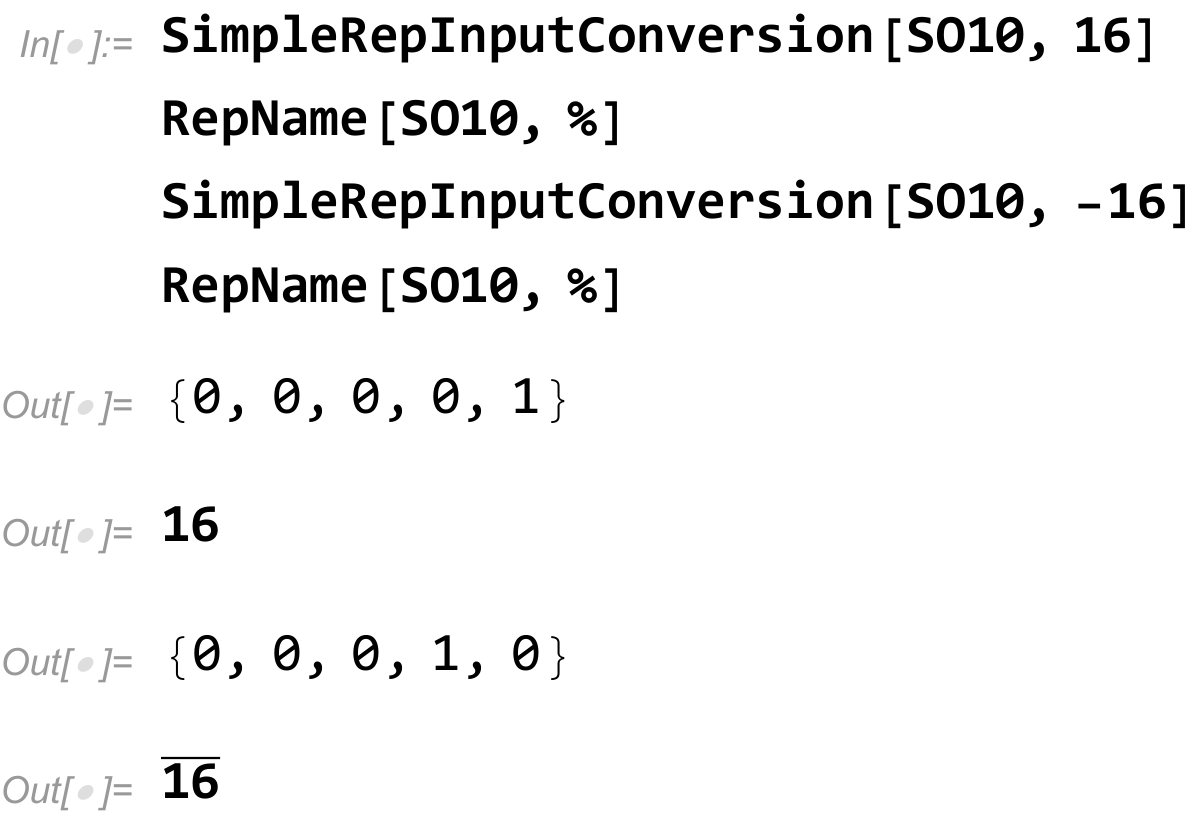}

\end{codeExample}

It is worth stressing that this simplified input notation does not
work for representations with primes in their names; the only way
of providing them as input is with Dynkin coefficients. As for $SO(8)$
representations which can have sub-indices, it is best to avoid the
simplified input notation since it is not straightforward to describe
which representation is being selected by the function \texttt{SimpleRepInputConversion}.

~

Finally, representations of the $SU(n)$ groups are often labeled
by partitions or Young tableaux. It is not possible to use directly
the latter objects to specify $SU(n)$ representation in \texttt{GroupMath},
but there are two functions which convert back and forth between partitions
and Dynkin coefficients of $SU(n)$ groups (\texttt{ConvertPartitionToDynkinCoef}
and \texttt{ConvertToPartitionNotation})\index{ConvertPartitionToDynkinCoef}\index{ConvertToPartitionNotation}.
They are described later.

\subsection{\label{subsec:PropertiesOfAGroup}Properties of a group}

The list of simple complex Lie algebras and many of their important
properties can be found in the literature. It might therefore seem
of little interest to include the computation of those properties
in a computer code. However, in some cases it might useful to have
an automated way of accessing this information.

An important feature of a simple Lie group is its adjoint representation.
Its Dynkin coefficients are given by the \texttt{Adjoint} function:\index{Adjoint}

\begin{codeSyntax}
Adjoint[<group>]
\end{codeSyntax}

\begin{codeExample}

\includegraphics[scale=0.62]{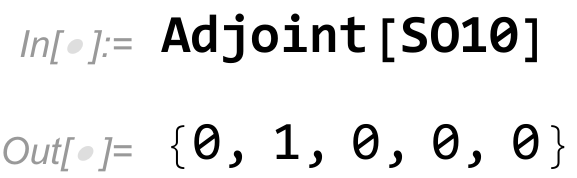}

\end{codeExample}

The positive roots are another important feature which is computed
by the following function:\index{PositiveRoots}

\begin{codeSyntax}
PositiveRoots[<group>]
\end{codeSyntax}

\begin{codeExample}

\includegraphics[scale=0.62]{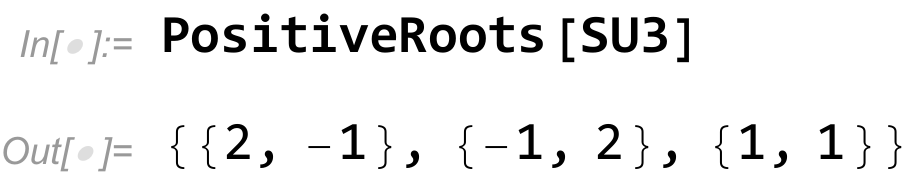}

\end{codeExample}

\subsection{\label{subsec:PropertiesOfRepresentations}Properties of a representation}

The list of all representations of a simple Lie group up to some size
can be obtained with the function \texttt{RepsUpToDimN}, while the
function \texttt{RepsUpToDimNNoConjugates} only returns those representations
with no bars in their name (for example the $\overline{\boldsymbol{3}}$
and $\overline{\boldsymbol{6}}$ of $SU(3)$ would be omitted). An
option \texttt{UseName} can be used with these two functions.

\index{RepsUpToDimN}

\begin{codeSyntax}
RepsUpToDimN[<simple Lie group>,<max dimension>]
\end{codeSyntax}

\begin{codeExample}

\includegraphics[scale=0.62]{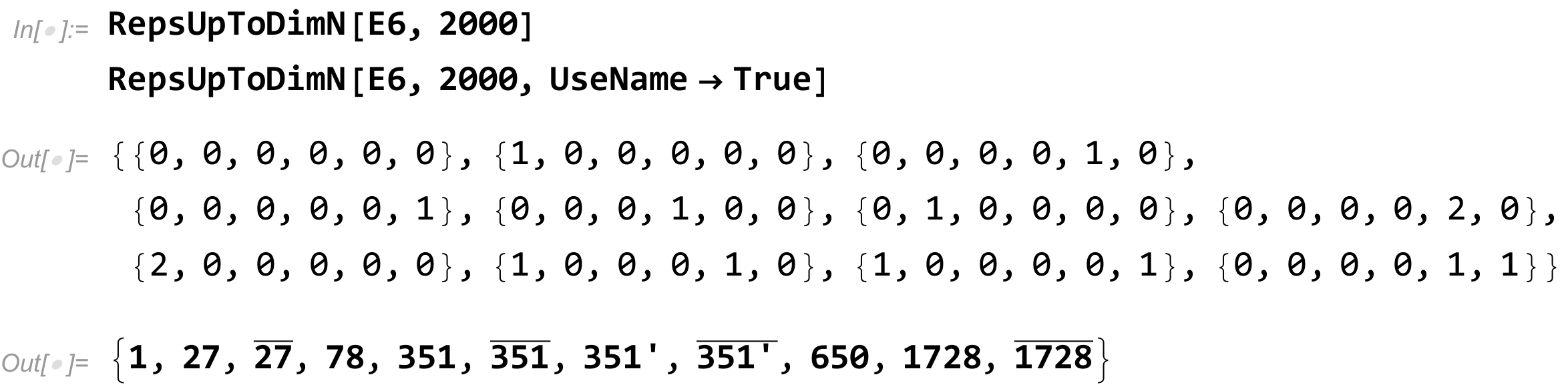}

\end{codeExample}

\index{RepsUpToDimNNoConjugates}

\begin{codeSyntax}
RepsUpToDimNNoConjugates[<simple Lie group>,<max dimension>]
\end{codeSyntax}

\begin{codeExample}

\includegraphics[scale=0.62]{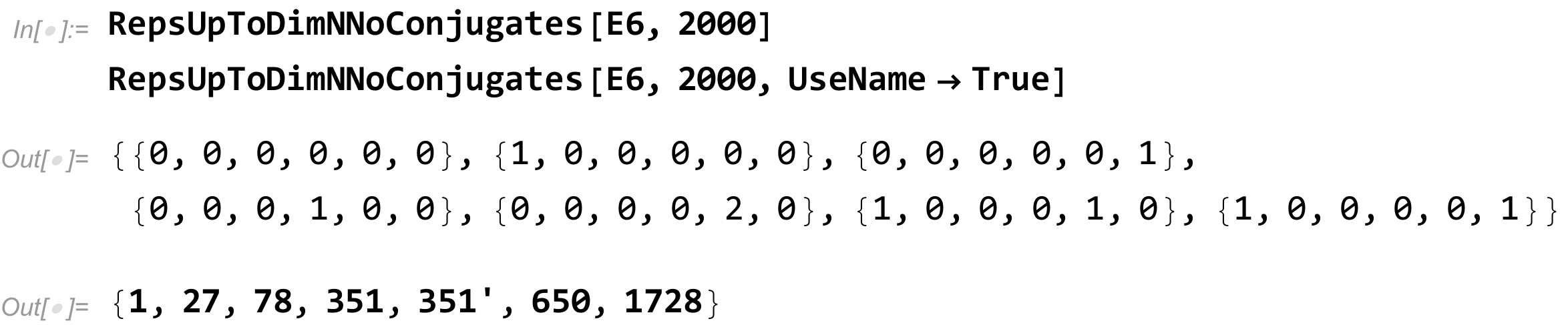}

\end{codeExample}

To compute the Dynkin coefficients of the complex conjugate of a representation,
one can use \texttt{ConjugateIrrep}:

\index{ConjugateIrrep}

\begin{codeSyntax}
ConjugateIrrep[<group>,<representation>]
\end{codeSyntax}

\begin{codeExample}

\includegraphics[scale=0.62]{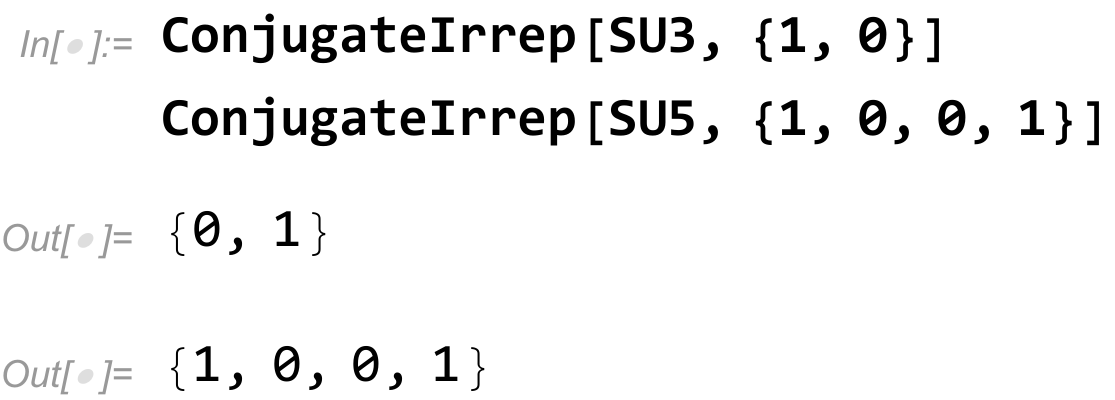}

\end{codeExample}

In general, a representation can be real (R), pseudo-real (PR) or
complex (C). This information can be obtained with the following function:

\index{TypeOfRepresentation}

\begin{codeSyntax}
TypeOfRepresentation[<group>,<representation>]
\end{codeSyntax}

\begin{codeExample}

\includegraphics[scale=0.62]{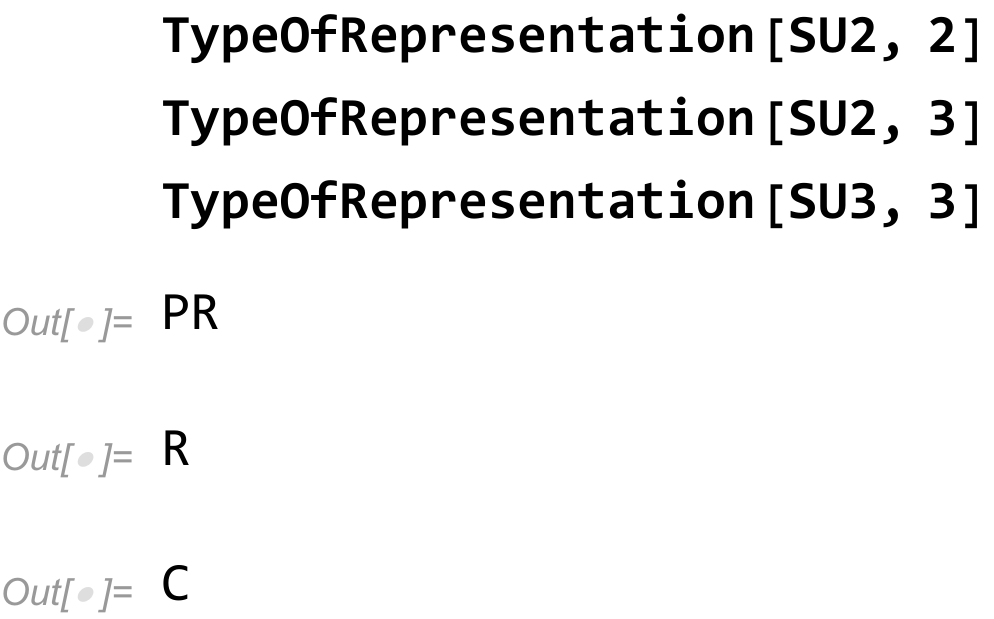}

\end{codeExample}

The \texttt{DimR} function returns the size of a representation of
a Lie group, using the Weyl character formula. If the group is a product
of several factors, this function computes a list with the sizes the
different representations of each factor group (rather than the overall
size of the representation).

\index{DimR}

\begin{codeSyntax}
DimR[<group>,<representation>]
\end{codeSyntax}

\begin{codeExample}

\includegraphics[scale=0.62]{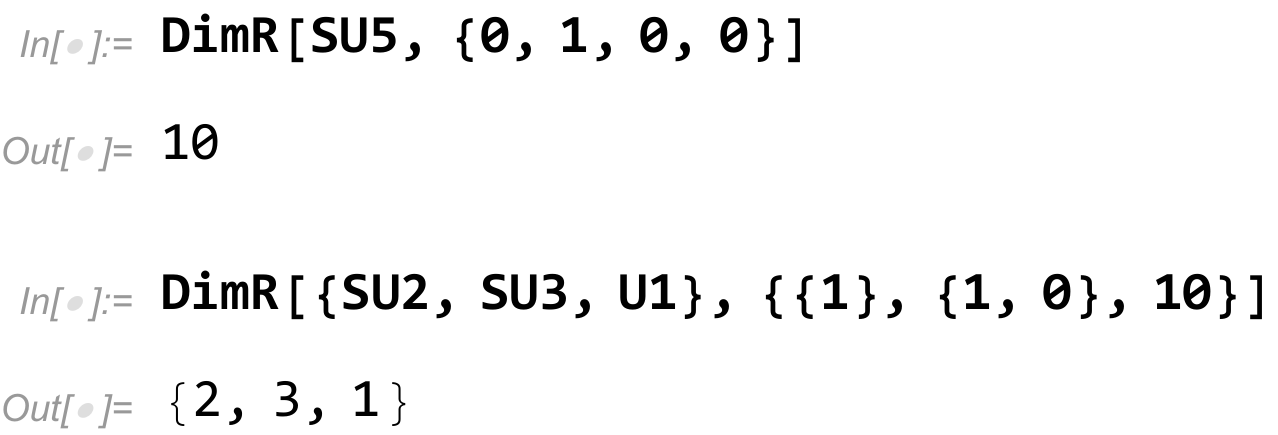}

\end{codeExample}

Two other important numbers which characterize a representation $R$
are the Casimir invariant $C\left(R\right)$ and the Dynkin index
$S\left(R\right)$. They are related to the algebra generator matrices
$T^{c}$ as follows:\footnote{Repeated indices are assumed to be summed over.}
\begin{align}
C\left(R\right)\mathbb{1} & =T^{c}T^{c}\,,\label{eq:Casimir}\\
S\left(R\right)\delta^{ab} & =\textrm{Tr}\left(T^{a}T^{b}\right)\,.\label{eq:DynkinIndex}
\end{align}
As such, $S\left(R\right)$ times the dimension of the algebra is
the same as $C\left(R\right)$ times the dimension of the representation
$R$. The functions \texttt{Casimir} and \texttt{DynkinIndex} compute
these two numbers (if the group is not simple, they operate analogously
to \texttt{DimR}). Nevertheless, it is worth noting that the answer
depends on the normalization taken for the algebra generators; the
one used by the program corresponds to $S\left(R\right)=1/2$ for
the fundamental representation of the $SU(N)$ groups.\footnote{More generally, the convention is that the biggest root $\alpha_{max}$
of a simple algebra has norm 1: $\left\langle \alpha_{max},\alpha_{max}\right\rangle =1$.}

\index{Casimir}

\begin{codeSyntax}
Casimir[<group>,<representation>]
\end{codeSyntax}

\index{DynkinIndex}

\begin{codeSyntax}
DynkinIndex[<group>,<representation>]
\end{codeSyntax}

\begin{codeExample}

\includegraphics[scale=0.62]{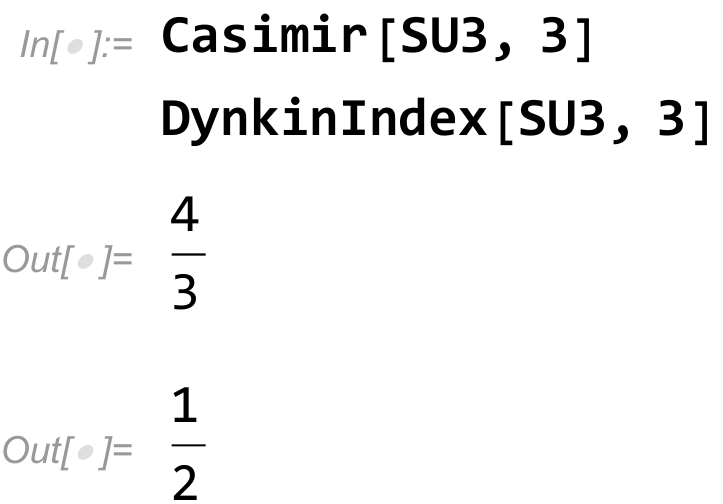}

\end{codeExample}

Another important quantity is $a\left(R\right)$ appearing in the
relation
\begin{equation}
a\left(R\right)d^{abc}=\textrm{Tr}\left(\left\{ T^{a},T^{b}\right\} T^{c}\right)\,,\label{eq:anomaly}
\end{equation}
where $d^{abc}$ is a symmetric tensor which is fixed for a given
group. The number $a\left(R\right)$ is important in the study of
anomalies in quantum field theory, as it appears in the expression
of amplitudes of 1-loop diagrams with three external gauge boson lines,
hence the name triangular anomaly. It can be computed with the function
\texttt{TriangularAnomalyValue}, which applies the algorithm described
in \cite{Okubo:1977sc}. If the group is not simple, $a$ is calculated
for each of the factor groups, and furthermore there might exist mixed
anomalies where one or more of the group generators in expression
(\ref{eq:anomaly}) correspond to a $U(1)$, while the others belong
to another factor group. Hence, if the group is not simple the function
\texttt{TriangularAnomalyValue} returns a list of numbers, corresponding
to each of these $a$ values. By using the option \texttt{Verbose
-> True}, the meaning of each number is spelled out.

\index{TriangularAnomalyValue}

\begin{codeSyntax}
TriangularAnomalyValue[<group>,<representation>]
\end{codeSyntax}

\begin{codeExample}

\includegraphics[scale=0.62]{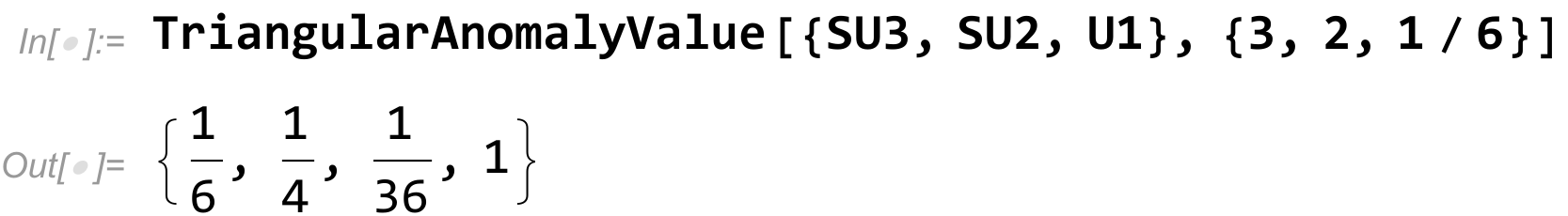}

\end{codeExample}

\begin{codeExample}

\includegraphics[scale=0.62]{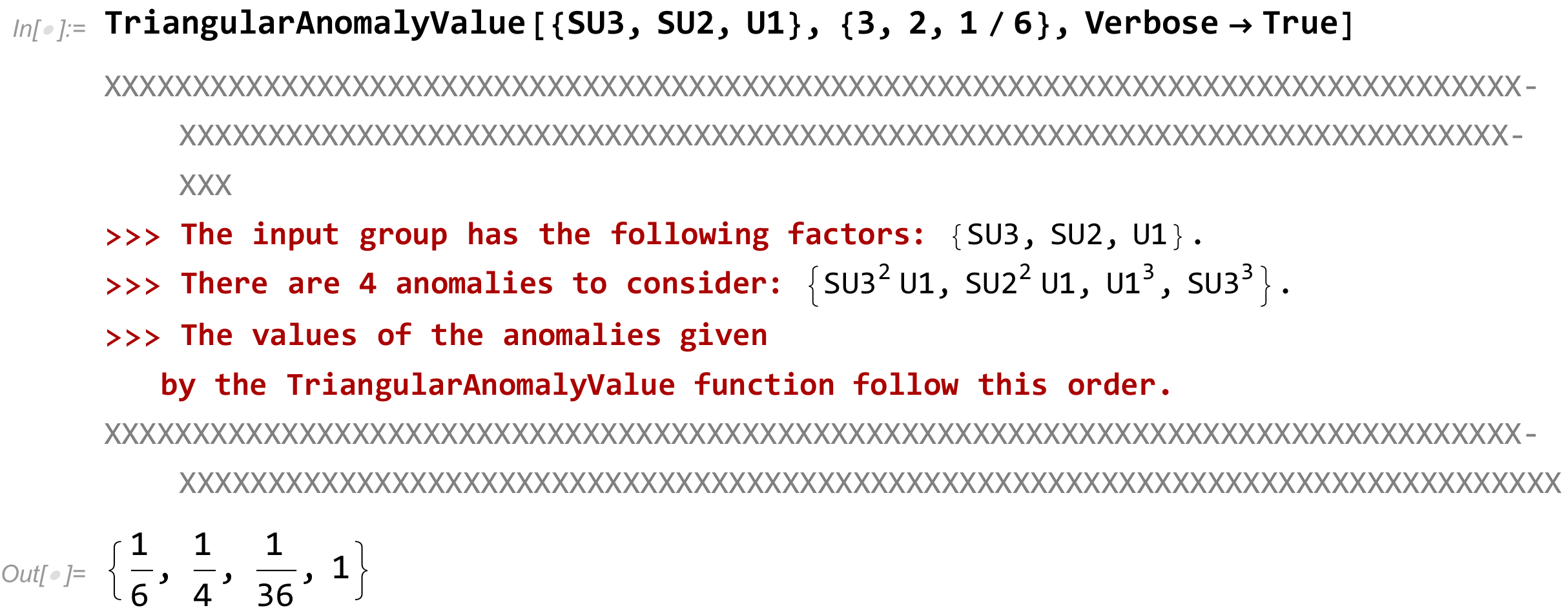}

\end{codeExample}

As explained in appendix, the weights $\omega$ of a simple Lie algebra
with rank $m$ can be seen as lists of $m$ integers $\omega_{i}=2\left\langle w,\alpha_{i}\right\rangle /\left\langle \alpha_{i},\alpha_{i}\right\rangle $.
They can be computed with the \texttt{Weights} functions, which returns
a list with elements \texttt{\{<weight>,<multiplicity>\}}. This format
is justified by the fact that weights can have degeneracy, with some
values appearing repeated.

\index{Weights}

\begin{codeSyntax}
Weights[<group>,<representation>]
\end{codeSyntax}

\begin{codeExample}

\includegraphics[scale=0.62]{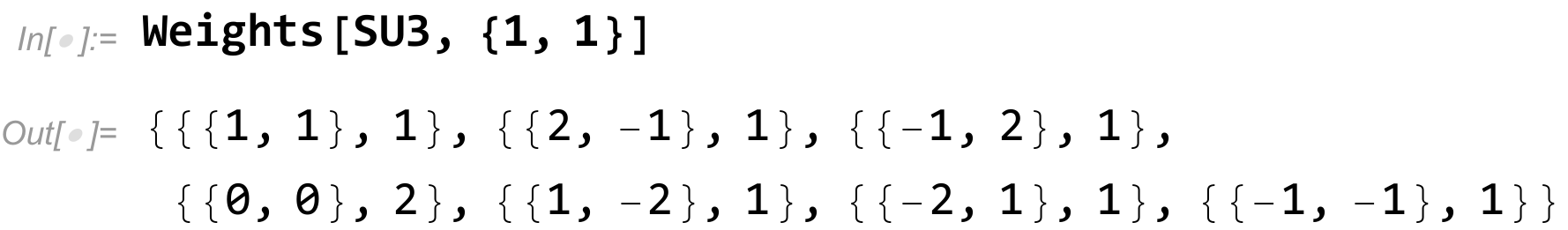}

\end{codeExample}

In the case of the \{1,2,3,4\} irrep of $SU(5)$, which has dimension
198540, we can check that the sum of the multiplicities of all weights
adds up to this number:

\begin{codeExample}

\includegraphics[scale=0.62]{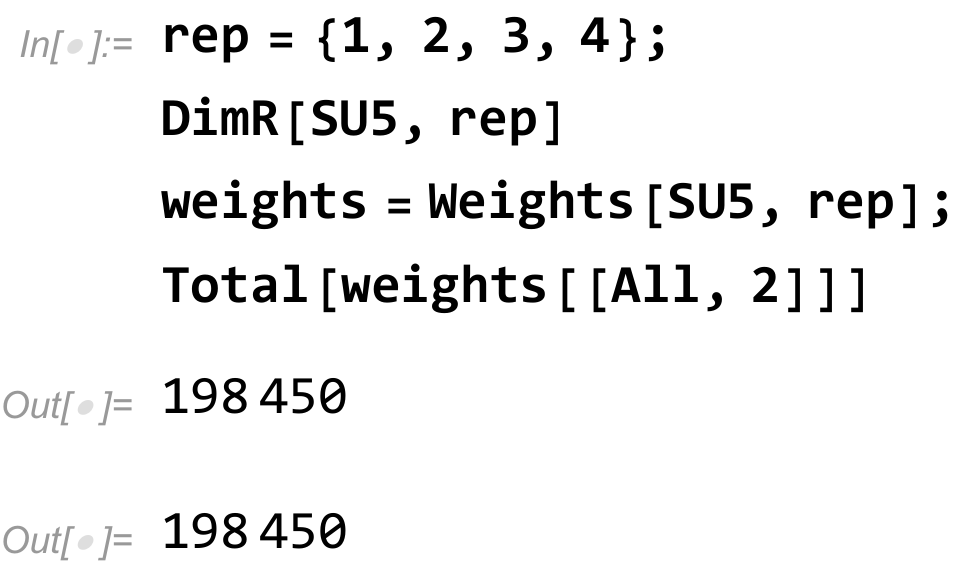}

\end{codeExample}

Out of all weights of a representation, there are a few whose $m$
coefficients $\omega_{i}$ are all non-negative: they are called dominant
weights. They can be obtained, as in the following example, by using
the \texttt{Weights} function.

\begin{codeExample}

\includegraphics[scale=0.62]{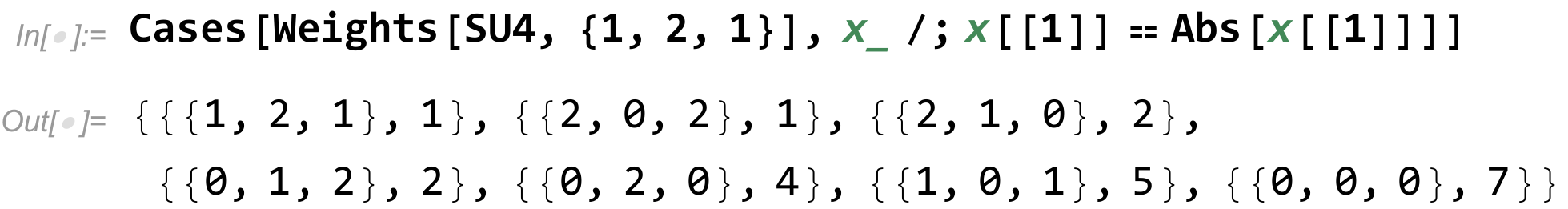}

\end{codeExample}

\noindent An alternative which is better suited for very large irreps
is to use \texttt{DominantWeights}:

\index{DominantWeights}

\begin{codeSyntax}
DominantWeights[<group>,<representation>]
\end{codeSyntax}

\begin{codeExample}

\includegraphics[scale=0.62]{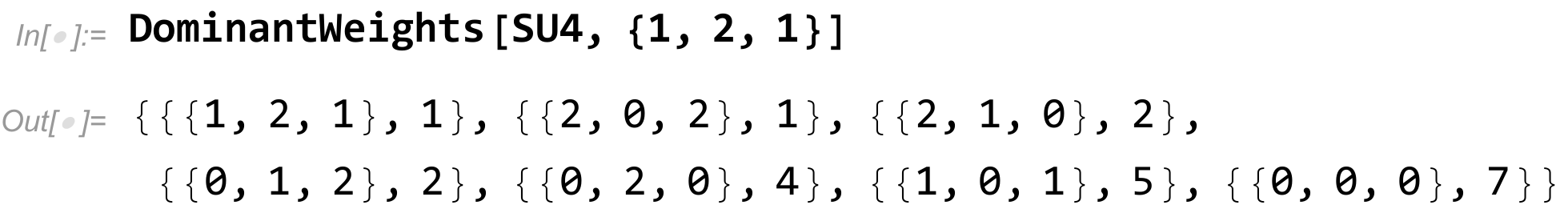}

\end{codeExample}

Through the action of the Weyl group (see the appendix), every weight
can be associated to a dominant one, which is computed by this function:

\index{DominantConjugate}

\begin{codeSyntax}
DominantConjugate[<group>,<weight>]
\end{codeSyntax}

\begin{codeExample}

\includegraphics[scale=0.62]{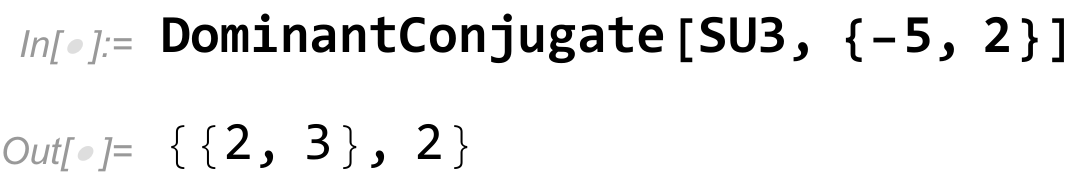}

\end{codeExample}

\noindent The first component of the output is the dominant weight,
and the second indicates the minimum number of Weyl reflections needed
to make the original weight dominant. On this topic, there are two
other functions worth mentioning: \texttt{ReflectWeight} and \texttt{WeylOrbit}.
The first performs the $m$ elementary reflections on a weight; the
second returns all weights in an orbit of the Weyl group which contains
some dominant weight $\omega$ indicated by the user.\footnote{The algorithms used by \texttt{DominantWeights}, \texttt{DominantConjugate},
\texttt{ReflectWeight} and \texttt{WeylOrbit} are essentially the
ones in \cite{Snow:1990,Snow:1993}.}

\index{ReflectWeight}

\begin{codeSyntax}
ReflectWeight[<group>,<weight>, <index i of the elementary reflection>]
\end{codeSyntax}

\begin{codeExample}

\includegraphics[scale=0.62]{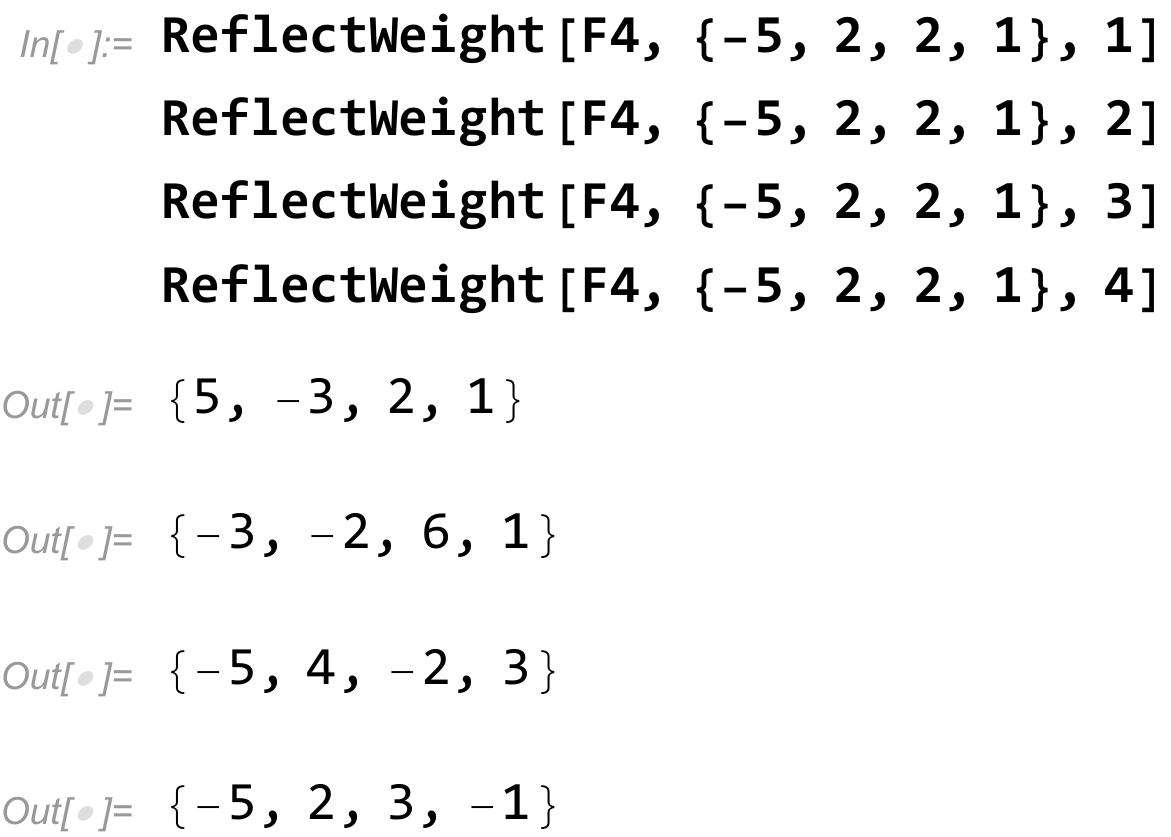}

\end{codeExample}

\index{WeylOrbit}

\begin{codeSyntax}
WeylOrbit[<group>,<dominant weight>]
\end{codeSyntax}

\begin{codeExample}

\includegraphics[scale=0.62]{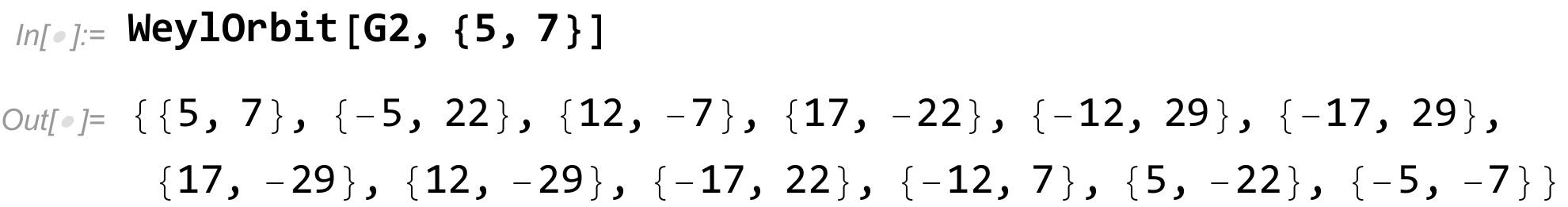}

\end{codeExample}

The weights of each irreducible representation of a simple Lie group
obey some modular equation (or two in the case of the $SO(2n)$ groups).
For example, in the case of $SU(3)$, the value of $\omega_{1}+2\omega_{2}$
(mod $3$) is the same for any weight $\left\{ \omega_{1},\omega_{2}\right\} $
of an irrep. Therefore, this number is characteristic of each irreducible
representation and it defines classes. This class-number can be computed
as follows:

\index{ConjugacyClass}

\begin{codeSyntax}
ConjugacyClass[<group>,<representation>]
\end{codeSyntax}

\begin{codeExample}

\includegraphics[scale=0.62]{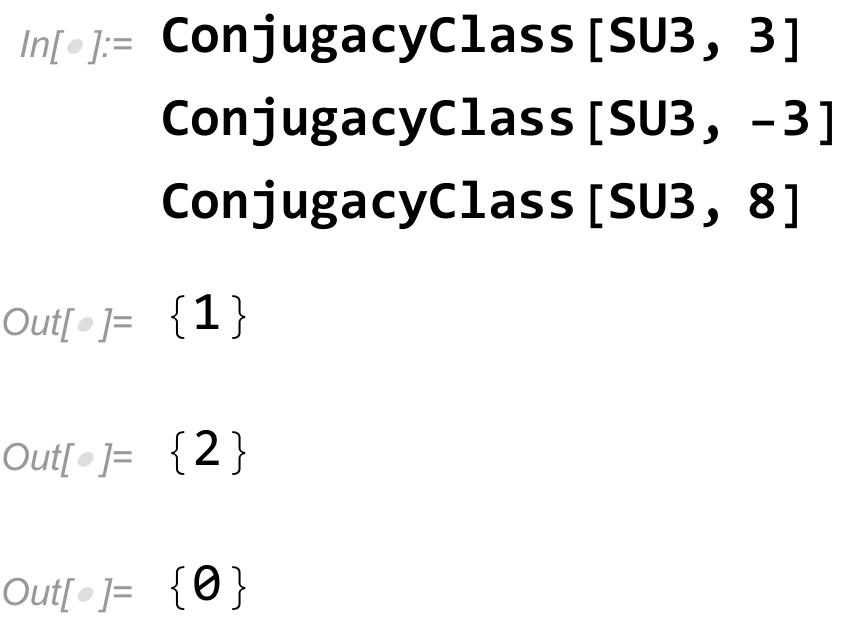}

\end{codeExample}

\noindent For consistency \texttt{ConjugacyClass} returns a list because,
for $SO(2n)$ groups, there are two class-numbers rather than one.
The modulus in the congruence relation(s) which define the various
classes (which is 3 in the case of $SU(3)$) depends only on the simple
Lie group, and it can be obtained with the following function:\index{ConjugacyClassGroupModIndices}

\begin{codeSyntax}
ConjugacyClassGroupModIndices[<group>]
\end{codeSyntax}

\begin{codeExample}

\includegraphics[scale=0.62]{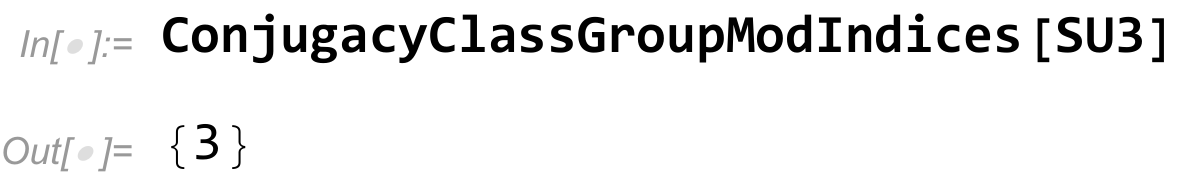}

\end{codeExample}

Let us now go over the decomposition of a product of irreps into irreducible
components of the group. For example, $\boldsymbol{2}\times\boldsymbol{2}=\boldsymbol{1}+\boldsymbol{3}$
in $SU(2)$ and $\boldsymbol{3}\times\boldsymbol{3}\times\boldsymbol{3}=\boldsymbol{1}+\boldsymbol{8}+\boldsymbol{8}+\boldsymbol{10}$
in $SU(3)$. Using the algorithm described in \cite{Snow:1993} \texttt{GroupMath}
is capable of quickly doing these operations, even for a high number
of factors and for very large irreps. The relevant function, \texttt{ReduceRepProduct},
works with simple and reductive groups, accepting the option \texttt{UseName->True}.

\index{ReduceRepProduct}

\begin{codeSyntax}
ReduceRepProduct[<group>,<list of representation>]
\end{codeSyntax}

\begin{codeExample}

\includegraphics[scale=0.62]{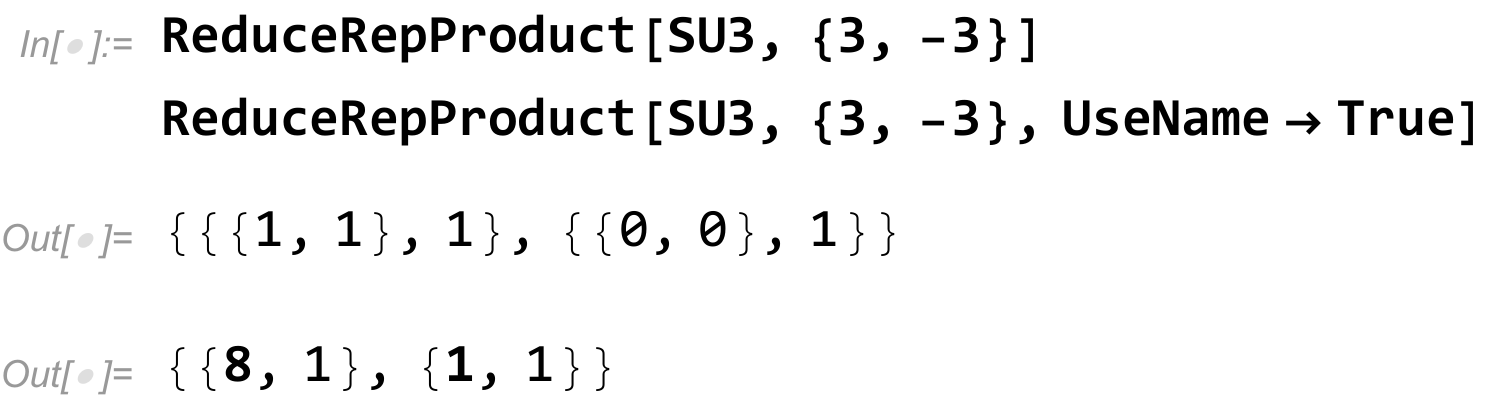}

\end{codeExample}

\noindent The output is a list with items of the form \texttt{\{<irrep>,
<multiplicity>\}}. There is no limit to the number of representations
to be multiplied. For example, the product $\boldsymbol{3}\times\boldsymbol{3}\times\boldsymbol{8}\times\boldsymbol{8}\times\boldsymbol{8}\times\overline{\boldsymbol{3}}\times\boldsymbol{8}$
of $SU(3)$ decomposes as follows

\begin{codeExample}

\includegraphics[scale=0.62]{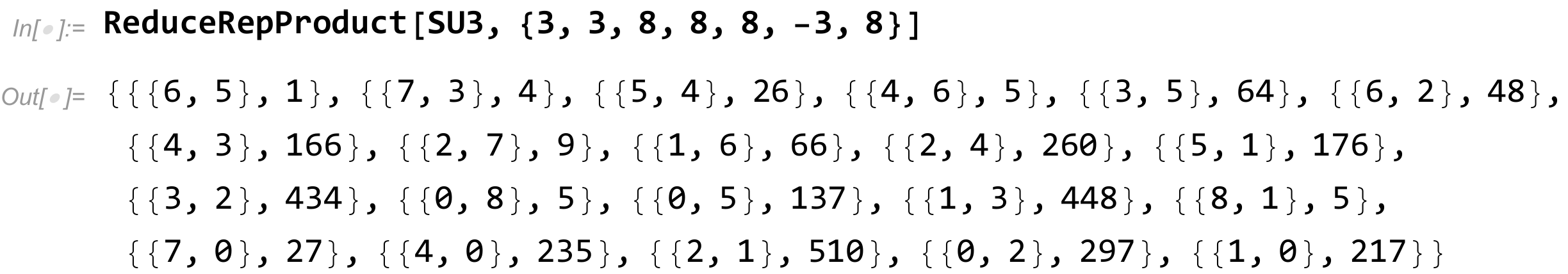}

\end{codeExample}

\noindent and even larger examples, such as the product of ten $\boldsymbol{120}$s
of $SO(10)$, can easily be handled (note that \{0,0,1,0,0\}=$\boldsymbol{120}$):

\begin{codeExample}

\includegraphics[scale=0.62]{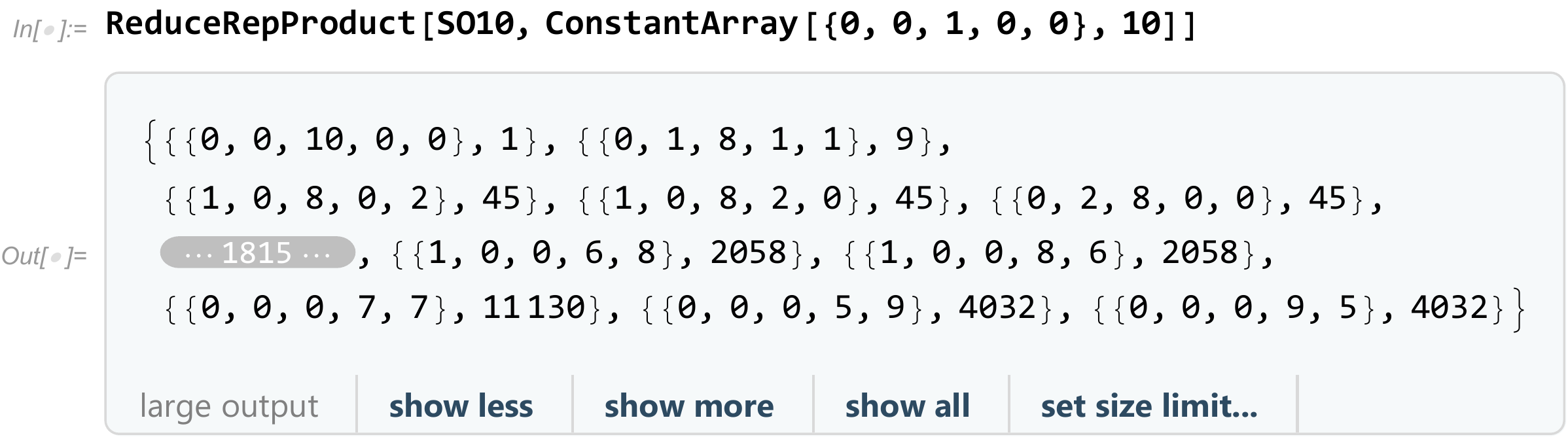}

\end{codeExample}

As noted already above, the function \texttt{ReduceRepProduct} also
works for semi-simple groups and $U(1)$'s. Here is one such case:

\begin{codeExample}

\includegraphics[scale=0.62]{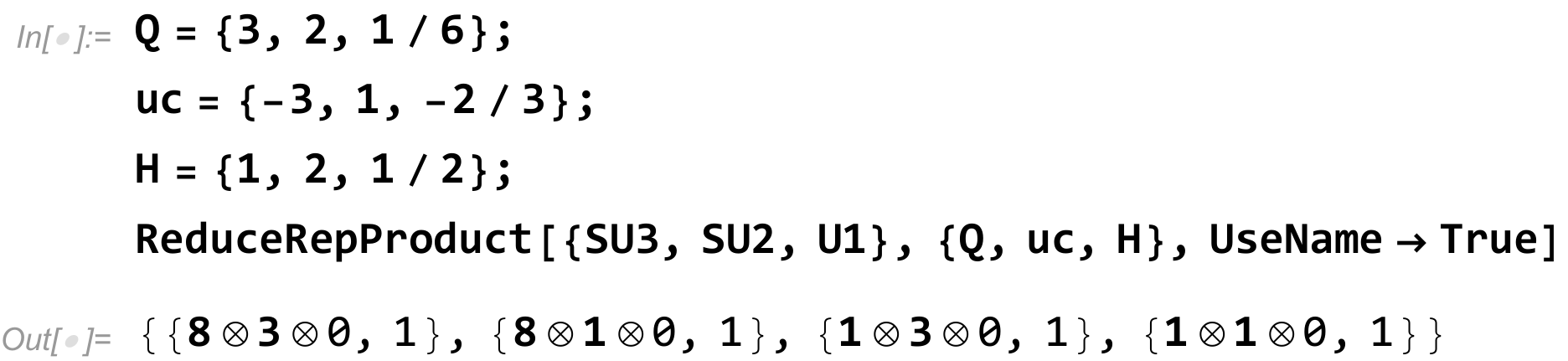}

\end{codeExample}

What \texttt{ReduceRepProduct} does not do is show the permutation
symmetry associated to each irreducible component of a product of
equal representations. For example, returning to the product of two
$SU(2)$ doublets, it is known that the singlet is anti-symmetric
under exchange of the two doublets, while the triplet is symmetric:
$\boldsymbol{2}\times\boldsymbol{2}=\boldsymbol{1}_{A}+\boldsymbol{3}_{S}$.
This extra information is computed by the functions \texttt{PermutationSymmetry}
and \texttt{PermutationSymmetryOfInvariants} \index{PermutationSymmetry}\index{PermutationSymmetryOfInvariants}described
in section \ref{sec:Permutation-groups}.

\subsection{\label{subsec:Basis-dependent-computations}Basis-dependent computations}

All properties discussed up to now do not depend on a choice of basis
for the representation matrices $T^{a}$. We shall now consider computations
performed by the program which are basis dependent.

It is important to keep in mind that if $T^{a}$ are unitary representation
matrices obeying the Casimir and Dynkin index relations (\ref{eq:Casimir})
and (\ref{eq:DynkinIndex}), then the same is true for the set of
matrices
\begin{equation}
O_{ab}U^{\dagger}T^{b}U
\end{equation}
if $O$ is an orthogonal matrix and $U$ is unitary. Therefore, a
basis choice implies picking an $O$ matrix for each simple group,
and a $U$ matrix for each of its irreducible representations. \texttt{GroupMath}'s
basis choice ensures that the $T^{a}$ have the following properties:
\begin{enumerate}
\item They are Hermitian: $T^{a}=\left(T^{a}\right)^{\dagger}$.
\item They obey equations (\ref{eq:Casimir}) and (\ref{eq:DynkinIndex})
with $C\left(R\right)$ and $S\left(R\right)$ as given by the \texttt{Casimir}
and \texttt{DynkinIndex} functions.\index{Casimir}\index{DynkinIndex}
\item The representation matrices of different irreps of a same group are
consistent, meaning that the structure constants $f_{abc}$ appearing
in the relation $\left[T^{a},T^{b}\right]=f_{abc}T^{c}$ are the same
for all irreducible representations.
\item There is a maximum number of diagonal matrices $T^{a}$ (their number
equals the group's rank). The matrices which are not diagonal come
in pairs of the form $L+L^{T}$ and $i\left(L-L^{T}\right)$, for
some strictly lower triangular matrices $L$.
\item The relation $T^{a}\left(R^{*}\right)=-\left[T^{a}\left(R\right)\right]^{*}$
holds for irreducible representations $R$ and $R^{*}$ which are
the complex conjugate of each other, ensuring that $\left\{ \exp\left[i\varepsilon_{a}T^{a}\left(R\right)\right]\right\} ^{*}=\exp\left[i\varepsilon_{a}T^{a}\left(R^{*}\right)\right]$.
\end{enumerate}
Note that even for real representations (for example the triplet of
$SU(2)$), we are allowed to choose a basis where $\left[\exp\left(i\varepsilon_{a}T^{a}\right)\right]^{*}\neq\exp\left[\exp\left(i\varepsilon_{a}T^{a}\right)\right]$.
In fact, a real basis would require the $T^{a}$ to be anti-symmetric,
and therefore non-diagonal; since the \texttt{GroupMath} basis always
ensures that there is a maximal number of diagonal $T^{a}$'s, we
may deduce that it is never a real basis.

With this said, arguably the most important basis-dependent quantity
that the program calculates are the $T^{a}$ matrices themselves:

\index{RepMatrices}

\begin{codeSyntax}
RepMatrices[<group>,<representation>]
\end{codeSyntax}

\begin{codeExample}

\includegraphics[scale=0.62]{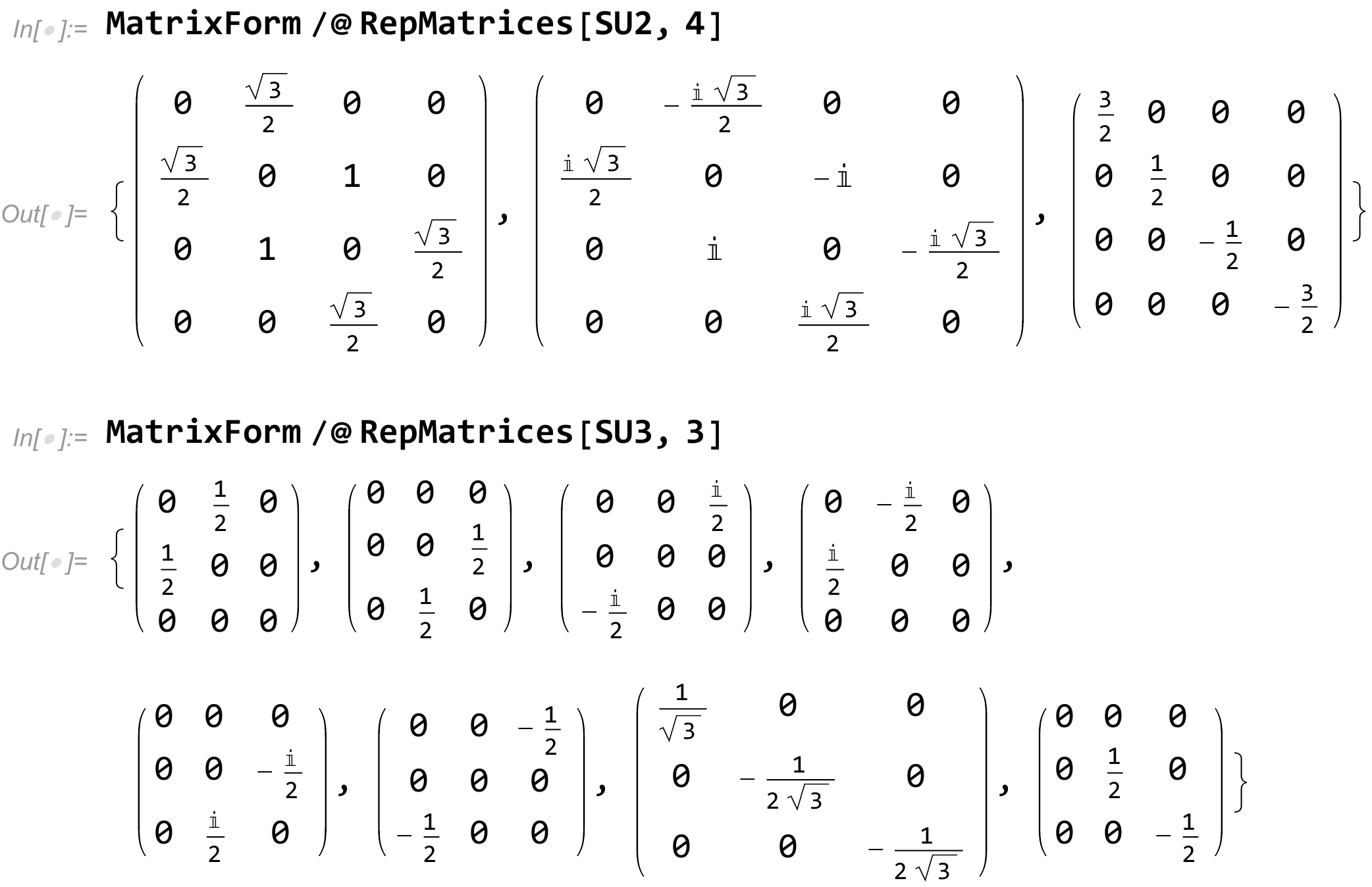}

\end{codeExample}

\noindent The output of the \texttt{RepMatrices} function is a list
of matrices which are not immediately displayed on screen because
they are in Mathematica's \texttt{SparseArray} format. In order to
view their entries one must use the \texttt{Normal} or \texttt{MatrixForm}
commands. Note also that in the second example shown above, the output
is similar but not exactly equal to one-half the Gell-Mann matrices.
That is a consequence of the fact that no hard-coding of special cases
was done: the \texttt{RepMatrices} function uses the same generic
algorithm to build the output independently of the group or the irreducible
representation (the algorithm is described in appendix B.1 of \cite{Fonseca:2013qka}).

The function \texttt{RepMatrices} also works with semi-simple and
$U(1)$ groups:

\begin{codeExample}

\includegraphics[scale=0.62]{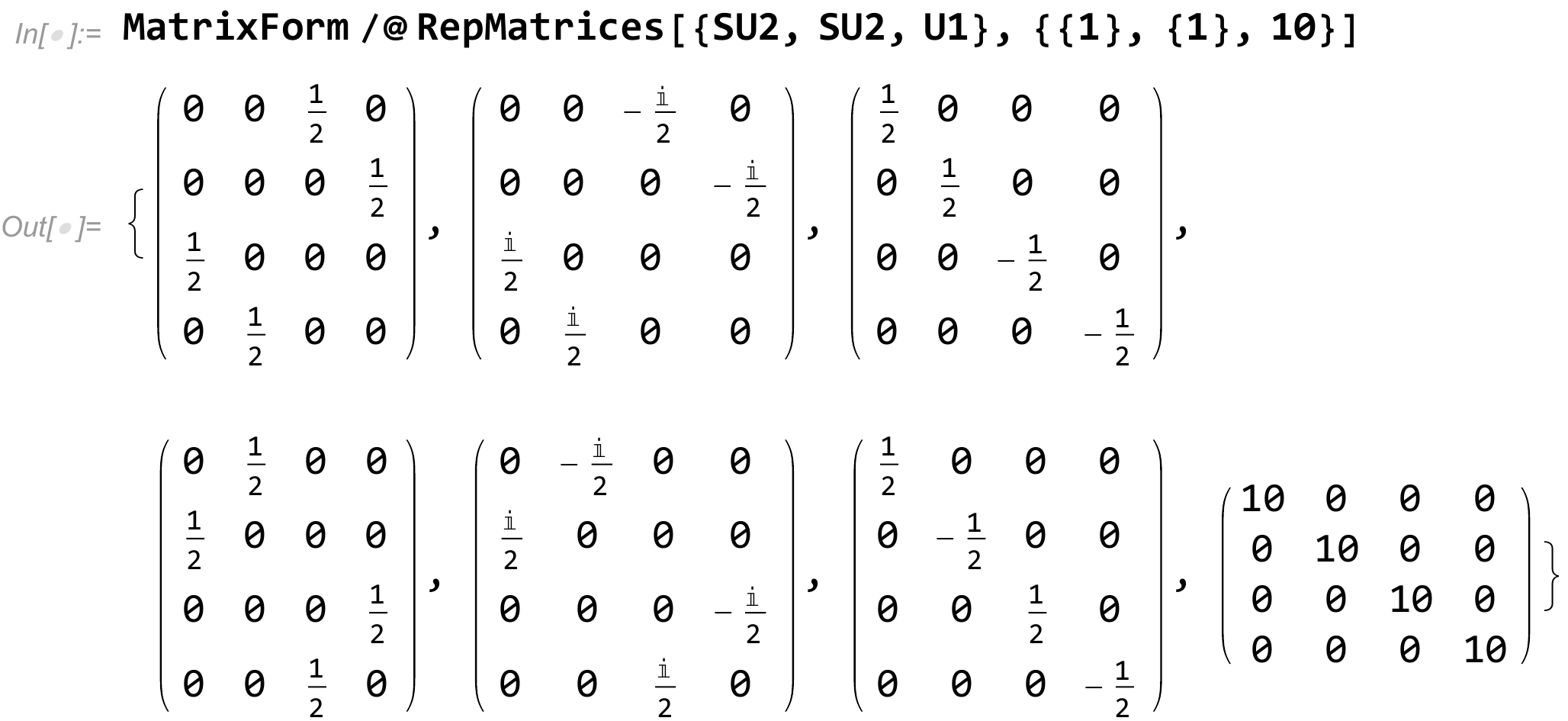}

\end{codeExample}

\noindent In this example, the irreducible representation is 4-dimensional
and the first three matrices correspond to the first $SU(2)$ factor,
the next three correspond to the second $SU(2)$ factor, and finally
the $U(1)$ representation matrix comes last.

Note that for real or pseudo-real representations, the simplified
input $d$ and $-d$ gets converted to the same Dynkin coefficients,
so the output of RepMatrices is unchanged by the minus sign:

\begin{codeExample}

\includegraphics[scale=0.62]{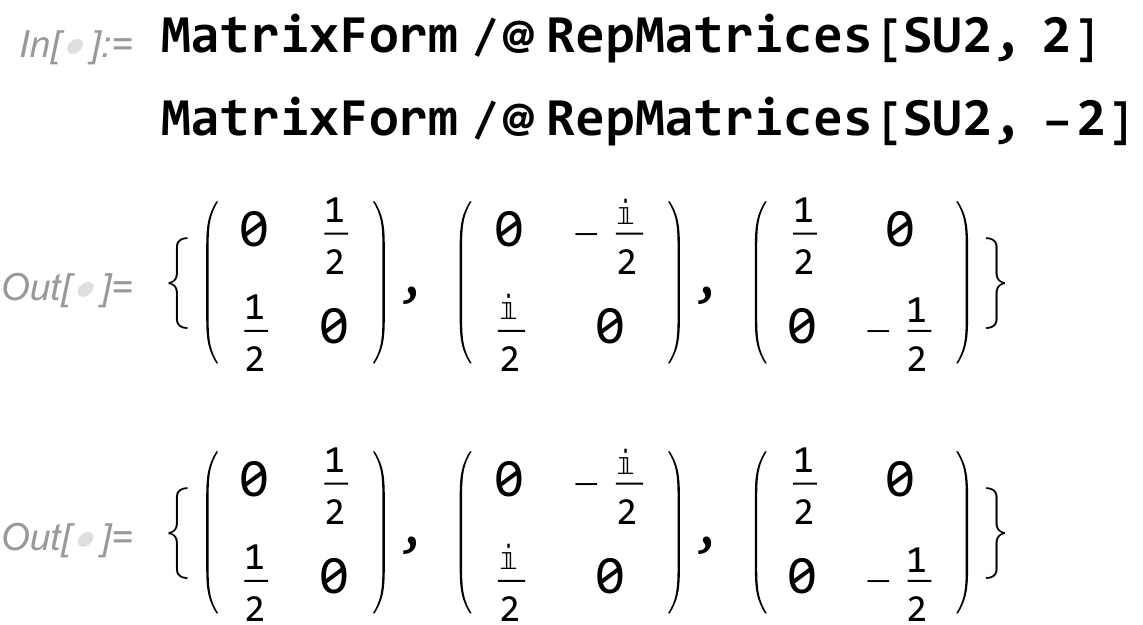}

\end{codeExample}

\noindent What this means is that conjugation of the matrices of real
or pseudo-real representations cannot be achieved with a minus sign
in the input. Instead, the user must manually do this operation ($T^{a}\rightarrow-\left(T^{a}\right)^{T}$):

\begin{codeExample}

\includegraphics[scale=0.62]{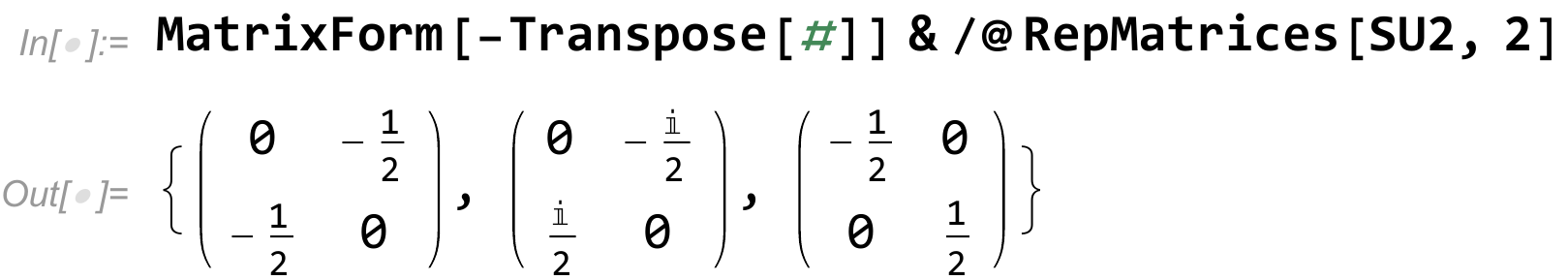}

\end{codeExample}

As mentioned in appendix, for a simple group with rank $r$ one can
find $3r$ elements $\left\{ e_{i},f_{i},h_{i}\right\} $ with $i=1,\cdots,r$
which obey the relations (\ref{eq:Chevalley-Serre-1})--(\ref{eq:Chevalley-Serre-3}).
The function \texttt{RepMinimalMatrices} computes explicit representation
matrices for them. Note that these matrices by themselves do not form
a basis of the algebra, and they do not obey the properties listed
earlier (which apply only to the output of \texttt{RepMatrices}).

\index{RepMinimalMatrices}

\begin{codeSyntax}
RepMinimalMatrices[<group>,<representation>]
\end{codeSyntax}

\begin{codeExample}

\hspace{-3mm}\includegraphics[scale=0.62]{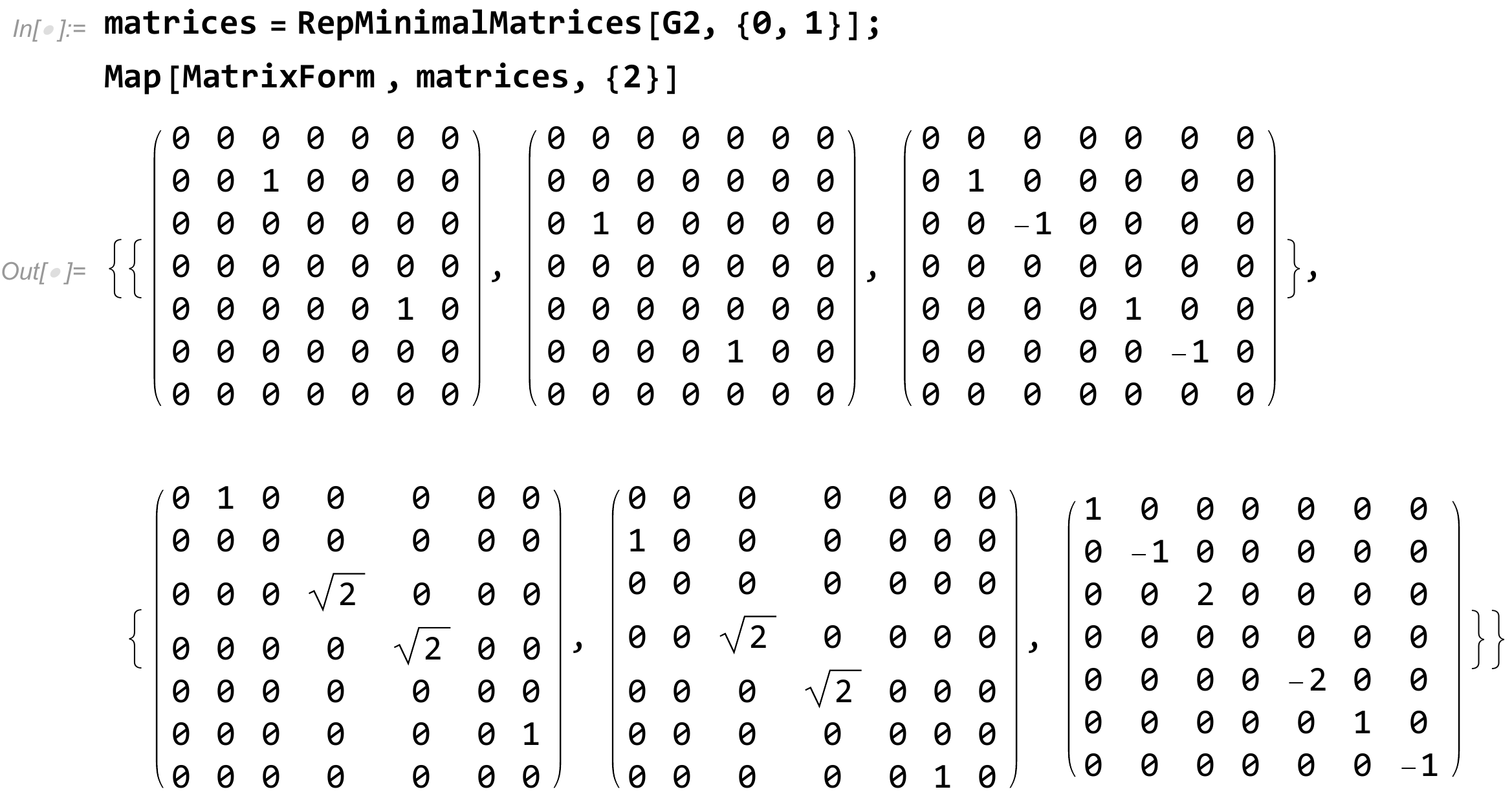}

\end{codeExample}

\noindent The output is a list of representation matrices for the
elements $\{\left\{ e_{1},f_{1},h_{1}\right\} ,\left\{ e_{2},f_{2},h_{2}\right\} ,$\allowbreak$\cdots\}$
using again the \texttt{SparseArray} format. In other words, this
function returns a list of size $r$ where each element itself is
a list of 3 matrices. Just like \texttt{RepMatrices}, \texttt{RepMinimalMatrices}
also works with reductive groups.

When it comes to the adjoint representation of a group, there is a
subtlety worth knowing. One can compute its matrices just as any other
representation: \index{RepMatrices}

\begin{codeExample}

\includegraphics[scale=0.62]{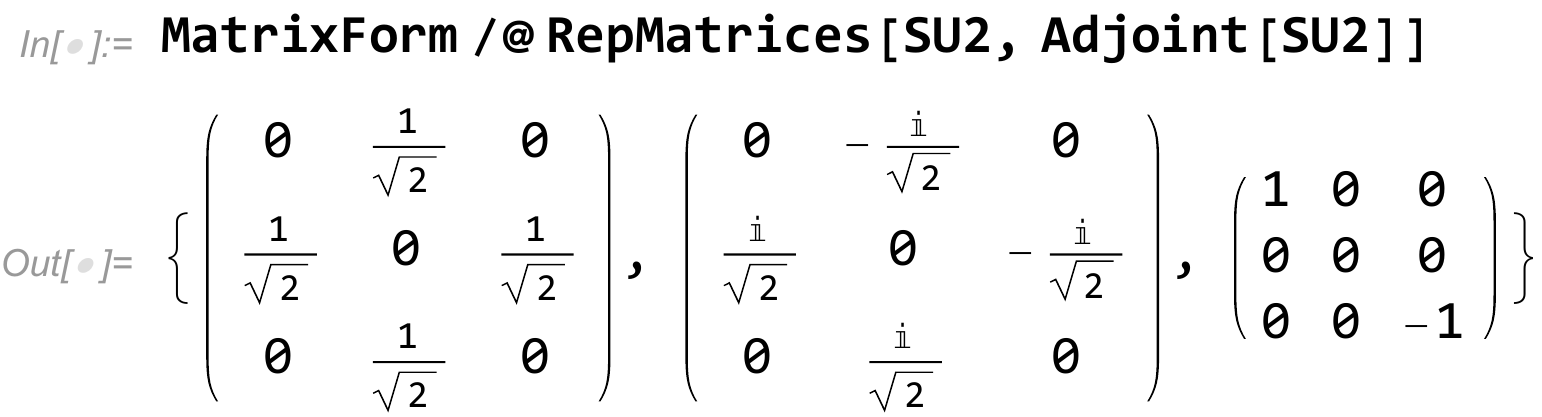}

\end{codeExample}

\noindent However, these matrices are not directly comparable with
the structure constants appearing in the relation $\left[T^{a},T^{b}\right]=f_{abc}T^{c}$
because of a basis difference. In particular, note that the matrices
$\left(\widetilde{T}^{a}\right)_{bc}\equiv-if_{abc}$ are never diagonal,
as opposed to the output of \texttt{RepMatrices}. In order to get
the $\widetilde{T}^{a}$ one can use instead the \texttt{GaugeRep}
function.

\index{GaugeRep}

\begin{codeSyntax}
GaugeRep[<group>]
\end{codeSyntax}

\begin{codeExample}

\includegraphics[scale=0.62]{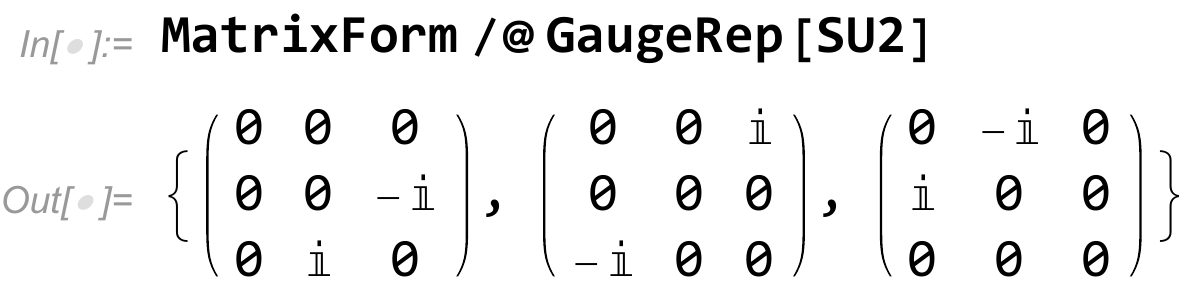}

\end{codeExample}

With these representation matrices, one can write explicitly the components
of an irreducible representation from the product of the components
of other representations. The numerical values appearing in these
relations are usually called Clebsch-Gordan coefficients --- at least
for the case of the $SU(2)$ group. The closely related functions
\texttt{Invariants} and \texttt{IrrepInProduct} compute these numbers.

Consider for example two doublets of $SU(2)$ with components labeled
\begin{equation}
\left(\begin{array}{c}
\mathtt{a\left[1\right]}\\
\mathtt{a\left[2\right]}
\end{array}\right)\textrm{ and }\left(\begin{array}{c}
\mathtt{b\left[1\right]}\\
\mathtt{b\left[2\right]}
\end{array}\right)\,.
\end{equation}
The combination \texttt{a{[}1{]}b{[}2{]}-a{[}2{]}b{[}1{]}} and any
multiple of it is invariant under $SU(2)$. These are the types of
expressions calculated by the \texttt{Invariants} function, which
works for reductive groups and for products of an arbitrary numbers
of representations.

\index{Invariants}

\begin{codeSyntax}
Invariants[<group>,<list of representations>]
\end{codeSyntax}

\begin{codeExample}

\includegraphics[scale=0.62]{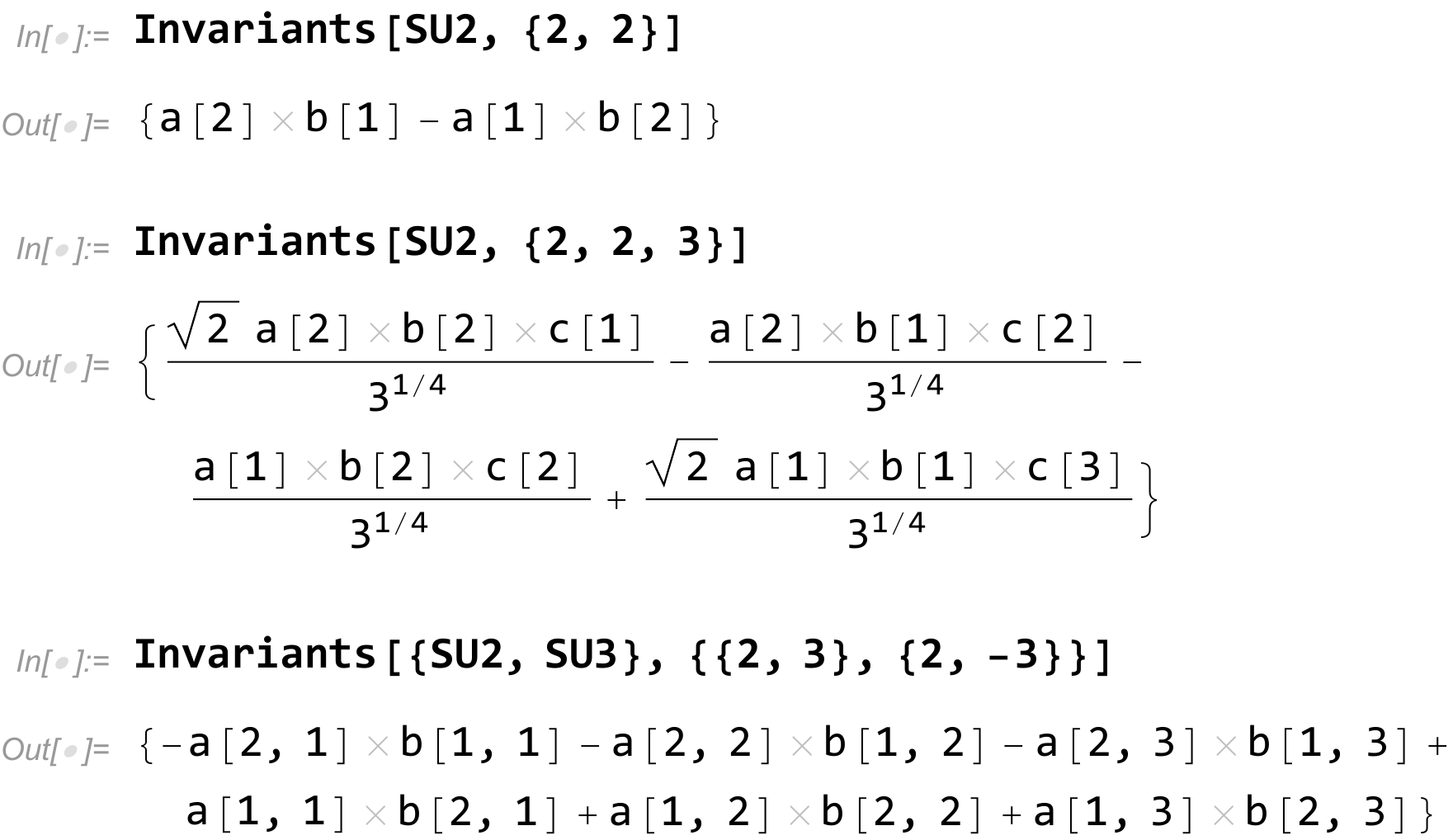}

\end{codeExample}

Several features of this function are noteworthy:
\begin{itemize}
\item Its output is a list of expressions (hence the curly brackets) because
in general one might have more than one independent invariant.
\item The letters \texttt{a}, \texttt{b}, \texttt{c}, \texttt{d}, ... are
used to label the components of the first, second, third, fourth,
... representations, as ordered by the user. The basis for each of
them is the one of \texttt{RepMatrices}.\index{RepMatrices} Numbers
are added to each letter \texttt{\small{}x} to denote sequential components
of the representation: \texttt{\small{}x{[}1{]}}, \texttt{\small{}x{[}2{]}},
... . There is one index for each simple factor group so the components
for the representation $\left(\boldsymbol{3},\boldsymbol{2},7\right)$
of $SU(3)\times SU(2)\times U(1)$ are \{\texttt{\small{}x{[}1,1{]}},
\texttt{\small{}x{[}1,2{]}}, \texttt{\small{}x{[}2,1{]}}, \texttt{\small{}x{[}2,2{]}},
\texttt{\small{}x{[}3,1{]}}, \texttt{\small{}x{[}3,2{]}}\} and they
are assumed to transform with the matrices \texttt{\small{}RepMatrices{[}\{SU3,SU2,U1\},\{3,2,7\}{]}}.
\item If some expression is a group invariant, so is any multiple of it.
More generally, if there are several invariants one can make arbitrary
linear combinations of them. In order to remove some of this arbitrariness,
\texttt{GroupMath} orthonormalizes a set of independent invariants
as follows:\footnote{Furthermore, when there are repeated representations being multiplied,
the \texttt{Invariants} function ensures that the output decomposes
in a very specific way into irreducible representations of the relevant
permutation group. More information on this feature can be found in
the built-in documentation of the function.} writing each invariant as $\kappa_{\alpha\beta\gamma...}^{(i)}\mathtt{a[}\alpha\mathtt{]b[}\beta\mathtt{]c[}\gamma\mathtt{]}\cdots$
(with $i$ indexing each of them), the $\kappa^{(i)}$ numerical coefficients
obey the relations
\begin{equation}
\left[\kappa_{\alpha\beta\gamma...}^{(i)}\right]^{*}\kappa_{\alpha\beta\gamma...}^{(j)}=\delta_{ij}\sqrt{dim\left(\mathtt{a}\right)dim\left(\mathtt{b}\right)dim\left(\mathtt{c}\right)\cdots}
\end{equation}
where $dim(\mathtt{x})$ is the dimension of the irreducible representation
$\mathtt{x}$. For example, in the case of $\boldsymbol{2}\times\boldsymbol{2}\times\boldsymbol{3}$
in $SU(2)$, the only invariant must have coefficients normalized
to $\left[\kappa_{\alpha\beta\gamma}^{(1)}\right]^{*}\kappa_{\alpha\beta\gamma}^{(1)}=2\sqrt{3}$,
and that is why we see an overall factor of $3^{1/4}$ in the output
of \texttt{Invariants}.
\end{itemize}
It is possible to obtain just the numerical coefficients $\kappa_{\alpha\beta\gamma...}^{(i)}$
rather than the expressions $\kappa_{\alpha\beta\gamma...}^{(i)}\mathtt{a[}\alpha\mathtt{]b[}\beta\mathtt{]c[}\gamma\mathtt{]}\cdots$
by using the option \texttt{TensorForm -> True}:

\begin{codeExample}

\includegraphics[scale=0.62]{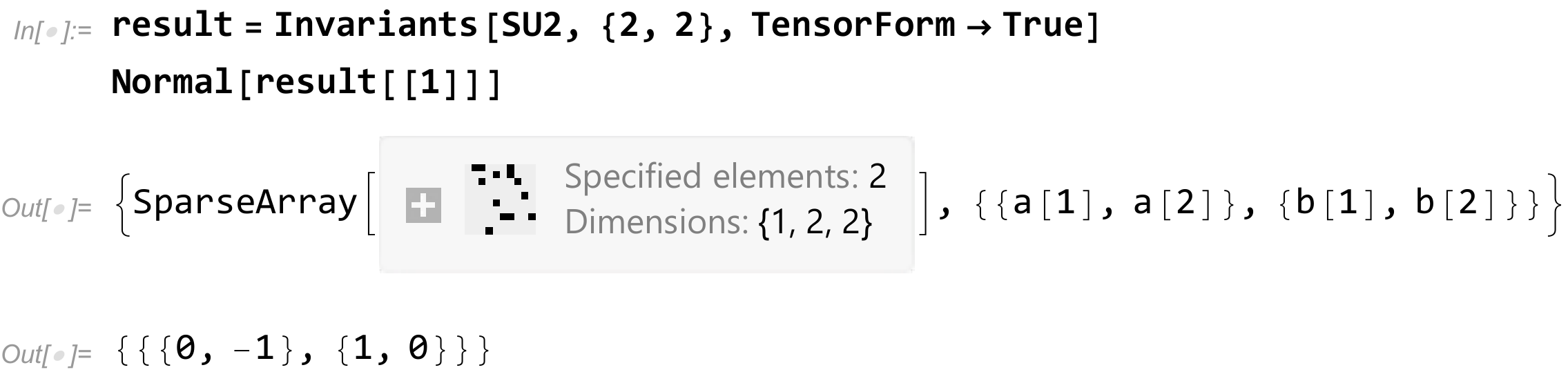}

\end{codeExample}

\noindent Using this options, the output is composed of two parts:
the first one are the $\kappa_{\alpha\beta\gamma...}^{(i)}$ coefficients,
and the second is just the list of representation components used
when \texttt{TensorForm -> False}.

For complex representations, conjugation is simple: one just needs
to indicate as input the conjugated irrep.

\begin{codeExample}

\includegraphics[scale=0.62]{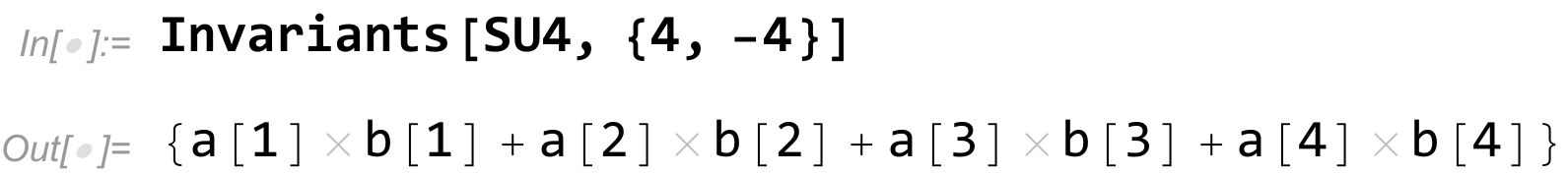}

\end{codeExample}

\noindent However, for real and pseudo-real irreps this does not work.
It is worth pointing out that, for those cases, using the simplified
input $-d$ instead of $d$ does not help either, as both correspond
to the same Dynkin coefficients:

\begin{codeExample}

\includegraphics[scale=0.62]{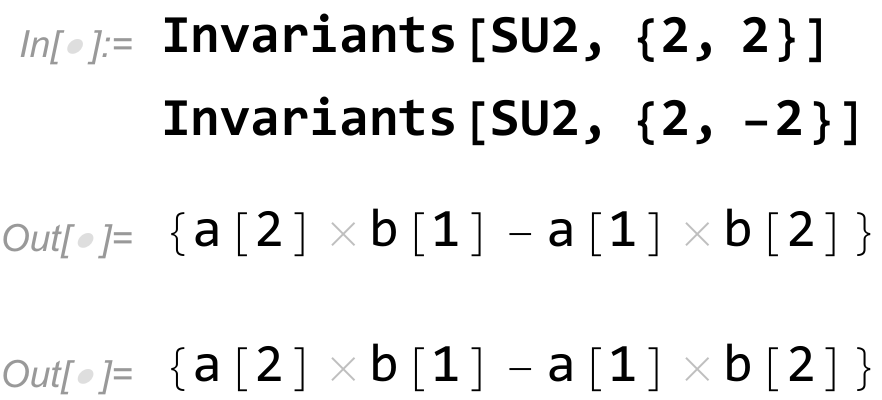}

\end{codeExample}

\noindent In order to conjugate real and pseudo-real irreps (or even
complex ones), there is an option \texttt{Conjugations -> \{<True
or False>, <True or False>, ...\}} which will instruct the program
to conjugate (\texttt{True}) or not (\texttt{False}) each of the irreps
provided as input:

\begin{codeExample}

\includegraphics[scale=0.62]{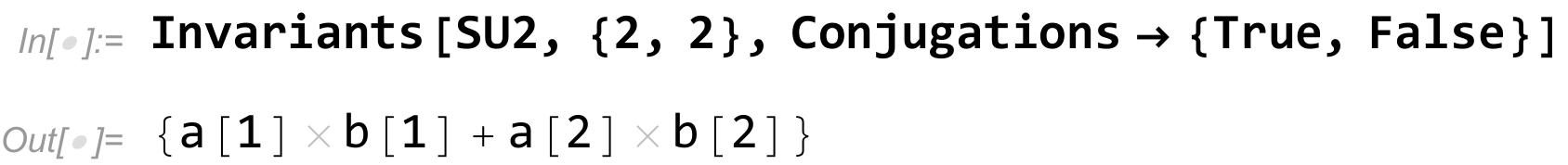}

\end{codeExample}

\noindent Here the vector $\left(\mathtt{a\left[1\right]},\mathtt{a\left[2\right]}\right)^{T}$
is assumed to transform with the matrices $-\left(T^{a}\right)^{T}$,
where the $T^{a}$ matrices are the ones returned by \texttt{RepMatrices{[}SU2,2{]}}
\index{RepMatrices} (equal to one-half of the Pauli matrices).

The computational time increases rapidly with the number of representations
being multiplied, as well as their size (memory usage can also be
a problem). Nevertheless there is no limit for either of these quantities:

\begin{codeExample}

\includegraphics[scale=0.62]{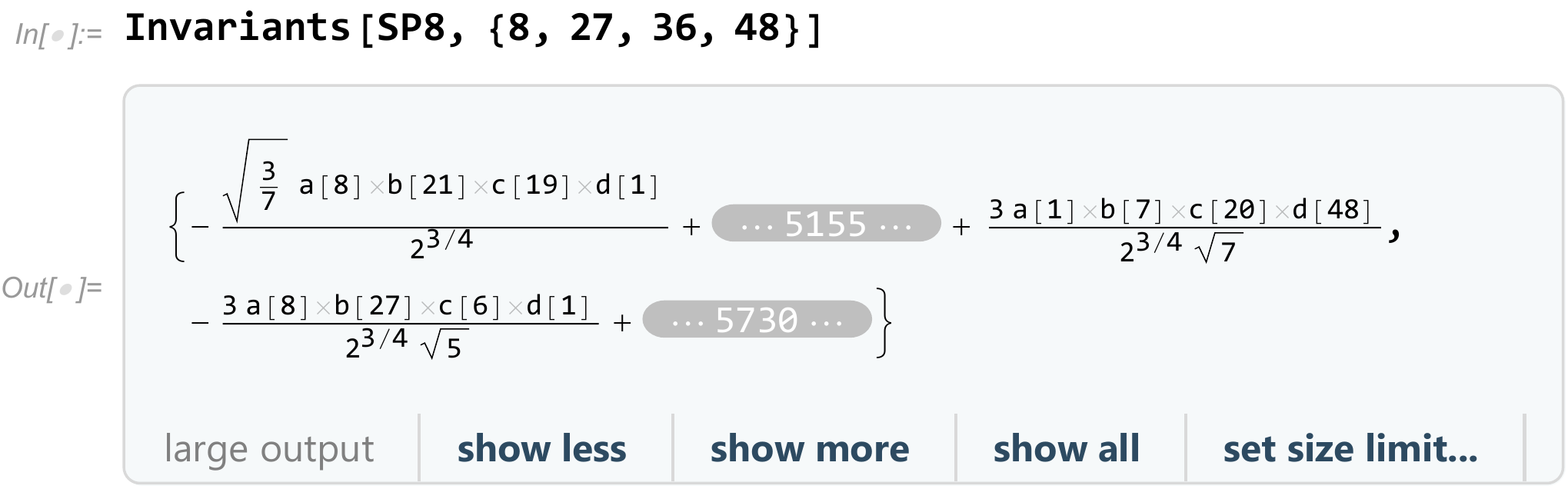}

\end{codeExample}

Let us now look at the function \texttt{\small{}IrrepInProduct}, which
is very similar to \texttt{\small{}Invariants}. Consider the $SU(2)$
product $\boldsymbol{2}\times\boldsymbol{2}\times\overline{\boldsymbol{3}}$:
if \texttt{\small{}$\sqrt{2}$} \texttt{\small{}a{[}1{]}} \texttt{\small{}b{[}1{]}}
\texttt{\small{}c{[}1{]}} + (\texttt{\small{}a{[}2{]}} \texttt{\small{}b{[}1{]}}
+ \texttt{\small{}a{[}1{]}} \texttt{\small{}b{[}2{]}}) \texttt{\small{}c{[}2{]}}
+ $\sqrt{2}$ \texttt{\small{}a{[}2{]}} \texttt{\small{}b{[}2{]}}
\texttt{\small{}c{[}3{]}} is an invariant, then surely the object
\begin{equation}
\left(\begin{array}{c}
\sqrt{2}\mathtt{a[1]}\mathtt{b[1]}\\
\mathtt{a[2]}\mathtt{b[1]}+\mathtt{a[1]}\mathtt{b[2]}\\
\sqrt{2}\mathtt{a[2]}\mathtt{b[2]}
\end{array}\right)
\end{equation}
transforms as a triplet $\boldsymbol{3}$ of $SU(2)$. The function
\texttt{\small{}IrrepInProduct} picks out such combinations which
transform irreducibly under a group.

\index{IrrepInProduct}

\begin{codeSyntax}
IrrepInProduct[<group>,<list of reprs in product>,<repr to be picked out>]
\end{codeSyntax}

\begin{codeExample}

\includegraphics[scale=0.62]{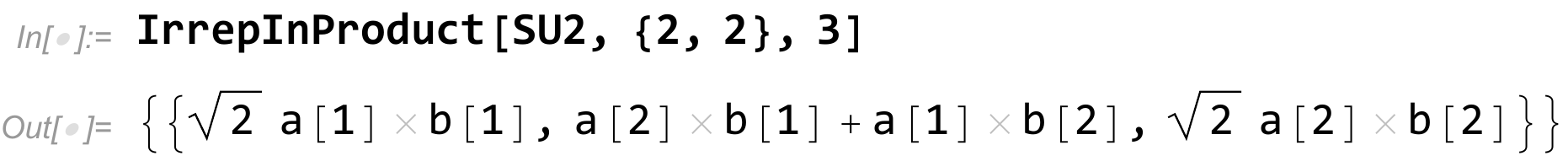}

\end{codeExample}

Given the similarity between the two functions, the comments made
in relation to \texttt{\small{}Invariants} apply to \texttt{\small{}IrrepInProduct}
as well. But there are two differences. The first one is that \texttt{\small{}IrrepInProduct}
does not perform a normalization of the expressions. The other difference
is that to conjugate the representations in the product (the two doublets
in the example above) one should use the option \texttt{\small{}ConjugateRepsInProduct},
while \texttt{\small{}ConjugateTargetRep} conjugates the representation
to be picked out. For instance, the combination of two doublets which
transforms as $\overline{\boldsymbol{3}}$ (as opposed to $\boldsymbol{3}$)
is the following:

\begin{codeExample}

\includegraphics[scale=0.62]{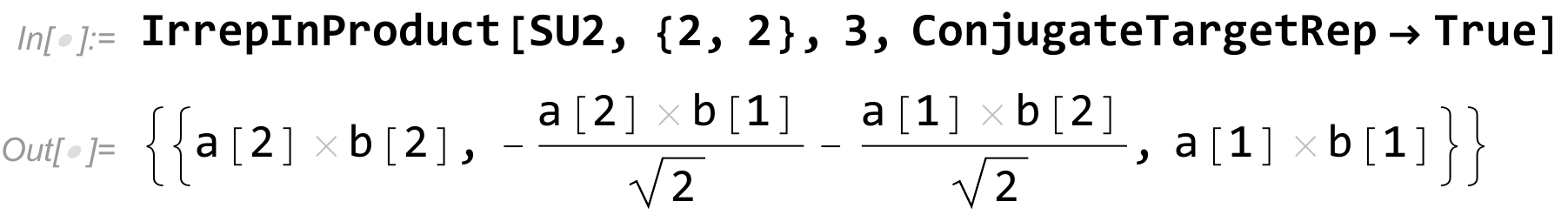}

\end{codeExample}

\noindent An even simpler example is this:

\begin{codeExample}

\hspace{-10mm}\includegraphics[scale=0.62]{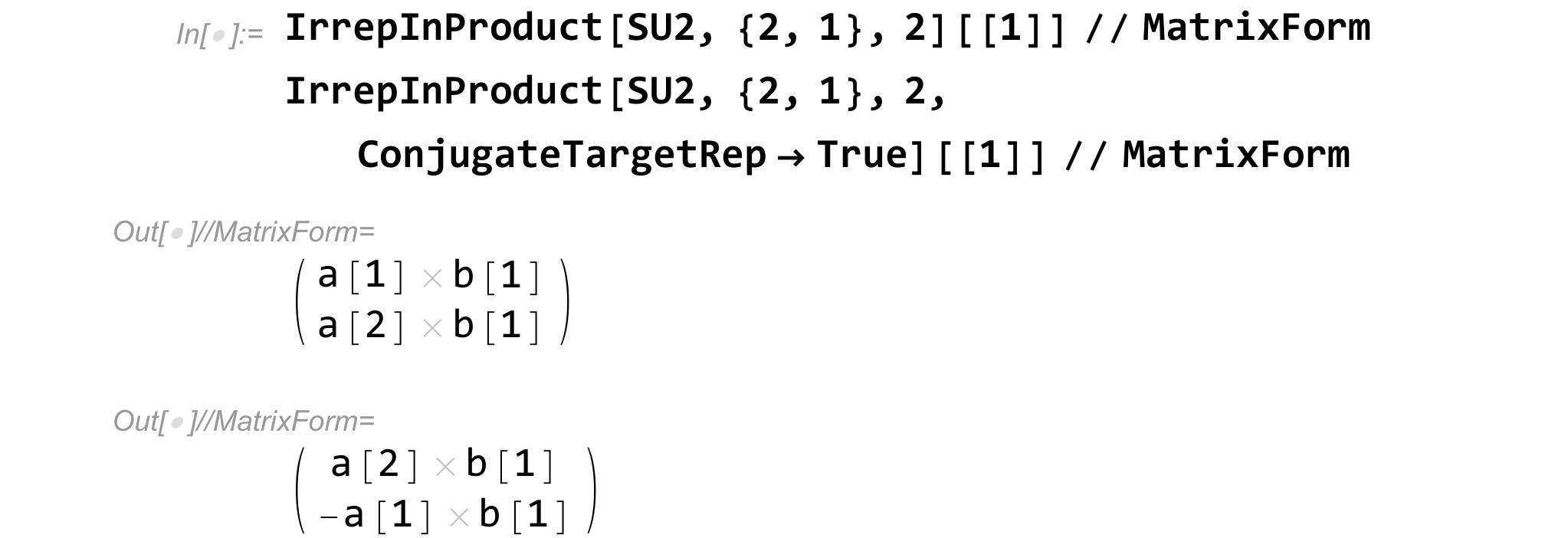}

\end{codeExample}

\subsection{\label{subsec:Symmetry-breaking}Symmetry breaking}

It is often important to know the subgroups $H$ of a Lie group $G$
and understand how the representations of $G$ behave when restricted
to the elements of $H$. In many situations, an irreducible representation
$R$ of $G$ decomposes into several irreducible representations of
$H$; this information is often referred to as the \textit{branching
rules} of $R$.

Concerning the groups themselves, $H$ is said to be a \textit{maximal}
subgroup of $G$ if there is no other subgroup of $G$ containing
$H$ (apart from $H$ and $G$ itself). For any subgroup $H$, by
tracing all chains of subgroups $G_{0}\equiv G\rightarrow G_{1}\rightarrow G_{2}\rightarrow\cdots\rightarrow G_{q}\equiv H$
such that $G_{i}$ is a maximal subgroup of $G_{i-1}$, one can find
all ways of embedding $H$ in $G$. \texttt{GroupMath} provides a
function \texttt{MaximalSubgroups} which lists the maximal subgroups
of a given group, and also \texttt{Embeddings} which can be used to
find all the non-equivalent embeddings of any group in a larger one
(obviating the need to track the chains of maximal subgroups just
described). The output of both these functions contains the so called
embedding \textit{projection matrices} $P$ which are important for
the calculation of branching rules. To be specific, a weight $\omega=\left(\omega_{1},\omega_{2},\cdots,\omega_{p}\right)$
of $G$ is associated with a weight $\omega^{\prime}=\left(\omega_{1}^{\prime},\omega_{2}^{\prime},\cdots,\omega_{p^{\prime}}^{\prime}\right)$
of $H$, with
\begin{equation}
\omega_{i}^{\prime}=P_{ij}\omega_{j}\,.
\end{equation}
Since the rank of $G$ ($=p$) is necessarily larger or equal to the
one of $H$ ($=p^{\prime}$), the projection matrix $P$ cannot have
more rows than columns.

Once a projection matrix matrix $P$ associated to a particular embedding
of $H$ in $G$ is known (with the help of the above functions, or
by any other means), it can be supplied to the function \texttt{DecomposeRep}
which finds the branching rules of any irreducible representation
of $G$.\footnote{It should be pointed out that projection matrices are not unique:
several of them can be associated to the same branching rules.}

\index{MaximalSubgroups}

\begin{codeSyntax}
MaximalSubgroups[<group>]
\end{codeSyntax}

\begin{codeExample}

\hspace{-3mm}\includegraphics[scale=0.61]{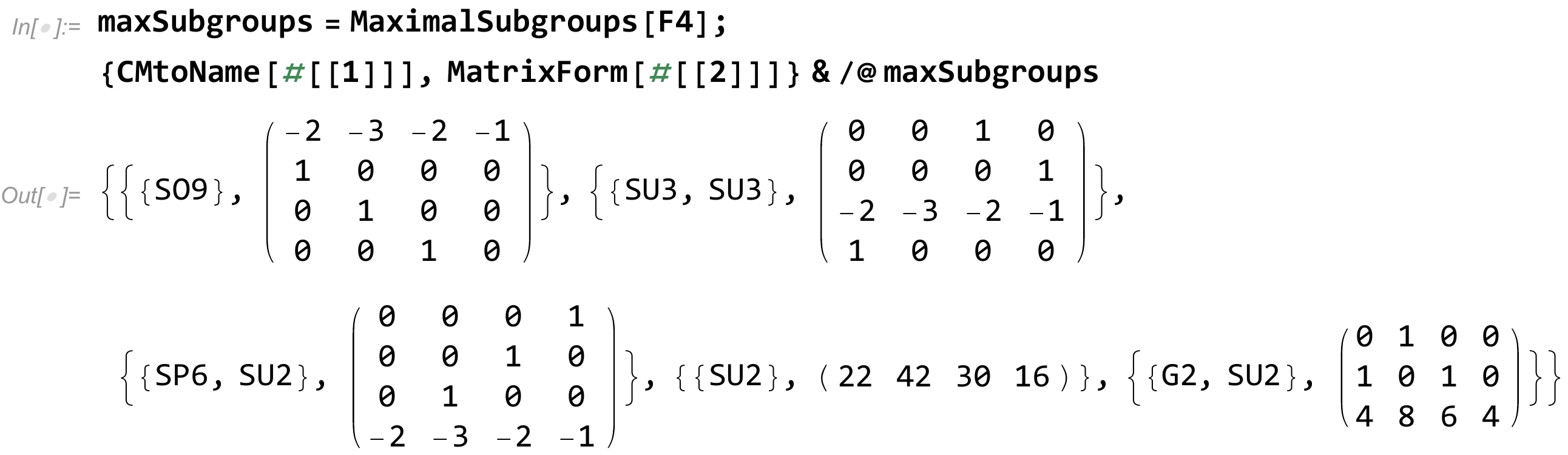}

\end{codeExample}

\noindent The output is a list with the format \texttt{\small{}\{<subgroup>
,<projection matrix>\}}. The \texttt{\small{}<subgroup>} is a list
of Cartan matrices, which can be converted to a human-friendly string
with the \texttt{\small{}CMtoName} function.\index{CMtoName} There
is also a function \texttt{DisplayEmbeddings} \index{DisplayEmbeddings}
which automatically prints a list with elements \texttt{\small{}\{<subgroup>,<projection
matrix>\}} in a grid:

\begin{codeExample}

\includegraphics[scale=0.62]{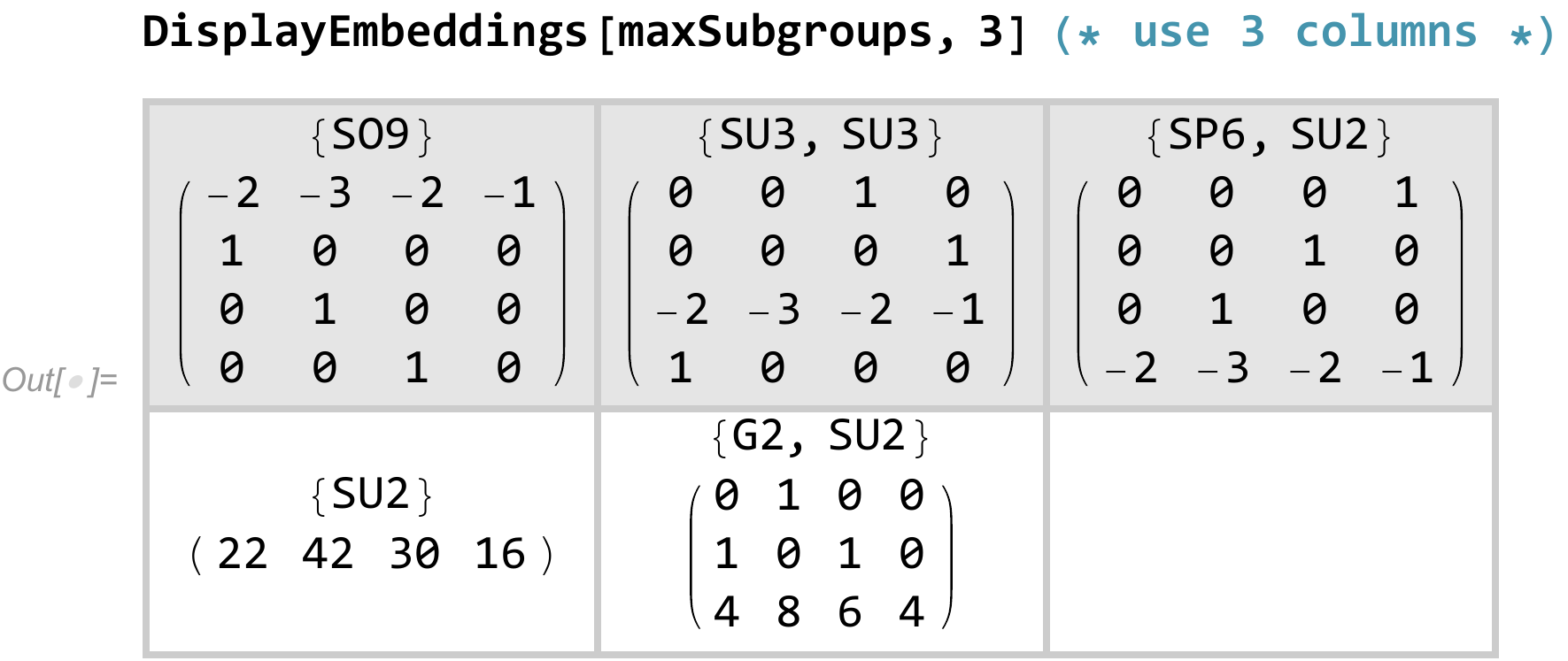}

\end{codeExample}

\index{Embeddings}

\begin{codeSyntax}
Embeddings[<group>,<subgroup>]
\end{codeSyntax}

\begin{codeExample}

\includegraphics[scale=0.62]{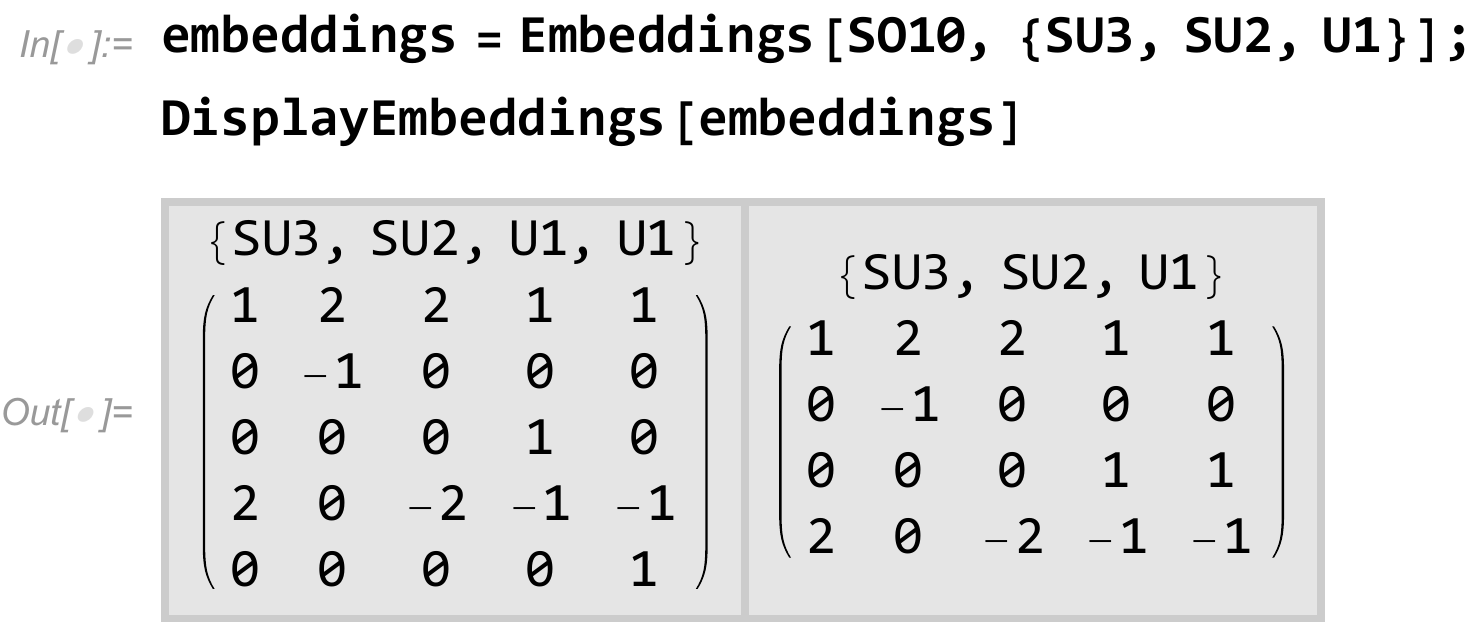}

\end{codeExample}

\noindent The output format of \texttt{Embeddings} is as the same
as the one of \texttt{\small{}MaximalSubgroups}. Concerning $U(1)$'s,
arbitrary linear combinations of their charges can be made to produce
the reduction $U(1)^{m}\rightarrow U(1)^{m^{\prime}\leq m}$, so this
process is controlled by continuous degrees of freedom (the coefficients
of the linear combinations). As such, \texttt{Embeddings} leaves this
to the user: the function will always return the maximum number of
allowed $U(1)$'s. In light of this, the only impact of indicating
\texttt{\small{}U1}'s as factors of the desired subgroup is that embeddings
which cannot accommodate a sufficient number of $U(1)$'s are ignored.
For example:

\begin{codeExample}

\includegraphics[scale=0.62]{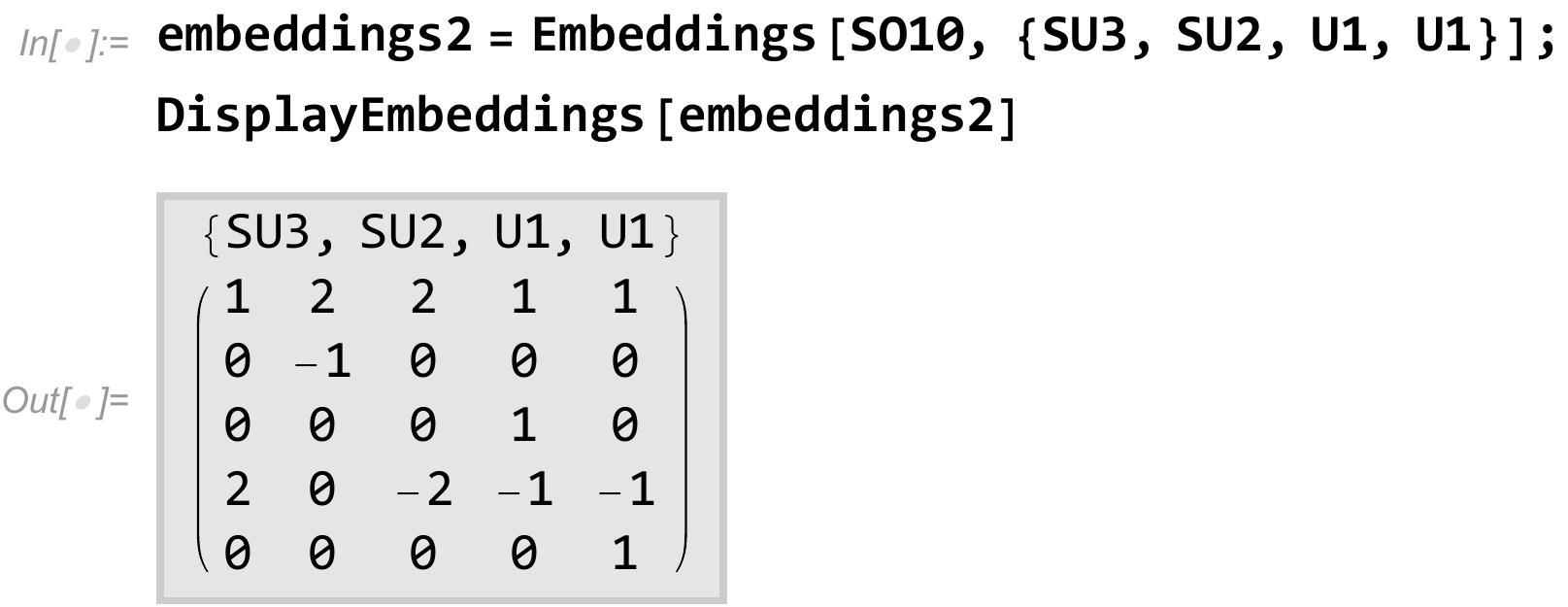}

\end{codeExample}

Note that the group $G$ does not need to be simple (an input such
as \texttt{Embeddings{[}\{SO10,E6\}, \{SU3,SU2,U1,U1\}{]}} is valid).

The embedding information (a subgroup and its associated projection
matrix) can then be passed on to \texttt{\small{}DecomposeRep}.

\index{DecomposeRep}

\begin{codeSyntax}
DecomposeRep[<group>,<representation of group>,<subgroup>,

<projection matrix>]
\end{codeSyntax}

\begin{codeExample}

\includegraphics[scale=0.62]{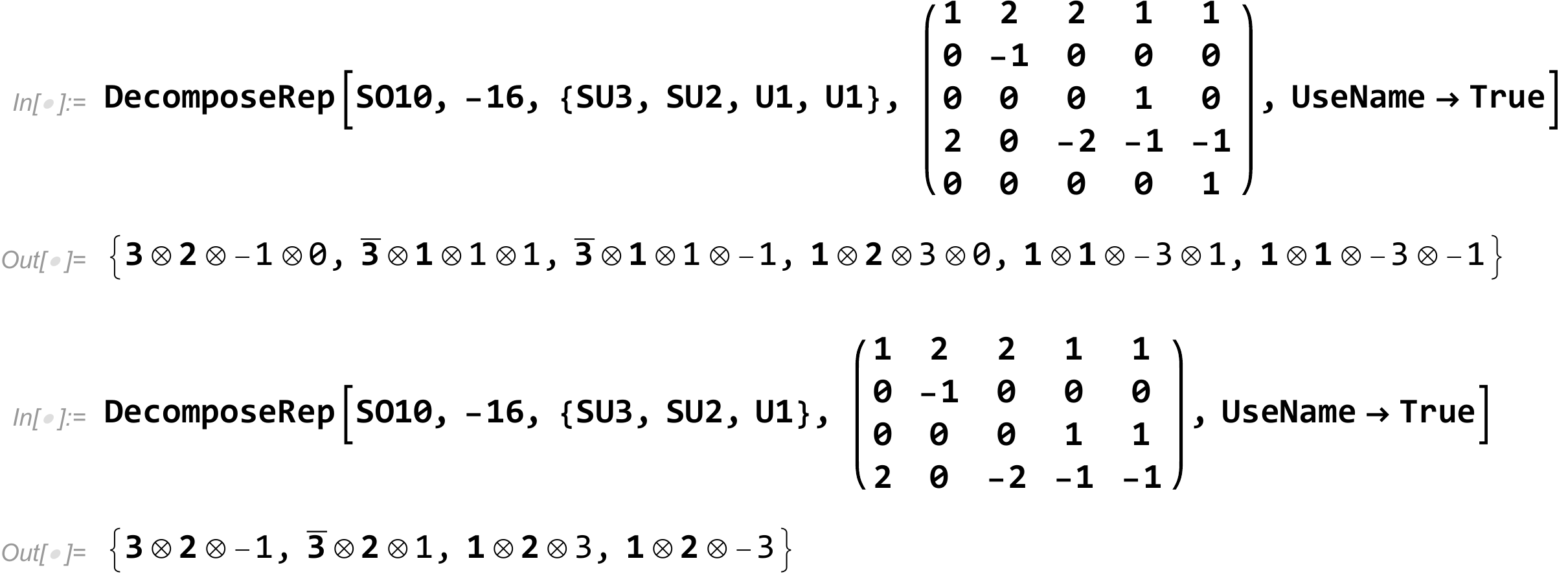}

\end{codeExample}

It is possible to omit the projection matrix. In that case, the function
\texttt{\small{}DecomposeRep} will not return any data; instead, it
will print a table with the decomposition of a given representation
under all possible embeddings of the subgroup:

\begin{codeExample}

\hspace{-3mm}\includegraphics[scale=0.61]{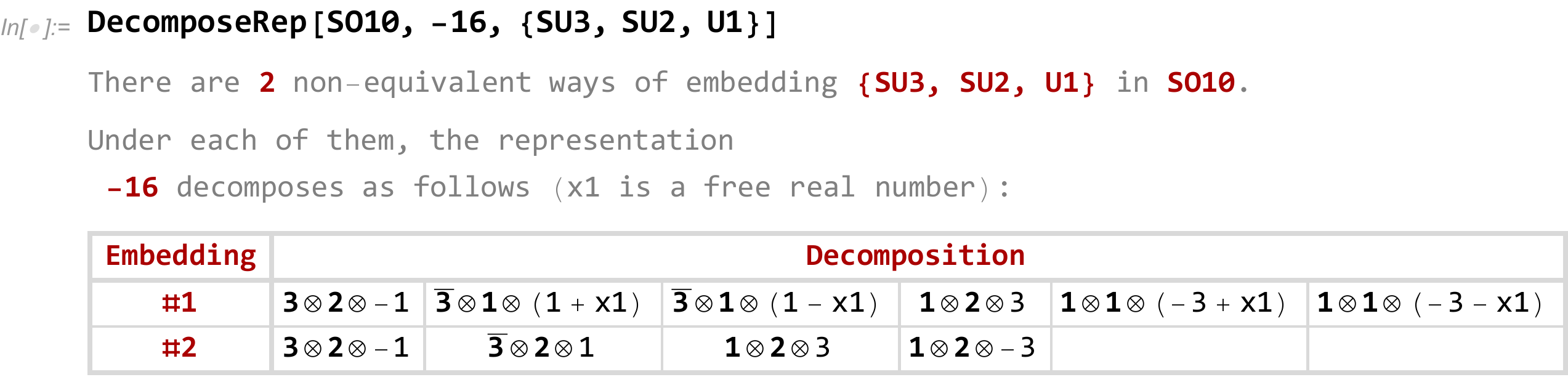}

\end{codeExample}

\noindent The \texttt{x1}, \texttt{x2}, ... variables which may appear
are free parameters which control the linear combinations of $U(1)$'s
to be preserved; they show up only when it is possible to embed more
$U(1)$'s than those requested by the user.

There are several important remarks to be made about the functions
just mentioned (in particular \texttt{\small{}Maximal}\texttt{\-}\texttt{\small{}Subgroups}
and \texttt{\small{}Embeddings}). Let us go through them one by one.

A maximal subalgebra $H$ of some $G$ can be classified as \textit{regular}
if its root space is generated by a subset of the root vectors of
$G$ and if the Cartan subalgebra of $H$ is contained in the one
of $G$.\footnote{Just by looking at the Dynkin diagram of $G$ and following simple
rules one can find all such subalgebras.} If this is not the case, $H$ is a \textit{special} maximal subalgebra.
By default the output of the function \texttt{MaximalSubgroups} \index{MaximalSubgroups}does
not distinguish these two types of subalgebras, but it will do so
if the option \texttt{RplusS -> True} is used.

Thanks to the seminar works \cite{Dynkin1957-1,Dynkin1957-2}, for
any Lie group $G$ it is rather straightforward to compile a list
of all subgroups $\left\{ H_{1},H_{2},\cdots,H_{i}\right\} $ which
can possibly be maximal subgroups. However, some of the $H_{i}$ in
such list might not be maximal (they might be contained in an another
subgroup $H_{j\neq i}$ in the list) and it is considerably harder
to establish which subgroups are actually maximal.

To appreciate this difficulty, note that it is not enough to compare
the size of various subgroups $H_{i}$. For example, even though it
is obviously possible to embed an $SU(2)$ group in $SU(2)\times U(1)$,
both are listed as maximal subgroups of $SU(3)$. The reason behind
this seemingly contradictory situation is that it is possible to embed
$SU(2)$ in $SU(3)$ in two inequivalent ways. In fact, under the
subgroup $SU(2)_{1}\subset SU(2)\times U(1)\subset SU(3)$ a triplet
of $SU(3)$ decomposes as a doublet plus a singlet ($\boldsymbol{3}\rightarrow\boldsymbol{2}+\boldsymbol{1}$),
while the triplet is still irreducible under $SU(2)_{2}\subset SU(3)$
($\boldsymbol{3}\rightarrow\boldsymbol{3}$).

The concept of \textit{inequivalent} embeddings of a subgroup which
was just invoked also requires further explanation. An embedding $\phi$
of a group $H$ in $G$ is a function which associates the elements
of $H$ to distinct elements of $G$ while preserving the structure
of $H$ ($\phi$ is said to be an injective homomorphism). Two embeddings
of a group $H$ in $G$ --- $\phi_{1}$ and $\phi_{2}$ --- are
said to be \textit{linearly equivalent} (henceforth just \textit{equivalent})
if for any representation $\rho$ of $G$ the maps $\rho\circ\phi_{1}$
and $\rho\circ\phi_{2}$ are isomorphic representations of $H$, which
is the same as saying that they lead to the same branching rules.\footnote{In practice, it is not necessary to check the branching rules for
all representations $\rho$ of $G$: it suffices to consider the defining
representation $\left(1,0,0,\cdots,0\right)$, except for $SO(2n)$
where it is also necessary to consider the branching rules of the
spinor representation $\left(0,\cdots,0,0,1\right)$.}

When compiling lists of embeddings with the functions \texttt{MaximalSubgroups}
and \texttt{Embeddings},\index{Embeddings} the program assumes this
definition of equivalence, with the following caveat. Two embeddings
might also be considered to be equivalent for the following reasons:
\begin{itemize}
\item A symmetry\footnote{The symmetries discussed here are given by the outer automorphism
group.} of the group or the subgroup. Consider for the example the embedding
of $SU(4)\times SU(3)$ in $SU(7)$ associated with the branching
rule $\boldsymbol{7}\rightarrow\left(\boldsymbol{4},1\right)+\left(\boldsymbol{1},\boldsymbol{3}\right)$.
Clearly, there is another embedding under which it is the $\overline{\boldsymbol{7}}$
branching in this way, and therefore $\boldsymbol{7}\rightarrow\left(\overline{\boldsymbol{4}},1\right)+\left(\boldsymbol{1},\overline{\boldsymbol{3}}\right)$.
Furthermore, by swapping the representations of each of the subgroup
factors --- $SU(4)$ and $SU(3)$ --- by their conjugates, we should
also have the embeddings $\boldsymbol{7}\rightarrow\left(\boldsymbol{4},\boldsymbol{1}\right)+\left(\boldsymbol{1},\overline{\boldsymbol{3}}\right)$
and $\boldsymbol{7}\rightarrow\left(\overline{\boldsymbol{4}},1\right)+\left(\boldsymbol{1},\boldsymbol{3}\right)$.
In this way, from one branching rule one can immediately infer the
existence of three others. We have complex conjugated representations,
and this operation is associated with a symmetry of the $SU(n)$ groups:
graphically, the Dynkin diagram is flipped by this operation. There
is also such a $Z_{2}$ symmetry for the groups $SO(2n)$ and $E(6)$,
although in the case of the $SO(4n)$ groups it is not associated
to complex conjugation. Note also that the Dynkin diagram of $SO(8)$
is symmetric under the larger $S_{3}$ group of permutations.
\item Permutations of equal factor groups. Consider the embedding of $SU(2)\times SU(2)$
in $Sp(6)$ associated with the branching rule $\boldsymbol{6}\rightarrow\left(\boldsymbol{4},\boldsymbol{1}\right)+\left(\boldsymbol{1},\boldsymbol{2}\right)$.
By swapping the two $SU(2)$'s of the subgroup, we may trivially derive
the embedding $\boldsymbol{6}\rightarrow\left(\boldsymbol{1},\boldsymbol{4}\right)+\left(\boldsymbol{2},\boldsymbol{1}\right)$.
An analogous consideration also applies to equal factors, if there
is any, of the group itself.
\end{itemize}
These are considered to be trivial variations of a single embedding
by \texttt{GroupMath}, hence they are factored out in the outputs.
The same criteria seems to have been adopted in reference \cite{Slansky:1981yr}
which lists the maximal subgroups of the simple Lie groups up to rank
8, and by \cite{Yamatsu:2015npn} which extends this list up to groups
of rank 20. A comparison of the output of \texttt{MaximalSubgroups}
\index{MaximalSubgroups}with these two references reveals the following
differences. With regards to the first reference \cite{Slansky:1981yr},
the program does not indicate $SU(2)\times SU(2)\times SU(2)$ as
a maximal subgroup of $SO(12)$ while $Sp(6)$ does appear on the
list of maximal subgroups of $Sp(14)$. In both instances, the problem
seems to lie with \cite{Slansky:1981yr} as have been reported in
the literature \cite{deGraaf:2019dni}. On the other hand, comparing
with table 226 of the second version of \cite{Yamatsu:2015npn}, \texttt{GroupMath}
indicates the followings additional special maximal subgroups: $SO(5)$
for the groups $Sp(20)$ and $SO(30)$, and $SO(8)$ for $SO(28)$.

The calculations performed by \texttt{MaximalSubgroups} are time consuming,
but fortunately this function only requires one input: a simple group
$G$. Hence \texttt{MaximalSubgroups} \index{MaximalSubgroups}was
pre-run on all $G$'s up to rank 25, and its results were saved to
a file which is loaded on startup (for larger groups which were not
pre-computed the running time can be long).

Up to version 1.0, \texttt{GroupMath} would find all embeddings of
some subgroup $H$ in a group $G$ by automating the algorithm which
is also used in computations done by hand: all chains of maximal subgroups
$G_{0}\equiv G\rightarrow G_{1}\rightarrow G_{2}\rightarrow\cdots\rightarrow G_{q}\equiv H$
would be calculated, and duplicated embeddings obtained in this way
would be eliminated. Unfortunately, this method can be very time consuming
when the ranks of $G$ and $H$ are very different. Therefore, the
function \texttt{Embeddings}\footnote{Previously called \texttt{FindAllEmbeddings}.}\index{Embeddings}
works differently in the latest version of the program. It implements
the idea described in \cite{Fonseca:2015aoa} for the particular case
$G=SU(n)$ and $H=SU(3)\times SU(2)\times U(1)$ but which can be
extended for any other classical group $G$, and any $H$. The algorithm
will not be explained in detail here; nevertheless we shall consider
the example $G=SU(14)$ and subgroup $H=SU(5)$. Rather than dealing
with an enormous number of chains of maximal subgroups from $G$ to
$H$, we might consider instead the possible branching rules of the
fundamental representation $\boldsymbol{14}$:
\begin{align}
\boldsymbol{14} & \rightarrow\boldsymbol{10}+4\times\boldsymbol{1}\,,\nonumber \\
 & \rightarrow2\times\boldsymbol{5}+4\times\boldsymbol{1}\,,\nonumber \\
 & \rightarrow\boldsymbol{5}+\overline{\boldsymbol{5}}+4\times\boldsymbol{1}\,,\nonumber \\
 & \rightarrow\boldsymbol{5}+9\times\boldsymbol{1}\,.\label{eq:SU(14)-SU(5)}
\end{align}
These are the only possibilities, factoring out the trivial variations
discussed earlier. Note that the $\boldsymbol{14}$ cannot branch
into 14 singlets of $SU(5)$ as that would imply that the algebra
generators $T^{a}$ of the subgroup are all null matrices.

Furthermore, the four possibilities in expression (\ref{eq:SU(14)-SU(5)})
correspond to valid embeddings. The reason is simple: the set of 24
matrices with dimensions $14\times14$ forming the indicated representations
of $SU(5)$ (which happen to be reducible) are certainly unitary and
all have unit determinant, therefore in all four cases they are a
subset of the $14^{2}-1$ matrices which form the fundamental representation
of $SU(14)$.

In this way, it is not hard to handle even bigger rank differences:
for example, there are 104 inequivalent ways of embeddings $SO(10)$
in $SU(100)$. Furthermore, from the branching rules of the fundamental
representation of $SU(n)$ it is straightforward to extract a projection
matrix which can then be used to decompose other representations.
The number of $U(1)$'s which commute with the subgroup $H$ is given
by the number irreps into which the fundamental representation of
$SU(n)$ decomposes, minus 1 (see \cite{Fonseca:2015aoa}). This means
that for the cases shown in expression (\ref{eq:SU(14)-SU(5)}), the
first embedding admits 4 extra $U(1)$'s --- i.e. $H=SU(5)\times U(1)^{4}$
would be possible --- while the second/third/forth embeddings can
accommodate up to 5/5/9 $U(1)$'s. Finally, with some adaptations
this technique can also be used on the other classical groups: $G=SO(n)$
and $Sp(2n)$.

~

There is one more function to be discussed in relation to symmetry
breaking, namely \texttt{\small{}Subgroup}\texttt{\-}\texttt{\small{}Coefficients}.\index{SubgroupCoefficients}
What it does, its input, as well as its output are rather elaborate,
so let us proceed in steps.

Consider an $SU(3)$ triplet $\left(t_{1},t_{2},t_{3}\right)^{T}$
and and anti-triplet $\left(\overline{t}_{1},\overline{t}_{2},\overline{t}_{3}\right)^{T}$
such that the combination $t_{i}\overline{t}_{i}$ is invariant under
this group. As mention above, it is possible to embed $SU(2)$ in
$SU(3)$ such that a triplet branches into a doublet plus a singlet.
To be explicit,
\begin{equation}
\left(\begin{array}{c}
t_{1}\\
t_{2}\\
t_{3}
\end{array}\right)=B_{1}\left(\begin{array}{c}
d_{1}\\
d_{2}\\
s
\end{array}\right)\quad\textrm{ and }\left(\begin{array}{c}
\overline{t}_{1}\\
\overline{t}_{2}\\
\overline{t}_{3}
\end{array}\right)=B_{2}\left(\begin{array}{c}
d_{1}^{\prime}\\
d_{2}^{\prime}\\
s^{\prime}
\end{array}\right)\,,
\end{equation}
where the $d_{i}$ and $d_{i}^{\prime}$ are the components of doublets,
while $s$ and $s^{\prime}$ are $SU(2)$ singlets; $B_{1}$ and $B_{2}$
are 3 by 3 unitary matrices. We may re-write the original $SU(3)$
invariant in terms of the new components, and from $SU(2)$ invariance
alone we know that
\begin{equation}
t_{1}\overline{t}_{1}+t_{2}\overline{t}_{2}+t_{3}\overline{t}_{3}=\alpha_{1}\left(d_{1}d_{2}^{\prime}-d_{2}d_{1}^{\prime}\right)+\alpha_{2}ss^{\prime}\,,
\end{equation}
and all that is left is to determine is $\alpha_{1}$ and $\alpha_{2}$.
These two numbers would be free if we were to ask only for invariance
under the $SU(2)$ subgroup; however, invariance under the full $SU(3)$
group fixes them (to be precise, it fixes their ratio). Note also
that the $\alpha_{i}$ depend on how we choose to normalize the $SU(2)$
and $SU(3)$ invariants, as well as on the matrices $B_{1}$ and $B_{2}$.
For example, a change $t_{i}\overline{t}_{i}\rightarrow2t_{i}\overline{t}_{i}$
would imply doubling the $\alpha_{i}$, while insertion of a minus
sign in the third column of $B_{1}$ would lead to a change of sign
for $\alpha_{2}$.

With these caveats in mind, \texttt{\small{}SubgroupCoefficients}
calculates the $\alpha_{i}$ coefficients and the $B_{i}$ matrices
for a product of representations, assuming the explicit representation
matrices returned by \texttt{\small{}RepMatrices} and the invariants
as normalized by \texttt{\small{}Invariants}. It works only for regular
embeddings.

\index{SubgroupCoefficients}

\begin{codeSyntax}
SubgroupCoefficients[<group>,<list of representations>,

<projection matrix>,<subgroup information>]
\end{codeSyntax}

Details on how to use this function can be found in the program's
built-in documentation files. For the example at hand, we would write

\begin{codeExample}

\includegraphics[scale=0.62]{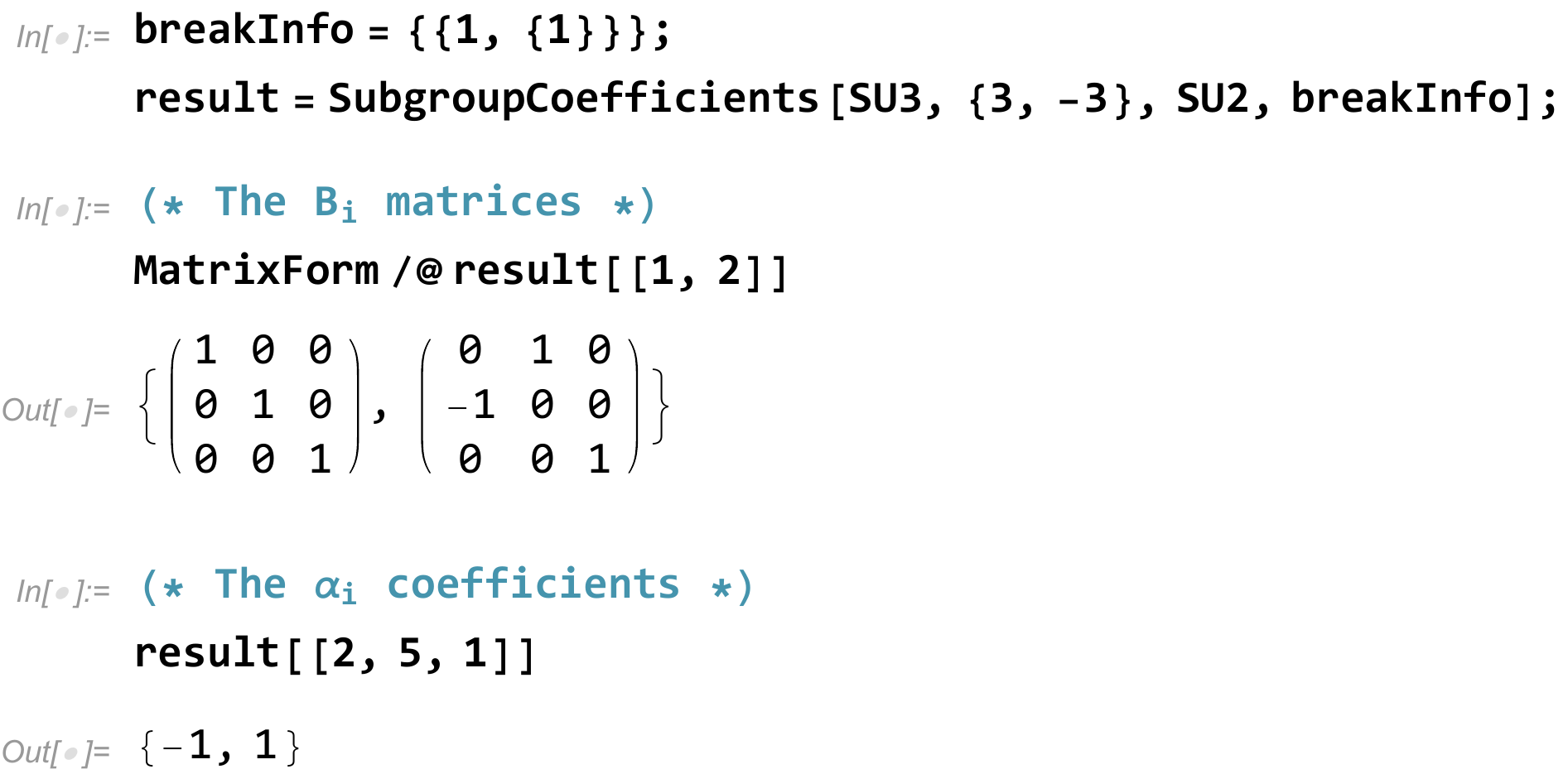}

\end{codeExample}

From the rest of the data saved in the variable \texttt{\small{}result}
--- which is not fully printed above --- one can tell that the first
number in the last output is the coefficient of the contraction of
the two doublets ($\alpha_{1}$) and the second number corresponds
to the singlets ($\alpha_{2}$). We can test this result with \texttt{\small{}Invariants}.
The relevant expressions are

\begin{codeExample}

\includegraphics[scale=0.62]{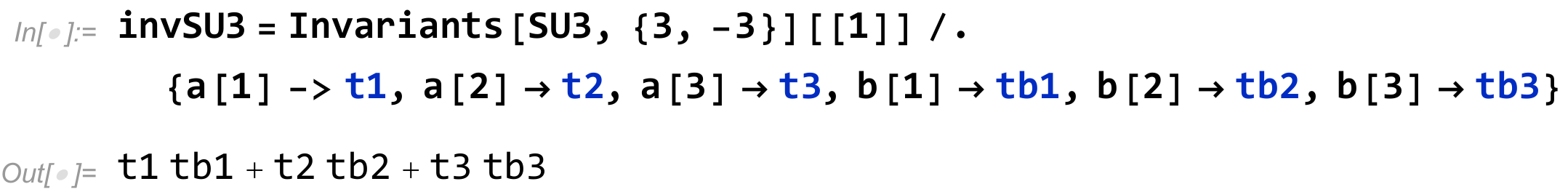}

\end{codeExample}

\begin{codeExample}

\includegraphics[scale=0.62]{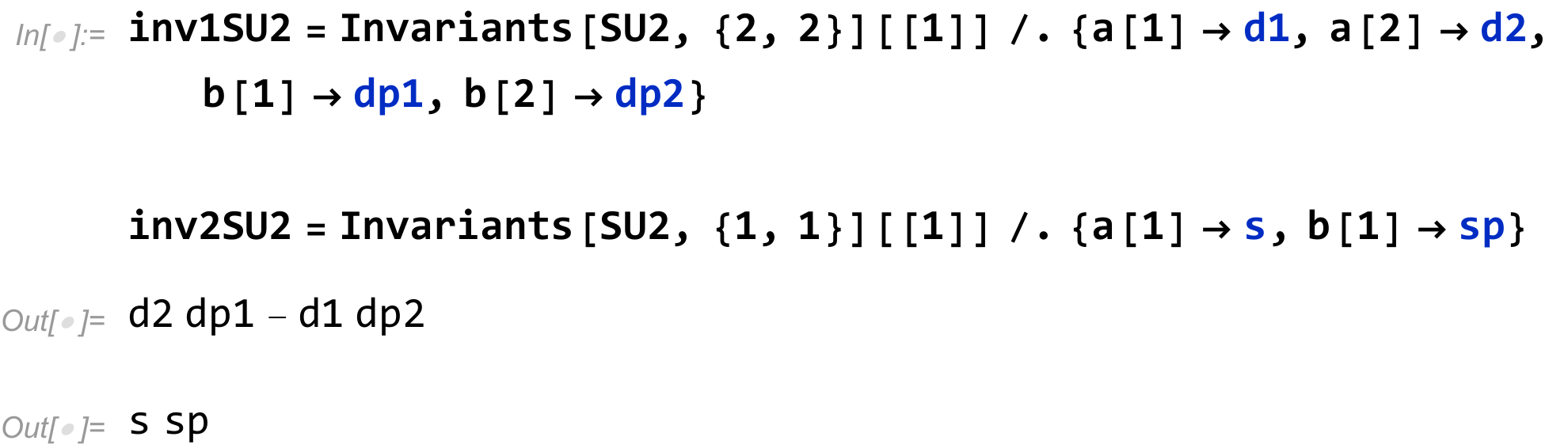}

\end{codeExample}

From the $B_{i}$ matrices printed above we get the dictionary \texttt{\small{}\{t1->d1,
t2->d2, t3->s, tb1->dp2, tb2->-dp1, tb3->sp\}} and with it we can
indeed make a correspondence between the $SU(3)$ and $SU(2)$ invariants,
using the coefficients $\left(\alpha_{1},\alpha_{2}\right)=\left(-1,1\right)$:

\begin{codeExample}

\includegraphics[scale=0.62]{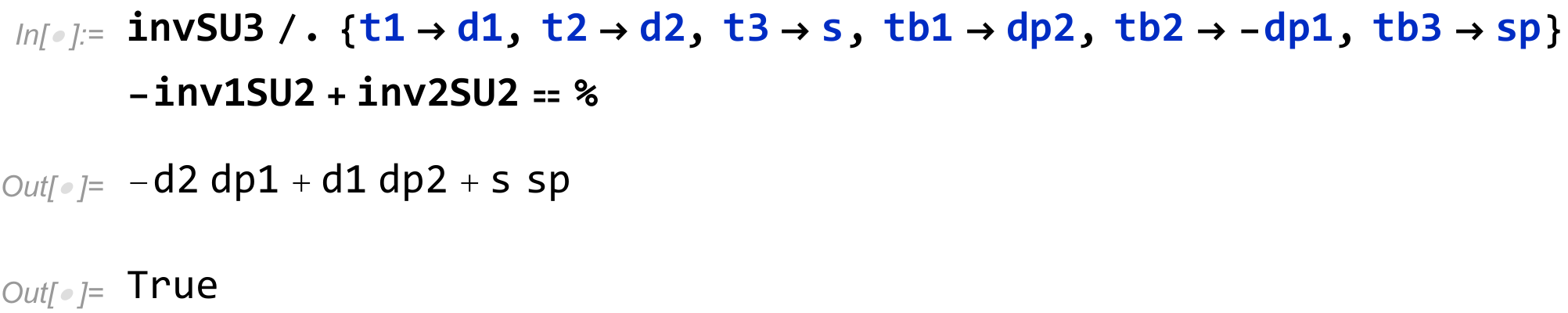}

\end{codeExample}

This is a simple example which can be worked out by hand; however,
more elaborate cases often need to be considered (see for example
the Georgi-Jarlskog relations \cite{Georgi:1979df}).\texttt{\small{} SubgroupCoefficients}
can handle products of an arbitrary number of representations, but
note that in all cases the interpretation of its output requires an
understanding of the corresponding output of the \texttt{RepMatrices}
and \texttt{\small{}Invariants} functions.\index{RepMatrices}\index{Invariants}

\section{\label{sec:Permutation-groups}Permutation groups}

The permutations of $m$ objects form a group which is usually called
$S_{m}$. The family of finite groups $S_{1}$, $S_{2}$, ... is important
in various contexts, including in the analysis of the decomposition
of direct products of Lie group representations (this is to be discussed
in subsection \ref{subsec:5.3}). Mostly because of this last connection,
\texttt{GroupMath} contains several functions related to the permutation
groups which are described in this section. The following is a quick
summary of the most important features of these groups.

The number of elements of the $S_{m}$ finite group (i.e. its order)
is $m!$. Each of its irreducible representations can be labeled with
a \textit{partition} $\lambda$ of $m$, which is a list of non-increasing
natural numbers $\lambda_{i}$ adding up to $m$.  For example, \{4\},
\{3,1\}, \{2,2\}, \{2,1,1\} and \{1,1,1,1\} are the possible partitions
of 4, therefore $S_{4}$ has 5 irreducible representations. A partition
$\lambda=\left\{ \lambda_{1},\lambda_{2},\cdots\right\} $ is very
often represented by a \textit{Young diagram} which contains $\lambda_{i}$
boxes in row $i$, with all rows left-aligned. For instance,\ytableausetup{centertableaux}\ytableausetup{boxsize=0.7em}
\begin{equation}
\{5,4,1,1\}=\ydiagram{5,4,1,1}\,.
\end{equation}

Partitions also play another role. As with any finite group, the elements
of $S_{m}$ can be organized in conjugacy classes and, in the case
of the permutation groups, the conjugacy classes themselves can be
labeled with partitions. Take $S_{4}$: using the cycle notation,
the 5 classes consist on the permutations of the form $\left(\,{\cdot}\,{\cdot}\,{\cdot}\,{\cdot}\,\right)$,
$\left(\,{\cdot}\,{\cdot}\,{\cdot}\,\right)\left(\,{\cdot}\,\right)$,
$\left(\,{\cdot}\,{\cdot}\,\right)\left(\,{\cdot}\,{\cdot}\,\right)$,
$\left(\,{\cdot}\,{\cdot}\,\right)\left(\,{\cdot}\,\right)\left(\,{\cdot}\,\right)$
and $\left(\,{\cdot}\,\right)\left(\,{\cdot}\,\right)\left(\,{\cdot}\,\right)\left(\,{\cdot}\,\right)$.\footnote{Consider four objects $a$, $b$, $c$ and $d$. The elements $\left(1423\right)$
and $\left(13\right)\left(24\right)$ correspond to the permutations
$\left\{ a,b,c,d\right\} \rightarrow\left\{ c,d,b,a\right\} $ and
$\left\{ a,b,c,d\right\} \rightarrow\left\{ c,d,a,b\right\} $.}

A \textit{Young tableau} with shape $\lambda$ is obtained by filling
the Young diagram of $\lambda$ with natural numbers. A \textit{standard
Young tableau} must be filled with the values 1, 2, ..., $m$ (assuming
$\lambda$ is a partition of $m$), and these numbers must be arranged
in such a way that they increase along each row (from left to right)
and along each column (from top to bottom). For example\ytableausetup{boxsize=1.3em}\begin{gather}\begin{ytableau}1 & 3 \\ 2&4\end{ytableau}\,.\end{gather}

Less stringent conditions apply to a \textit{semi-standard Young tableau}.\footnote{It is important to note that there is no consensus in the literature
on the definition of this type of tableaux.} It is filled with the natural numbers up to some value $c$ which
does not need to be related to the shape $\lambda$, and repetitions
are allowed. Numbers must increase along each column and they cannot
decrease along rows. For instance, there are three such tableaux for
$c=2$ and $\lambda=\left\{ 3,1\right\} $:\begin{gather}\begin{ytableau}1 & 1 & 1 \\ 2\end{ytableau}\,,\;\begin{ytableau}1 & 1 & 2 \\ 2\end{ytableau}\,,\;\begin{ytableau}1 & 2 & 2 \\ 2\end{ytableau}\;.\label{eq:tableaux-3-1}\end{gather}It
turns out that the number of standard Young tableaux with shape $\lambda$
is the same as the dimension of the irreducible representation (of
some permutation group) associated to this partition. On the other
hand, the number of semi-standard Young tableau of shape $\lambda$
and using the natural numbers up to $c$ is given by the so-called
Hook content formula (see \cite{Stanley:1999}).\footnote{This number also corresponds to the size of the irreducible representation
of $SU(c)$ associated to the partition or Young diagram $\lambda$:
for example $\lambda=\left\{ 3,1\right\} $ is associated to the triplet
representation of $SU(2)$ if $c=2$. The relation between partitions
and representations of the special unitary groups is explored in section
\ref{subsec:5.3}.}

The group $S_{m_{1}}\times S_{m_{2}}\times\cdots$ is contained in
the larger $S_{m}$ group with $m=\sum_{i}m_{i}$, and it might be
important to know how do the representations of these two groups relate.
For example the irreducible representation of $S_{5}$ associated
to $\lambda=\left\{ 4,1\right\} $ is reducible when restricted to
the subgroup $S_{2}\times S_{2}\times S_{1}$: it decomposes as two
copies of $\left(\left\{ 2\right\} ,\left\{ 2\right\} ,\left\{ 1\right\} \right)$
plus one copy of $\left(\left\{ 1,1\right\} ,\left\{ 2\right\} ,\left\{ 1\right\} \right)$
and $\left(\left\{ 2\right\} ,\left\{ 1,1\right\} ,\left\{ 1\right\} \right)$.
We can also ask what are the representations of $S_{5}$ which contain
the irreducible representation $\left(\left\{ 2\right\} ,\left\{ 2\right\} ,\left\{ 1\right\} \right)$
of $S_{2}\times S_{2}\times S_{1}$ (for example). The generic answer
to this type of question is provided by the well known Littlewood-Richardson
rule \cite{Littlewood-Richardson}. In this particular case, it provides
the following answer involving the functions $s_{\lambda}$ (these
are the Schur functions):
\begin{equation}
s_{\left\{ 2\right\} }s_{\left\{ 2\right\} }s_{\left\{ 1\right\} }=s_{\{5\}}+2s_{\{4,1\}}+2s_{\{3,2\}}+s_{\{2,2,1\}}+s_{\{3,1,1\}}\,.
\end{equation}
The definition and importance of the functions appearing here will
not be discussed further. Instead, I will simply mention that the
equation above can be interpreted as saying that $\left(\left\{ 2\right\} ,\left\{ 2\right\} ,\left\{ 1\right\} \right)$
is contained once in the irreducible representations $\left\{ 5\right\} $,
$\left\{ 2,2,1\right\} $ and $\left\{ 3,1,1\right\} $ of $S_{5}$,
and twice in both $\{4,1\}$ and $\{3,2\}$.

After this quick review of the main features of the permutation group
$S_{m}$, we move on to a description of the \texttt{GroupMath} functions
connected to it.

\subsection{Partitions, Young diagrams and Young tableaux}

It is possible to draw the Young diagram associated to a partition
$\lambda$ with the function \texttt{YoungDiagram}.

\index{YoungDiagram}

\begin{codeSyntax}
YoungDiagram[<partition>]
\end{codeSyntax}

\begin{codeExample}

\includegraphics[scale=0.62]{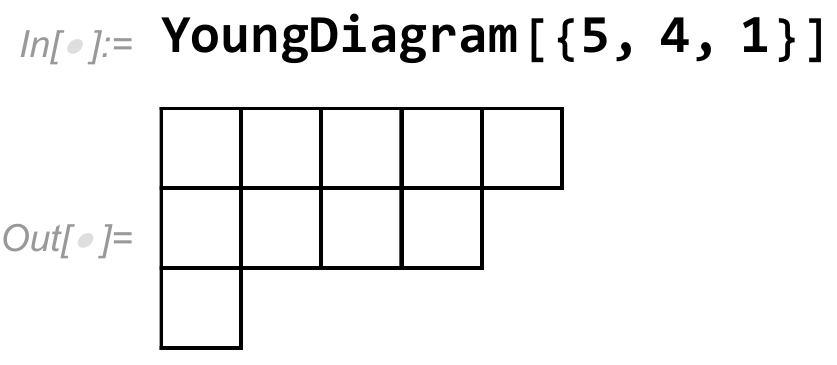}

\end{codeExample}

The standard Young diagrams associated with a given partition can
be computed by calling \texttt{GenerateStandardTableaux}. By default,
each tableau is given as a list of lists, following the format \texttt{\{<numbers
in row 1>, <numbers in row 2>, ...\}}. With the option \texttt{Draw}
set to \texttt{True}, they are instead printed as Young tableaux.

\index{GenerateStandardTableaux}

\begin{codeSyntax}
GenerateStandardTableaux[<partition>]
\end{codeSyntax}

\begin{codeExample}

\includegraphics[scale=0.62]{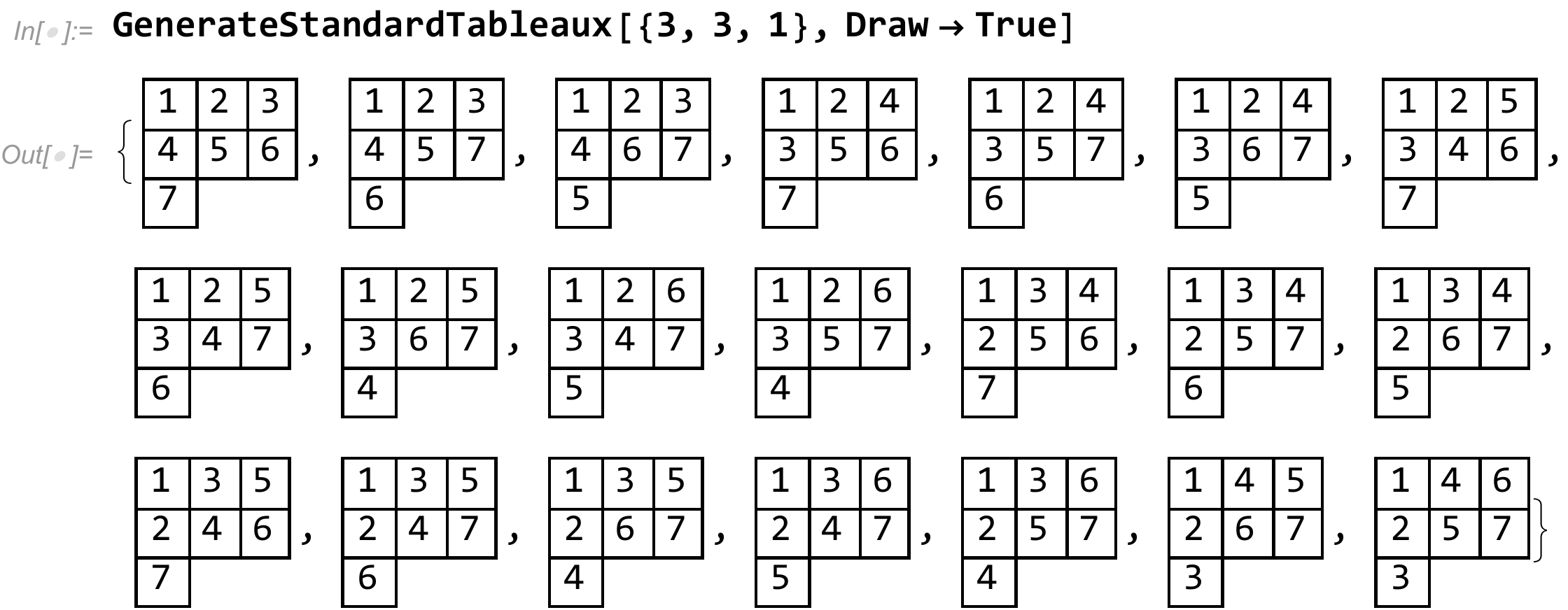}

\end{codeExample}

One can transpose Young diagrams and tableaux by converting their
rows into columns and vice-versa. The functions \texttt{TransposePartition}
and \texttt{TransposeTableaux} perform such operation.

\index{TransposePartition}

\begin{codeSyntax}
TransposePartition[<partition>]
\end{codeSyntax}

\index{TransposeTableaux}

\begin{codeSyntax}
TransposeTableaux[<tableaux>]
\end{codeSyntax}

\begin{codeExample}

\includegraphics[scale=0.62]{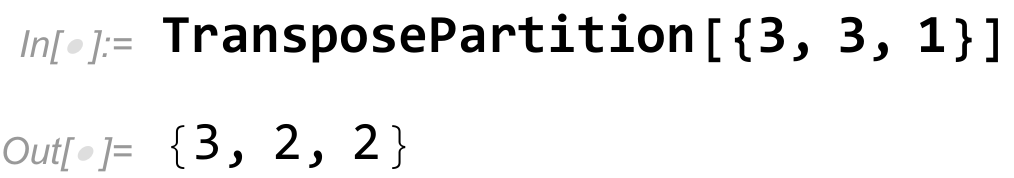}

\end{codeExample}

\begin{codeExample}

\includegraphics[scale=0.62]{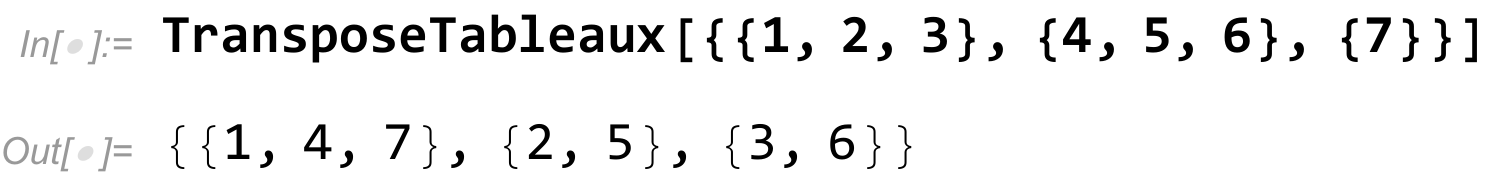}

\end{codeExample}

The number of semi-standard Young tableau with some shape $\lambda$
filled with the natural numbers $1,2,\cdots,c$ can be computed with
the function \texttt{HookContentFormula}.

\index{HookContentFormula}

\begin{codeSyntax}
HookContentFormula[<partition>,<c>]

\end{codeSyntax}

\begin{codeExample}

\includegraphics[scale=0.62]{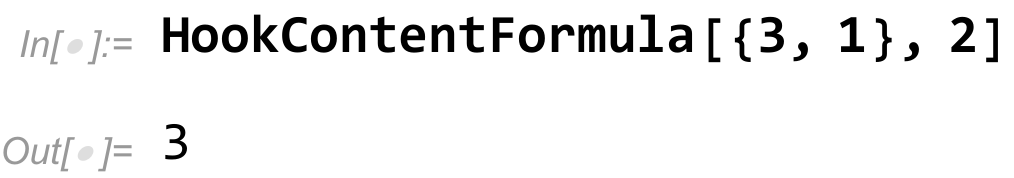}

\end{codeExample}

\noindent Note that the number of tableaux shown in (\ref{eq:tableaux-3-1})
is indeed 3.

\subsection{$S_{m}$ representations}

We will now consider functions involving the representations (and
classes) of the $S_{m}$ group. The dimension of an irreducible representation
given by a partition $\lambda$ can be calculated with \texttt{SnIrrepDim};
\texttt{SnClassOrder} computes the order (the number of elements)
of a class given by a partition $\mu$ and \texttt{SnClassCharacter}
calculates the character of a class $\mu$ in an irreducible representation
$\lambda$.

\index{SnIrrepDim}

\begin{codeSyntax}
SnIrrepDim[<irrep partition>]
\end{codeSyntax}

\begin{codeExample}

\includegraphics[scale=0.62]{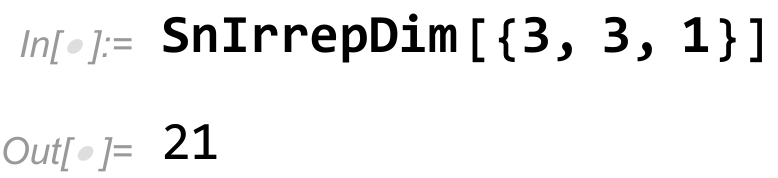}

\end{codeExample}

\index{SnClassOrder}

\begin{codeSyntax}
SnClassOrder[<class partition>]
\end{codeSyntax}

\begin{codeExample}

\includegraphics[scale=0.62]{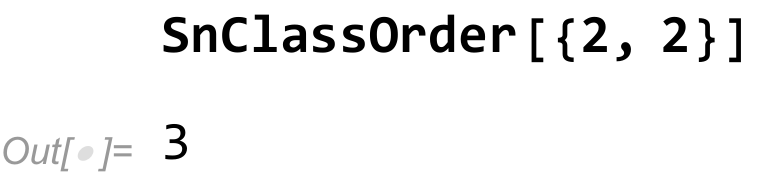}

\end{codeExample}

\index{SnClassCharacter}

\begin{codeSyntax}
SnClassCharacter[<irrep partition>,<class partition>]
\end{codeSyntax}

\begin{codeExample}

\includegraphics[scale=0.62]{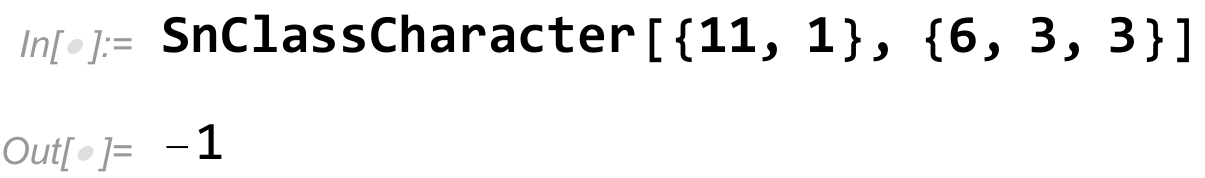}

\end{codeExample}

\noindent Here is a short code which prints the character table of
$S_{4}$ using \texttt{SnClassCharacter}:

\begin{codeExample}

\includegraphics[scale=0.62]{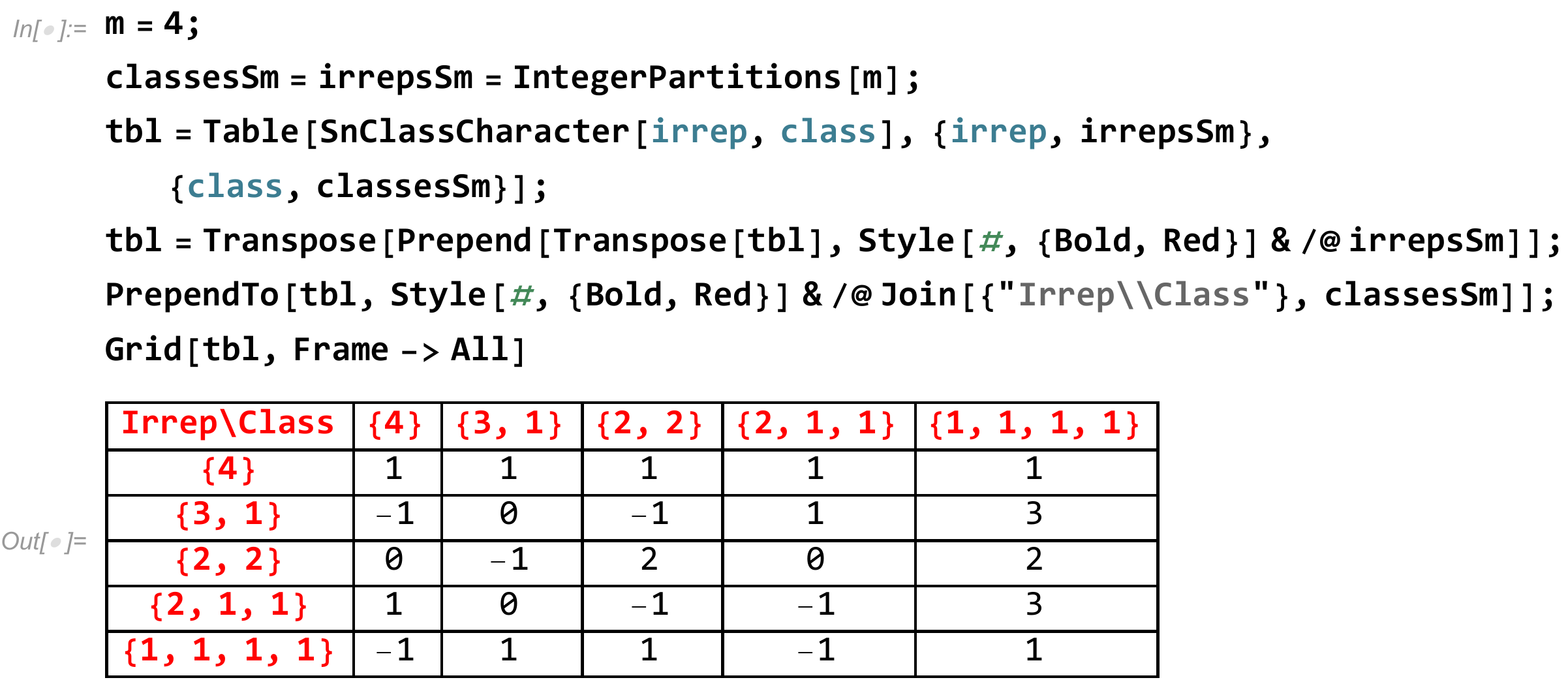}

\end{codeExample}

The matrices associated with an irreducible representation $\lambda$
can be calculated explicitly --- in a particular basis --- with
\texttt{SnIrrepGenerators}. However, note that the number of elements
of $S_{m}$ ($=m!$) grows rapidly with $m$, which means that even
for small $m$'s it would be unwise to provide the representation
matrices for each element. On the other hand, it is known that the
full group can be generated by just the transposition $\left(12\right)$
(exchange of the first two objects) and the cyclic permutation $\left(12\cdots m\right)$.\footnote{The remaining elements of $S_{m}$ can be obtained by repeated multiplication
of just these two.} Therefore \texttt{SnIrrepGenerators} provides only the matrices for
these two elements of the group. They are always real and orthogonal
(if this last property is not required, the option \texttt{OrthogonalizeGenerators->False}
can be used to significantly speed up the calculation).

\index{SnIrrepGenerators}

\begin{codeSyntax}
SnIrrepGenerators[<partition>]
\end{codeSyntax}

\begin{codeExample}

\includegraphics[scale=0.62]{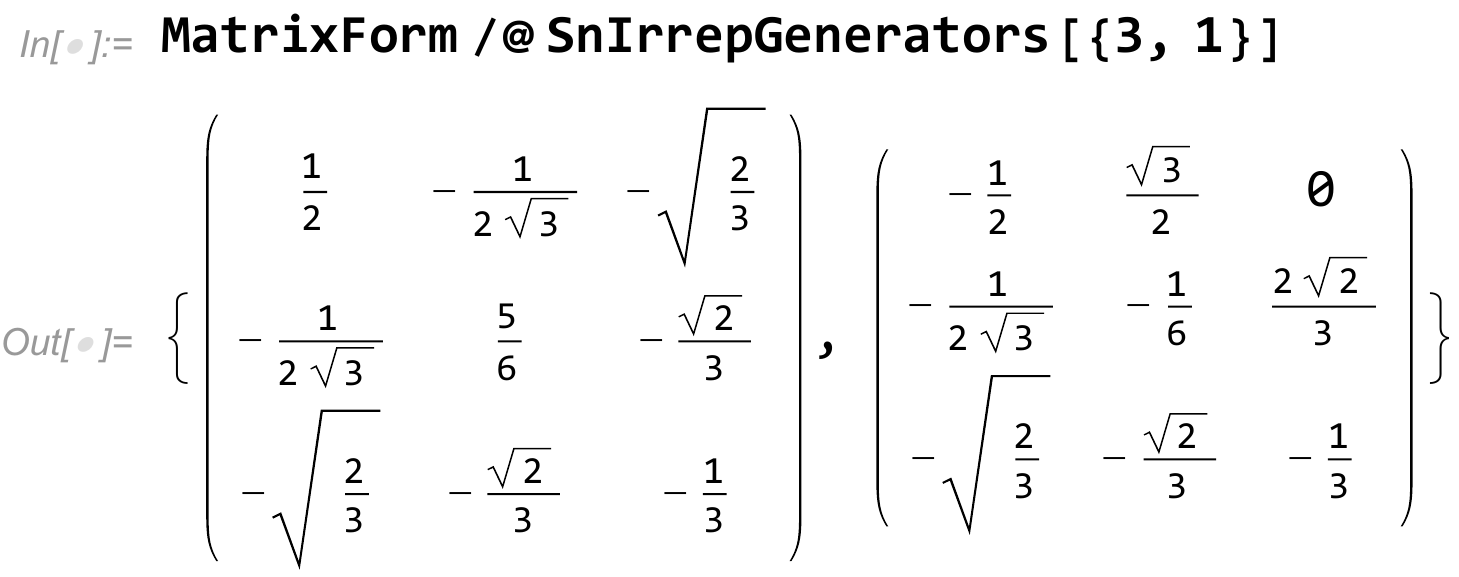}

\end{codeExample}

\begin{codeExample}

\includegraphics[scale=0.62]{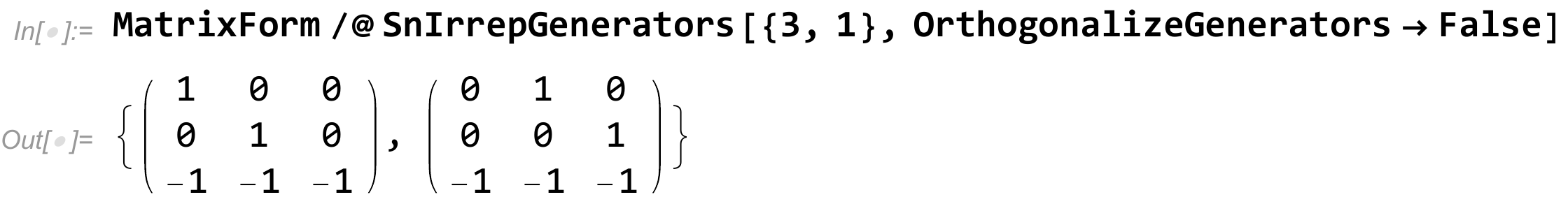}

\end{codeExample}

The product of various irreducible representations $\lambda_{i}$
of a permutation group can be decomposed into irreducible parts with
\texttt{DecomposeSnProduct}. The output is a list of the irreducible
representations contained in the product, and their multiplicity.
There is no limit to the number of representations being multiplied.

\index{DecomposeSnProduct}

\begin{codeSyntax}
DecomposeSnProduct[<list of partitions>]
\end{codeSyntax}

\begin{codeExample}

\includegraphics[scale=0.62]{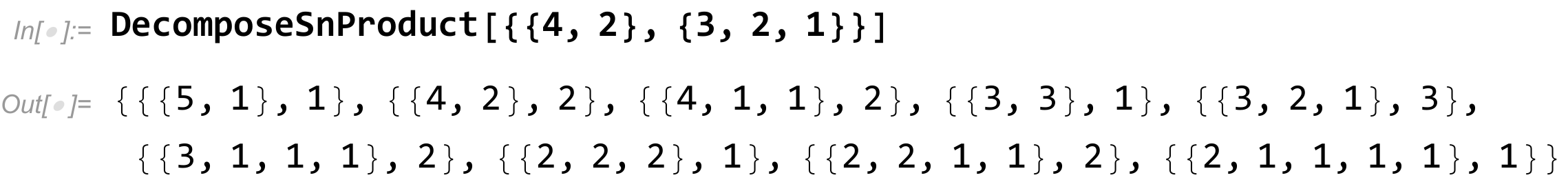}

\end{codeExample}

\noindent In this particular example, the output affirms that the
$S_{6}$ product $\left\{ 4,2\right\} \times\left\{ 3,2,1\right\} $
decomposes as $1\left\{ 5,1\right\} +2\left\{ 4,2\right\} +\cdots+1\left\{ 2,1,1,1,1\right\} $.

As mentioned in the brief summary of the features of permutations
groups, we may analyze what happens to $S_{m}$'s representations
under the subgroup $S_{m_{1}}\times S_{m_{2}}\times\cdots$ with $m_{1}+m_{2}+\cdots=m$.
The function \texttt{SnBranchingRules} calculates the decomposition
of an irreducible representation of $S_{m}$, while \texttt{LittlewoodRichardsonCoefficients}
computes all irreps of $S_{m}$ which contain a given representation
of the subgroup.

\index{SnBranchingRules}

\begin{codeSyntax}
SnBranchingRules[<Sm irrep partition>,<\{m1,m2,...\}>]
\end{codeSyntax}

\begin{codeExample}

\includegraphics[scale=0.62]{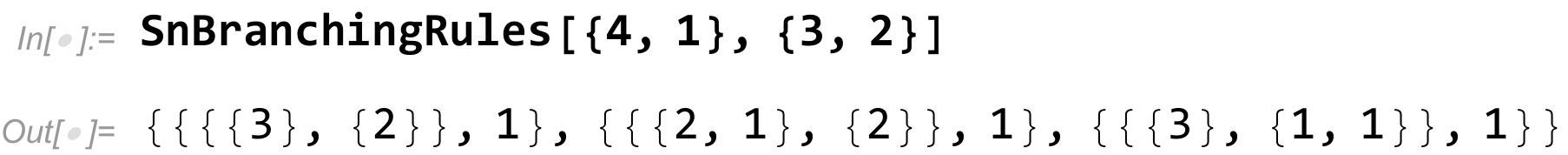}

\end{codeExample}

\noindent The first argument of this function (\texttt{\{4,1\}}) must
be a partition which identifies an irreducible representation of some
$S_{m}$ group. The second argument (\texttt{\{3,2\}}) is simply a
list of the indices $m_{i}$ of the subgroup $S_{m_{1}}\times S_{m_{2}}\times\cdots$.
The output for the example shown here implies that the irreducible
representation $\left\{ 4,1\right\} $ of $S_{5}$ decomposes as $1\left(\left\{ 3\right\} ,\left\{ 2\right\} \right)+1\left(\left\{ 2,1\right\} ,\left\{ 2\right\} \right)+1\left(\left\{ 3\right\} ,\left\{ 1,1\right\} \right)$
under $S_{3}\times S_{2}$.

\index{LittlewoodRichardsonCoefficients}

\begin{codeSyntax}
LittlewoodRichardsonCoefficients[<list of partitions>]
\end{codeSyntax}

\begin{codeExample}

\includegraphics[scale=0.62]{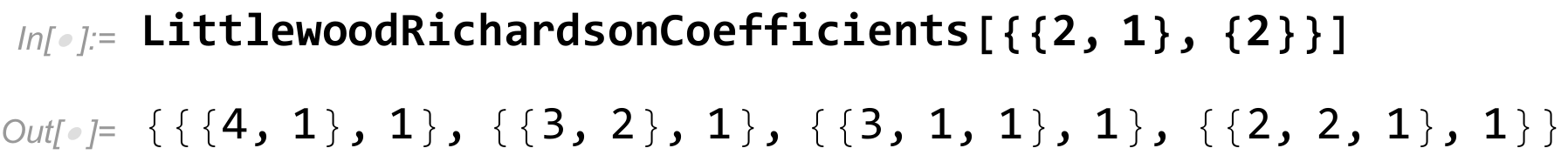}

\end{codeExample}

\noindent In the code above, the irreducible representation $\left(\left\{ 2,1\right\} ,\left\{ 2\right\} \right)$
of $S_{3}\times S_{2}$ was passed to the function \texttt{LittlewoodRichardsonCoefficients}.
The output indicates that it can be found once in each of the following
$S_{5}$ irreps: $\left\{ 4,1\right\} $ (in agreement with the previous
example where the \texttt{SnBranchingRules} function was used), $\left\{ 3,2\right\} $,
$\left\{ 3,1,1\right\} $ and $\left\{ 2,2,1\right\} $.

\subsection{Permutation symmetry of products of Lie group representations\label{subsec:5.3}}

The decomposition of products of Lie group representations can be
performed with the function \texttt{ReduceRepProduct}\index{ReduceRepProduct}
described earlier. However, when there are repeated representations
in these products, it is possible to say more about their decomposition
in irreducible representations of the Lie group.

Consider these two examples:
\begin{alignat}{1}
\boldsymbol{2}\times\boldsymbol{2} & =\boldsymbol{1}_{A}+\boldsymbol{3}_{S}\quad[SU(2)]\\
\boldsymbol{10}\times\boldsymbol{10} & =\boldsymbol{1}_{S}+\boldsymbol{45}_{A}+\boldsymbol{54}_{S}\quad[SO(10)]
\end{alignat}
In both cases the subscripts indicate what happens to an irrep under
permutations of the representations on the left. More generally, we
may think of the direct product
\[
V^{m}=\underbrace{V\times V\times\cdots V}_{m}
\]
of a vector space $V$ with dimension $n$. We take $\left|1\right\rangle ,\left|2\right\rangle ,\cdots,\left|n\right\rangle $
to be basis vectors of $V$, in which case the set of all $\left|i_{1}i_{2}\cdots i_{m}\right\rangle \equiv\left|i_{1}\right\rangle \left|i_{2}\right\rangle \cdots\left|i_{m}\right\rangle $
form a basis of $V^{m}$. Under permutations $\pi$ of the $V$ factors,
\begin{equation}
\left|i_{1}i_{2}\cdots i_{m}\right\rangle \overset{\pi}{\rightarrow}\left|i_{\pi\left(1\right)}i_{\pi\left(2\right)}\cdots i_{\pi\left(m\right)}\right\rangle \,,\label{eq:pi-operation}
\end{equation}
the $n^{m}$-dimensional vector space $V^{m}$ decomposes into irreducible
subspaces $V_{\lambda}^{m}$ labeled by partitions $\lambda$ of $m$.
This means that under permutations the elements of $V_{\lambda}^{m}$
do not mix with those of another $V_{\lambda^{\prime}\neq\lambda}^{m}$.
For example, in the case of $m=2$ and $n=3$ we have a 9-dimensional
vector space $V^{2}$ which decomposes into $V_{\{2\}}^{2}\equiv V_{S}^{2}$
and $V_{\left\{ 1,1\right\} }^{2}\equiv V_{A}^{2}$ (the symmetric
and the anti-symmetric subspaces). The first space, $V_{\{2\}}^{2}$,
is generated by the six vectors
\begin{equation}
\left\{ \left|11\right\rangle ,\left(\left|12\right\rangle +\left|21\right\rangle \right)/\sqrt{2},\left(\left|13\right\rangle +\left|31\right\rangle \right)/\sqrt{2},\left|22\right\rangle ,\left(\left|23\right\rangle +\left|32\right\rangle \right)/\sqrt{2},\left|33\right\rangle \right\} 
\end{equation}
while
\begin{equation}
\left\{ \left(\left|12\right\rangle -\left|21\right\rangle \right)/\sqrt{2},\left(\left|13\right\rangle -\left|31\right\rangle \right)/\sqrt{2},\left(\left|23\right\rangle -\left|32\right\rangle \right)/\sqrt{2}\right\} 
\end{equation}
forms a basis of $V_{\left\{ 1,1\right\} }^{2}$. Permutations do
not mix the two subspaces.

On the other hand, any linear transformation on $V$ induces a linear
transformation $T$ on $V^{m}$ and crucially the action of $T$ commutes
with the operation $\pi$ shown in (\ref{eq:pi-operation}). Hence
$V^{m}$ is decomposable into irreducible subspaces under the action
of $T\times\pi$. Since this is true for any transformation $T$ induced
by a linear transformation on $V$, it is certainly true if we consider
those forming a Lie group.

In the case of $m=3$ and $n=6$, the relevant subspaces are $V_{\{3\}}^{3}$
(of dimension 56), $V_{\{2,1\}}^{3}$ (of dimension 140) and $V_{\{1,1,1\}}^{3}$
(of dimension 20). Now take for example the 6-dimensional irreducible
representation of $SU(3)$ acting on $V$: $V_{\{3\}}^{3}$ decomposes
under $SU(3)$ as $\boldsymbol{1}+\boldsymbol{27}+\overline{\boldsymbol{28}}$,
$V_{\{2,1\}}^{3}$ decomposes as $2\left(\boldsymbol{8}+\boldsymbol{27}+\overline{\boldsymbol{35}}\right)$
and $V_{\{1,1,1\}}^{3}$ decomposes as $\boldsymbol{10}+\overline{\boldsymbol{10}}$.
It is no coincidence that $V_{\{2,1\}}^{3}$ contains two copies of
the $SU(3)$ representations $\boldsymbol{8}$, $\boldsymbol{27}$
and $\overline{\boldsymbol{35}}$: it is a consequence of the fact
that under $S_{3}$ permutations the entire subspace $V_{\{2,1\}}^{3}$
transforms as the 2-dimensional irreducible representation $\{2,1\}$.
In other words, $V_{\{2,1\}}^{3}$ transforms under the irreducible
representations $\left(\boldsymbol{8},\left\{ 2,1\right\} \right)$,
$\left(\boldsymbol{27},\left\{ 2,1\right\} \right)$ and $\left(\overline{\boldsymbol{35}},\left\{ 2,1\right\} \right)$
of $SU(3)\times S_{3}$. Using subscripts for the $S_{3}$ irreps,
the full decomposition reads
\begin{equation}
\boldsymbol{6}\times\boldsymbol{6}\times\boldsymbol{6}=\boldsymbol{1}_{\{3\}}+\boldsymbol{27}_{\{3\}}+\overline{\boldsymbol{28}}_{\{3\}}+\boldsymbol{8}_{\{2,1\}}+\boldsymbol{27}_{\{2,1\}}+\overline{\boldsymbol{35}}_{\{2,1\}}+\boldsymbol{10}_{\{1,1,1\}}+\overline{\boldsymbol{10}}_{\{1,1,1\}}\,.
\end{equation}

It is often important to have this kind of data (see for instance
\cite{Fonseca:2019yya}). In those cases, the function \texttt{ReduceRepProduct}\index{ReduceRepProduct}
is insufficient. However, at least in principle it is easy to recover
the permutation group information. Consider some representation $R$
of $G$ with weights $\omega_{i}$, $i=1,\cdots,n$. The $n^{2}$-dimensional
reducible representation $R^{2}$ has weights $\omega_{i}+\omega_{j}$,
with $i$ and $j$ taking all values from 1 to $n$. However, $R^{2}$
splits in a symmetric subspace $R_{\{2\}}^{2}$ and an anti-symmmetric
one $R_{\{1,1\}}^{2}$. The weights associated with the former are
$\omega_{i}+\omega_{j}$ with $1\leq i\leq j\leq n$, while the weights
associated with the latter are $\omega_{i}+\omega_{j}$ with $1\leq i<j\leq n$.
More generally, the rule is as follows. Weights $\omega_{i}$ are
a reference to the entries of the diagonal elements of a Lie algebra
$G$ in some representation $R$, hence the character (i.e. trace)
of the actual group transformation matrices are the sum $\sum_{j}\exp\left(i\omega_{j}\right)\equiv\sum_{j}x_{j}$.
Then, the character of $R_{\lambda}^{m}$, where $\lambda$ is a partition
of $m$, are given by the Schur polynomial $s_{\lambda}\left(x_{1},x_{2},\cdots,x_{n}\right)$.
For example, $s_{\{2\}}\left(x_{1},x_{2},\cdots,x_{n}\right)=\sum_{1\leq i\leq j\leq n}x_{i}x_{j}$
hence from each monomial we extract the weights $\omega_{i}+\omega_{j}$
with $1\leq i\leq j\leq n$, which matches what was expected for $R_{\{2\}}^{2}$.
On the other hand, $s_{\{1,1\}}\left(x_{1},x_{2},\cdots,x_{n}\right)=\sum_{1\leq i<j\leq n}x_{i}x_{j}$
which yield the weights $\omega_{i}+\omega_{j}$ with $1\leq i<j\leq n$
for $R_{\{1,1\}}^{2}$.

Since Schur polynomials can be readily calculated, is quite easy to
calculate the decomposition of $R^{m}$ in irreducible spaces under
the joint action of the Lie group and the permutation group, somewhat
in analogy with what can be done to decompose $R^{m}$ disregarding
the permutation group. However a brute force approach will only work
as long as the dimension of $R^{m}$ in not too large. Fortunately,
the authors of the computer code \texttt{LIE} describe in the program's
manual \cite{LieProgram} an algorithm do to the computation efficiently.
Following the nomenclature used in LIE, this algorithm was implemented
in a function called \texttt{Plethysms}.

\index{Plethysms}

\begin{codeSyntax}
Plethysms[<simple Lie group>,<representation>,<partition>]
\end{codeSyntax}

\begin{codeExample}

\includegraphics[scale=0.62]{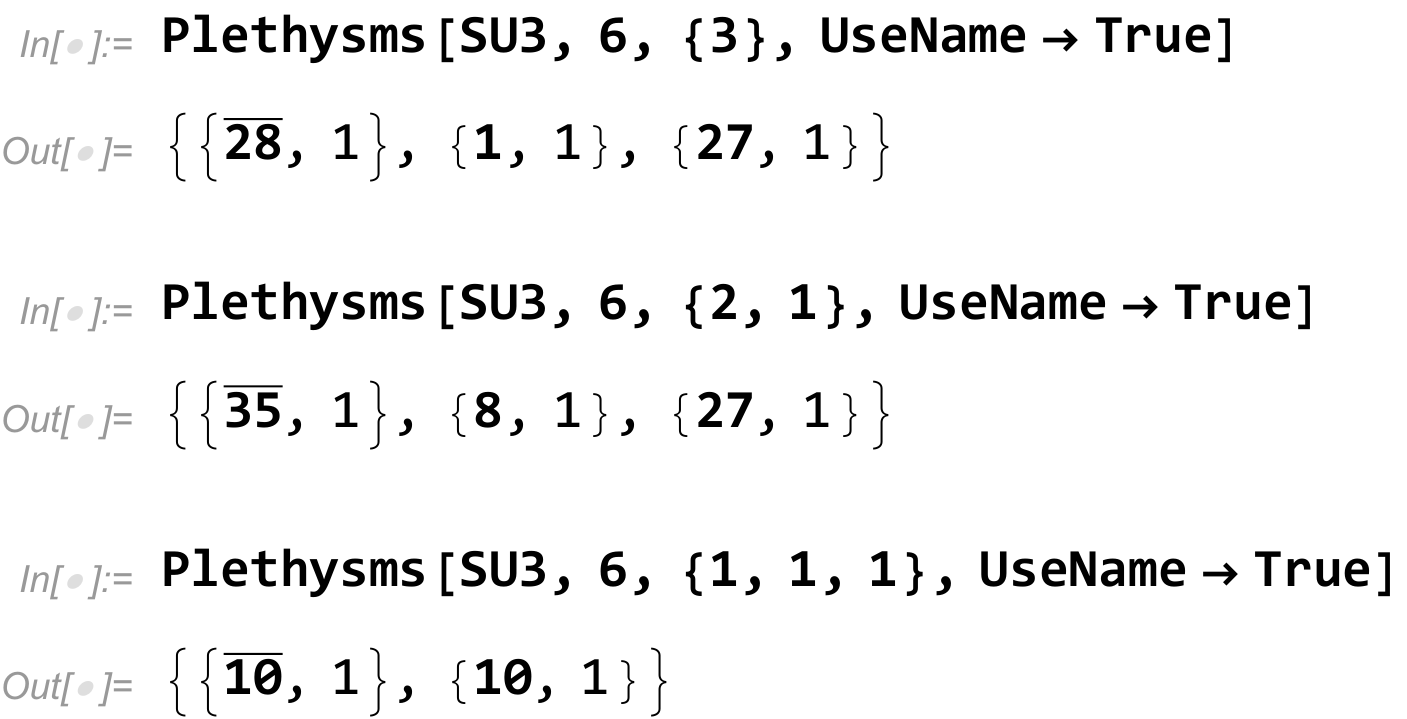}

\end{codeExample}

\noindent The user needs to provide a simple Lie group $G$, a representation
$R$ of that group, and a partition $\lambda$ of some number $m$,
and the output is a list of the representations of $G$ (with multiplicity)
contained in $R_{\lambda}^{m}$. The option \texttt{UseName -> True}
can be used to convert Dynkin coefficients into names of representations
in the output.

In practical situations, it might be necessary to consider (a) products
of different representations, with some of them repeated, and (b)
non-simple Lie groups. Relying on \texttt{Plethysms}, the function
\texttt{PermutationSymmetry} handles these more complicated cases,
being comparable in this respect to \texttt{ReduceRepProduct}.\index{ReduceRepProduct}
The difference between the two is that \texttt{Permutation\-Symmetry}
also calculates the relevant permutation symmetries; in other words,
it provides the decomposition of products of representations of a
semi-simple Lie group $G$ in irreducible representations of the group
$G\times S_{m}\times S_{m^{\prime}}\times\cdots$.\footnote{The cost of this extra information is a longer computation time.}
The value of the integers $m$, $m^{\prime}$ depends on the product
considered. For example, in the $SU(2)$ product of $\boldsymbol{2}\times\boldsymbol{2}\times\boldsymbol{3}\times\boldsymbol{3}$
the relevant permutation group is $S_{2}\times S_{2}$, while in $\boldsymbol{2}\times\boldsymbol{2}\times\boldsymbol{2}\times\boldsymbol{3}$
it is $S_{3}\times S_{1}$.

\index{PermutationSymmetry}

\begin{codeSyntax}
PermutationSymmetry[<Lie group>, <list of representations>]
\end{codeSyntax}

\begin{codeExample}

\hspace{-1mm}\includegraphics[scale=0.62]{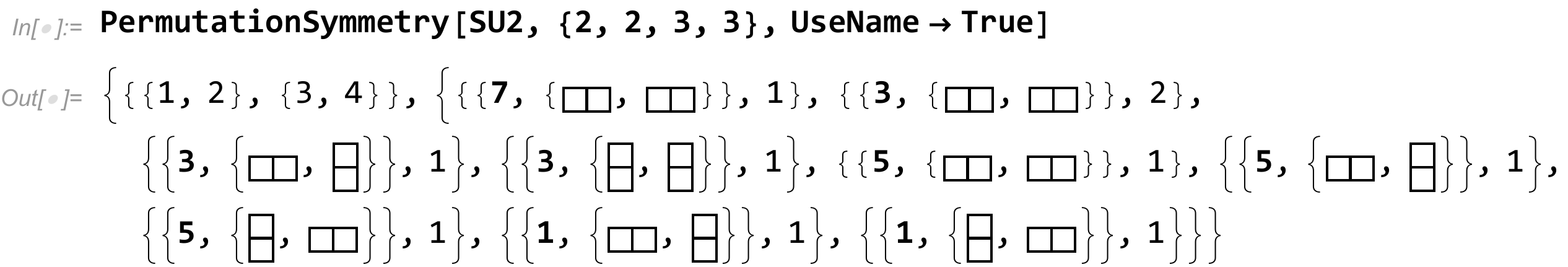}

\end{codeExample}

\begin{codeExample}

\includegraphics[scale=0.62]{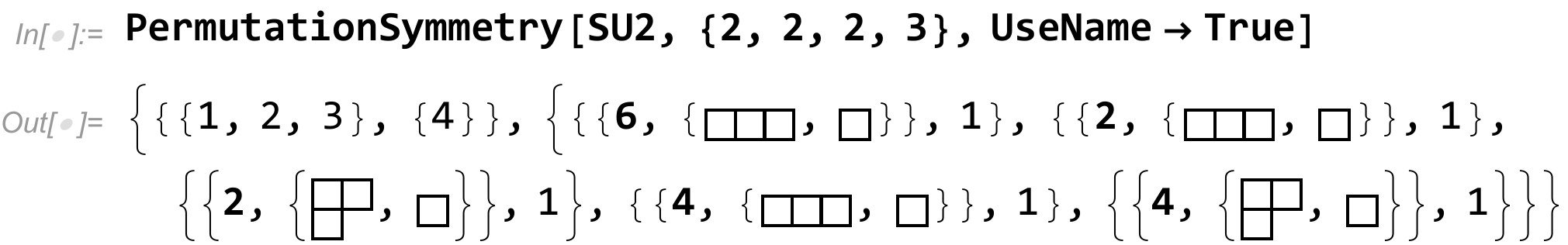}

\end{codeExample}

\begin{codeExample}

\includegraphics[scale=0.62]{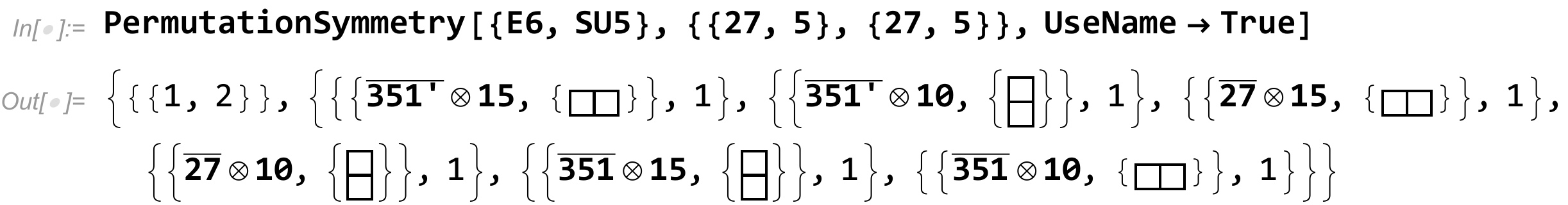}

\end{codeExample}

\noindent The output of the function consists of a list with two parts:
\texttt{\{<part1>, <part2>\}}. The content of the first part is rather
trivial: it identifies the position of repeated representations. For
example, in the product $\boldsymbol{2}\times\boldsymbol{2}\times\boldsymbol{3}\times\boldsymbol{3}$
the first and second representations are the same, and so are the
last two, hence the first part reads \texttt{\{\{1,2\},\{3,4\}\}}.
The second part of the output is the non-trivial one: it consists
on the list of the irreducible representations of the group $G\times S_{m}\times S_{m^{\prime}}\times\cdots$
in the format \texttt{\{<representation of G>,\{<partition of m>,<partition
of m'>,...\},<multiplicity>\}}. It is possible to use the option \texttt{UseName
-> True} to convert the representation of $G$ to a name and at the
same time draw Young diagrams for all the partitions.

The function \texttt{PermutationSymmetryOfInvariants} operates in
a similar fashion, except that it only displays the invariants under
$G$ in a product. Therefore the second part of the output of this
function does not contain the piece of information \texttt{<representation
of G>} mentioned earlier.

\index{PermutationSymmetryOfInvariants}

\begin{codeSyntax}
PermutationSymmetryOfInvariants[<Lie group>,<list of representations>]
\end{codeSyntax}

\begin{codeExample}

\includegraphics[scale=0.62]{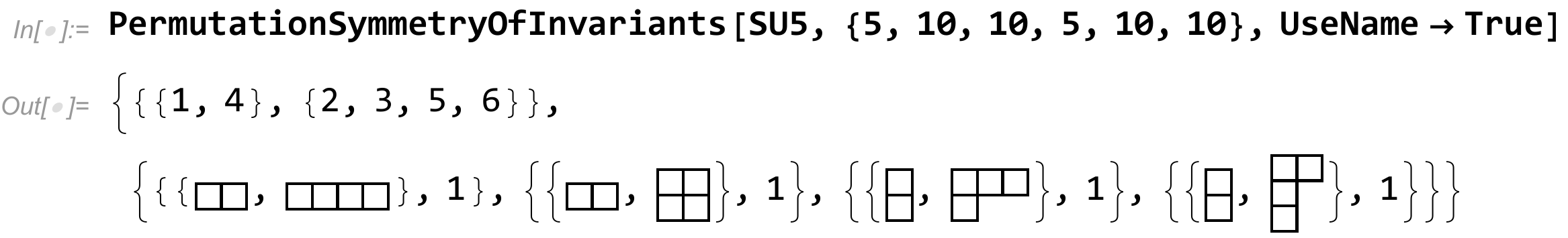}

\end{codeExample}

\subsection{\label{subsec:SU-n}The special unitary groups and $S_{m}$}

There is a particularly close relation between the special unitary
groups and the permutation groups. Indeed, it is possible to specify
a representation of $SU(n)$ with a Young tableaux, or equivalently
a partition $\lambda$ (which in turn can be associated to an irreducible
representation of a permutation group). That is because (a) the product
of $m$ copies of the fundamental representation $SU(n)$ with a permutation
symmetry $\lambda$ transforms irreducibly under $SU(n)$ and (b)
all irreducible representations of $SU(n)$ can be obtained in this
way.

As such, the irreducible representations of special unitary groups
can be labeled with either Dynkin coefficients or partitions, and
the conversion between the two notations can be performed with the
functions \texttt{Convert\-ToPartitionNotation} and \texttt{ConvertPartitionTo\-DynkinCoef}.
It is worth noting that to get Dynkin coefficients from a partition
it is necessary to know the value of the $n$ in $SU(n)$.

\index{ConvertToPartitionNotation}

\begin{codeSyntax}
ConvertToPartitionNotation[<SU(n) representation>]
\end{codeSyntax}

\begin{codeExample}

\includegraphics[scale=0.62]{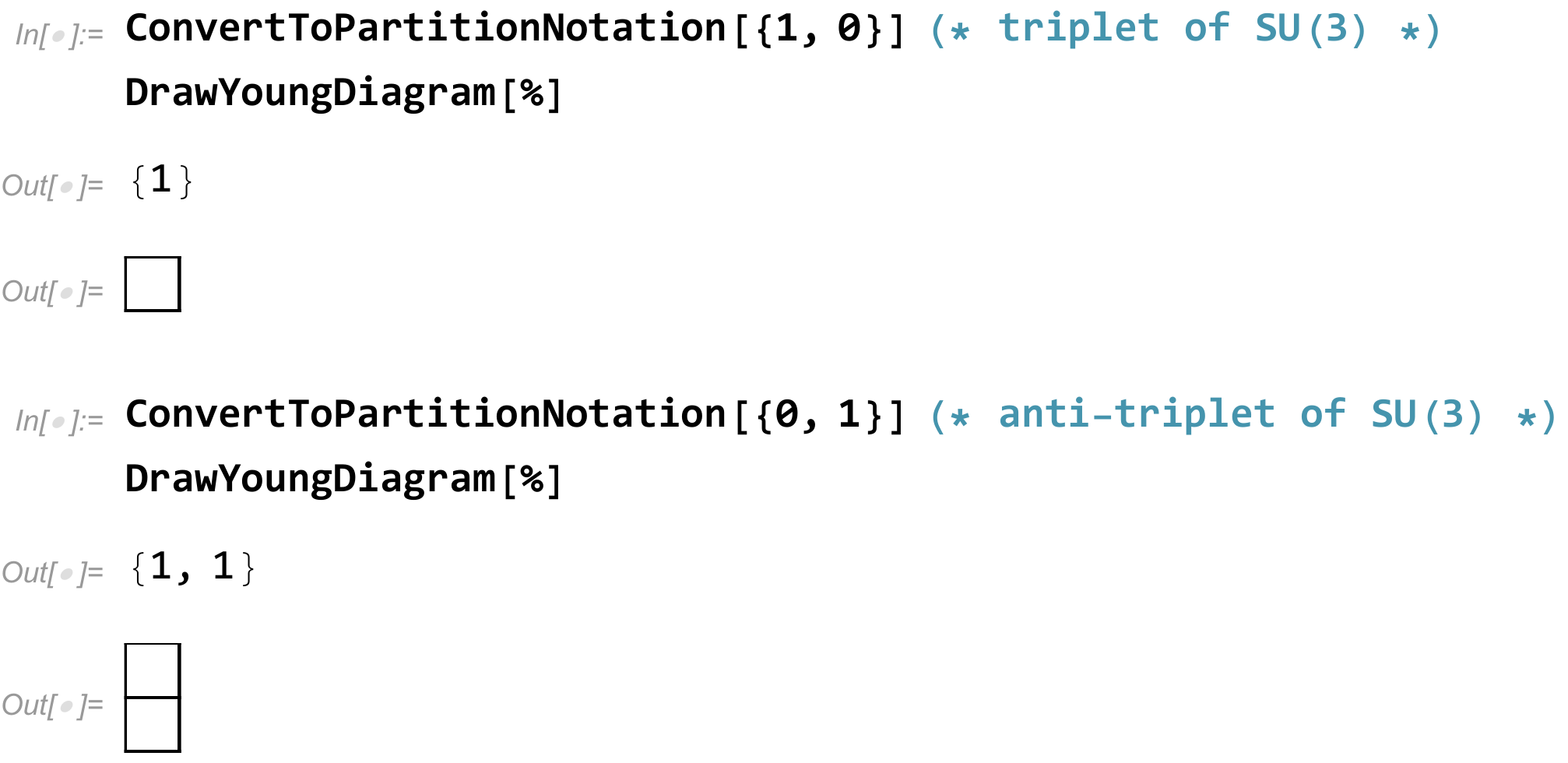}

\end{codeExample}

\index{ConvertPartitionToDynkinCoef}

\begin{codeSyntax}
ConvertPartitionToDynkinCoef[<n>,<partition>]
\end{codeSyntax}

\begin{codeExample}

\includegraphics[scale=0.62]{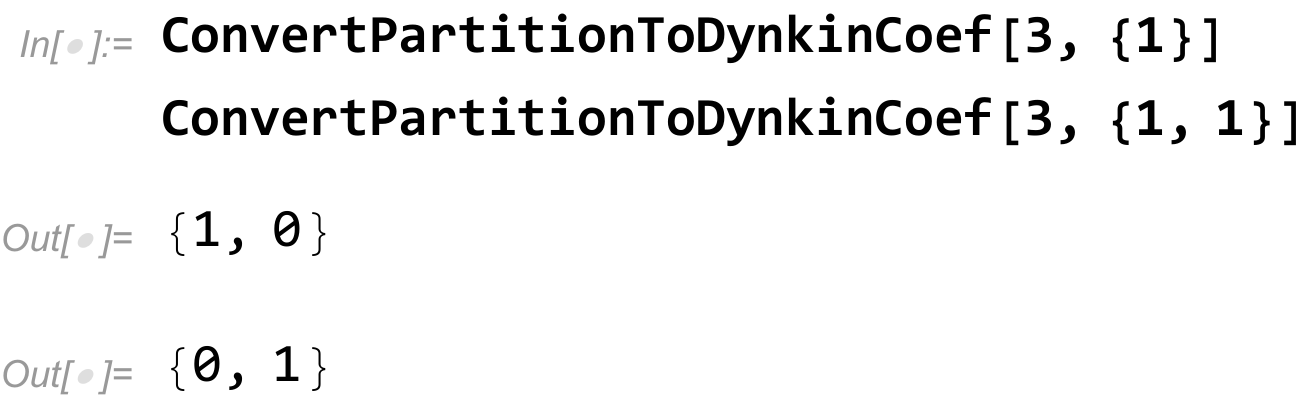}

\end{codeExample}

\section{\label{sec:Summary}Concluding remarks}

The various functions described here will hopefully constitute a useful
aid for performing computations with Lie and permutation groups. \texttt{GroupMath}'s
built-in documentation files include information on some details and
lesser important functions which were not mentioned in this text.

Future versions of the program are likely to include code for systematically
expressing $SU(n)$ invariants in the tensor formalism, using the
Levi-Civita $\epsilon$ and the Kronecker $\delta$ tensors, since
such a notation is simple and convenient. Another possibility under
consideration is the addition of a limited amount of code related
to finite groups beyond the ones describing permutations.

\section*{Acknowledgments}

I acknowledge the financial support from the Grant Agency of the Czech
Republic (GA\v{C}R) through contract number 20-17490S and from the
Charles University Research Center UNCE/SCI/013.

\appendix

\clearpage

\section{Brief review of Lie algebras}

\subsection{Simple, semi-simple and reductive Lie algebras}

A \textit{Lie algebra} $L$ is a vector space endowed with a binary
operation usually called the Lie bracket (or commutator) $[\cdot,\cdot]$
which is (i) linear, (ii) anti-symmetric and (iii) obeys the Jacobi
identity. The \textit{dimension} $d$ of the algebra is simply the
dimension of the vector space, so that all elements of $L$ can be
written as a linear combination of $d$ \textit{generators}.

Consider now the 2-dimensional Lie algebra $L$ generated by $x$
and $y$ and such that $\left[x,y\right]=x$. There is something peculiar
about this example: by taking the commutator among all elements of
$L$ we do not get back the full algebra $L$ but rather only the
elements proportional to $x$, so $\left[L,L\right]\neq L$. This
algebra is said to be \textit{solvable} because, starting with $L^{(1)}\equiv L$,
the sequence of algebras $L^{(i)}\equiv\left[L^{(i-1)},L^{(i-1)}\right]$
become equal to the trivial algebra $\left\{ 0\right\} $ for a large
enough value of $i$. In contrast to these cases, the \textit{semi-simple
Lie algebras} are those for which $\left[L,L\right]=L$. This definition
excludes elements $x\in L$ which commute with all the algebra (i.e.
$\left[x,L\right]=\left\{ 0\right\} $). Lie algebras $A$ exclusively
composed of such elements, $\left[A,A\right]=\left\{ 0\right\} $,
are called \textit{commutative} and they are associated to $U(1)^{d}$\textit{
abelian} groups (with $d$ being the dimension of the algebra $A$).
The set of \textit{reductive Lie algebras} consist of the semi-simple
Lie algebras and their direct sum with abelian algebras.

For any Lie algebra $L$ we must have $\left[L,L\right]\subset L,$
and any subspace $S$ of $L$ which also closes under the Lie bracket,
$\left[S,S\right]\subset S$, is a \textit{subalgebra}. Furthermore,
a subspace $I$ of $L$ with the property $\left[I,L\right]\subset I$
is called an \textit{ideal} (which is also a subalgebra). A non-abelian
Lie algebra $L$ is called \textit{simple} if it does not have any
ideals apart from the obvious ones: $L$ itself and $\left\{ 0\right\} $.
Semi-simple Lie algebras can be decomposed into a direct sum of simple
ones, hence they are the algebra of Lie groups $G_{1}\times G_{2}\times\cdots$
where the $G_{i}$ are called simple groups.

~

If is therefore enough to study simple Lie algebras in order to understand
the semi-simple and reductive ones. The rest of this appendix discusses
some properties of semi-simple Lie algebras which, without loss of
generality, can be seen as being simple.

\subsection{Cartan subalgebra, root vectors and roots.}

A maximum of $r$ of the algebra generators commute among themselves.
The vector space generated by them is called a \textit{Cartan subalgebra}
and the number $r$ is called the \textit{rank} of the algebra.

The remaining generators $e_{\alpha}$ (\textit{root vectors}) can
be organized in such a way that $\textrm{ad}\left(h\right)$ acts
diagonally on them for any $h$ in the Cartan subalgebra, meaning
that
\begin{equation}
\left[h,e_{\alpha}\right]=\alpha\left(h\right)e_{\alpha}
\end{equation}
for some number $\alpha\left(h\right)$. The function $\alpha$ itself
is called a \textit{root}. Since this function is linear, it is fully
determined by the values $\alpha\left(h_{i}\right)$ for some basis
$\left\{ h_{1},h_{2},\cdots,h_{r}\right\} $ of the Cartan subalgebra.
Therefore pragmatically one can identify a root with a list of $r$
numbers. The set of all roots forms the \textit{root system} $\Delta$.

\subsection{Positive and simple roots}

If $\alpha$ is a root so is $-\alpha$, therefore the root vectors
come in pairs $\left\{ e_{\alpha},e_{-\alpha}\right\} $. It is then
appropriate to distinguish positive from negative roots. This can
be done by selecting $r$ independent roots $\alpha_{i}$ and expressing
each root as a linear combination $c_{i}\alpha_{i}$; the \textit{positive
roots} are those for which the first non-zero $c_{i}$ coefficient
is positive. \textit{Simple roots} are positive roots which cannot
be expressed as the sum of two other positive roots, and there are
always as many of them as the rank $r$ of the algebra.

\subsection{Inner product of roots, the Cartan matrix and Dynkin diagrams}

Any root $\alpha$ can be associated to an element $h_{\alpha}$ of
the Cartan subalgebra through the relation
\begin{equation}
\alpha\left(h^{\prime}\right)\equiv\textrm{Tr}\left[\textrm{ad}\left(h^{\prime}\right)\textrm{ad}\left(h_{\alpha}\right)\right]=\sum_{\gamma\in\Delta}\gamma\left(h^{\prime}\right)\gamma\left(h_{\alpha}\right)
\end{equation}
which is valid for any element $h^{\prime}$ of the Cartan subalgebra.
An inner product $\left\langle \cdot,\cdot\right\rangle $ among roots
is often defined with this $\alpha\leftrightarrow h_{\alpha}$ correspondence:
\begin{equation}
\left\langle \alpha,\beta\right\rangle \equiv\beta\left(h_{\alpha}\right)\,.
\end{equation}
For a set of simple roots $\alpha_{i}$, the \textit{Cartan matrix}
$A$ has entries
\begin{equation}
A_{ij}=\frac{2\left\langle \alpha_{i},\alpha_{j}\right\rangle }{\left\langle \alpha_{j},\alpha_{j}\right\rangle }
\end{equation}
and it fully characterizes a complex simple Lie algebra $L$. It can
be represented graphically as a \textit{Dynkin diagram}. It turns
out that there is a limited number of possibilities for $A$ and consequently
for $L$: $SU(n)$ with $n\geq2$, $SO(n)$ with $n=5$ or $\geq7$,
$Sp(2n)$ with $n\geq3$, plus the exceptional Lie algebras $G_{2}$,
$F_{4}$, $E_{6}$, $E_{7}$ and $E_{8}$.

\subsection{Chevalley-Serre relations}

With the simple roots $\alpha_{i}$ we may define a particularly interesting
set of $3r$ algebra elements:
\begin{equation}
e_{i}\propto e_{\alpha_{i}}\,,\;f_{i}\propto e_{-\alpha_{i}}\,\textrm{ and }h_{i}\equiv\frac{2}{\left\langle \alpha_{i},\alpha_{i}\right\rangle }h_{\alpha_{i}}
\end{equation}
such that the following \textit{Chevalley-Serre relations} hold:
\begin{align}
\left[e_{i},f_{j}\right] & =\delta_{ij}h_{j}\,,\label{eq:Chevalley-Serre-1}\\
\left[h_{i},e_{j}\right] & =A_{ji}e_{j}\,,\\
\left[h_{i},f_{j}\right] & =-A_{ji}f_{j}\,.\label{eq:Chevalley-Serre-3}
\end{align}
Except for $SU(2)$, these $3r$ elements do not generate the full
algebra. However, the missing generators can easily be obtained by
repeatedly computing the commutator of the various $e$'s and $f$'s
above: $\left[e_{i},e_{j}\right]$, $\left[\left[e_{i},e_{j}\right],e_{k}\right]$,
..., and the same for the $f$'s.

\subsection{Representations, weights and Dynkin coefficients}

A representation $R$ is a linear and structure-preserving map which
associates a linear transformation over some vector space $\Phi$
to each Lie algebra element. The vector space is usually $\mathbb{C}^{n}$
and linear transformations over $\mathbb{C}^{n}$ are just $n\times n$
matrices, so it is often convenience to think of a representation
as a set of matrices (equal in number to the dimension $d$ of the
algebra). The integer $n$ is the \textit{dimension} of the representation
(assumed here to be finite).

If it is not possible to simultaneously block-diagonalize all matrices
of $R$ with a basis change, this representation is said to be \textit{irreducible}
(an \textit{irrep} for short); otherwise it is \textit{reducible}.
The latter are direct sums of irreps.

There is a particularly convenient way of labeling the $n$ components
of $\Phi$. All elements of the Cartan subalgebra can be represented
as diagonal matrices (in the same basis); for example the $h_{i}$
of the Chevalley-Serre relations act on some component $\omega$ of
$\Phi$ as follows:
\begin{equation}
R\left(h_{i}\right)\Phi^{\omega}=\omega\left(h_{i}\right)\Phi^{\omega}\,.
\end{equation}
Just like with roots, $\omega$ (called a \textit{weight}) can be
viewed as either a linear function or an $r$-dimensional vector.
It is then possible to assign a Cartan subalgebra element $h_{\omega}$
to each weight $\omega$ in analogy to the procedure used for roots,
and in this way the inner product $\left\langle \cdot,\cdot\right\rangle $
mentioned earlier can be extended to weights as well. The last equation
becomes
\begin{equation}
R\left(h_{i}\right)\Phi^{\omega}=2\frac{\left\langle \omega,\alpha_{i}\right\rangle }{\left\langle \alpha_{i},\alpha_{i}\right\rangle }\Phi^{\omega}\,.
\end{equation}
It should be noted that two or more components of $\Phi$ might be
associated to the same weight.

Weights can be sorted as follows. Each weight is written as a linear
combination of the simple roots $\alpha_{i}$ and then the coefficients
are compared: $\omega=c_{i}\alpha_{i}$ is considered to be higher/larger
than $\omega^{\prime}=c_{i}^{\prime}\alpha_{i}$ if the first non-zero
value of $c_{i}-c_{i}^{\prime}$ for $i=1,\cdots,r$ is positive;
otherwise the weight $\omega$ is lower/smaller than $\omega^{\prime}$.
With this understanding, there will always be a weight $\Lambda$
which is the highest in every irreducible representation, and it can
be used to unambiguously label it. In particular, irreps are often
specified in terms of their \textit{Dynkin coefficients}:
\begin{equation}
\Lambda_{i}=2\frac{\left\langle \Lambda,\alpha_{i}\right\rangle }{\left\langle \alpha_{i},\alpha_{i}\right\rangle }\,.
\end{equation}

\subsection{The Weyl group, orbits, and dominant weights}

For any representation, if $\omega$ is a weight so is
\begin{equation}
s_{\alpha}\left(\omega\right)\equiv\omega-2\frac{\left\langle \omega,\alpha\right\rangle }{\left\langle \alpha,\alpha\right\rangle }
\end{equation}
for any root $\alpha$. The $s_{\alpha}$ transformations, which can
be seen as reflecting the weight vectors $\omega$, generate a group
called the \textit{Weyl group} W. The \textit{orbit} of $\omega$
in W consists of the set of weights which can be obtained from $\omega$
by performing on it one or more reflections. Weights in the same orbit
are said to be \textit{conjugate} to one-another. It turns out that
in all Weyl group orbits there is a single \textit{dominant weight}
with non-negative Dynkin coefficients.

\renewcommand\indexname{List of GroupMath functions}\printindex{}

\clearpage{}

\end{document}